\documentclass[a4paper,11pt]{article}
\pdfoutput=1 

\usepackage{jheppub} 

\usepackage{dsfont}

\def\mc{\mathcal}
\def\beq{\begin{equation}}
\def\eeq{\end{equation}}
\def\beqn{\begin{eqnarray}}
\def\eeqn{\end{eqnarray}}
\def\MS{\mc M(\mc S)}
\def\dx{{\rm d}^2x}
\def\mf{\mathfrak}
\def\Tr{{\rm Tr}}
\def\X{\!\!&\times&\!\!}
\def\={\!\!&=&\!}
\def\+{\!\!&+&\!}
\def\-{\!\!&-&\!}
\def\eqq{\!\!&\equiv&\!}

\newcommand{\ntt}{${\mathcal N}=(2,2)\,$}
\newcommand{\ntwo}{${\mathcal N}=2\,$}

\title{\boldmath On exact correlation functions of chiral ring operators in  \boldmath{$2d$ $\mathcal{N}=(2, 2)$} SCFTs via localization}

\author{Jin Chen}

\affiliation{CAS Key Laboratory of Theoretical Physics, Institute of Theoretical Physics,\\ Chinese Academy of Sciences, Beijing 100190, China}

\emailAdd{jinchen@itp.ac.cn}

\abstract{We study the extremal correlation functions of (twisted) chiral ring operators via superlocalization in \ntt\ superconformal field theories (SCFTs) with central charge $c\geq 3$, especially for SCFTs with Calabi-Yau geometric phases. We extend the method in arXiv:\,1602.05971 with mild modifications, so that it is applicable to disentangle operators mixing on $S^2$ in nilpotent (twisted) chiral rings of $2d$ SCFTs. With the extended algorithm and technique of localization, we compute exactly the extremal correlators in $2d$ \ntt\ (twisted) chiral rings as non-holomorphic functions of marginal parameters of the theories. Especially in the context of Calabi-Yau geometries, we give an explicit geometric interpretation to our algorithm as the Griffiths transversality with projection on the Hodge bundle over Calabi-Yau complex moduli. We also apply the method to compute extremal correlators in K\"{a}hler moduli, or say twisted chiral rings, of several interesting Calabi-Yau manifolds. In the case of complete intersections in toric varieties, we provide an alternative formalism for extremal correlators via localization onto Higgs branch. In addition, as a spinoff we find that, from the extremal correlators of the top element in twisted chiral rings, one can extract chiral correlators in A-twisted topological theories.}


\arxivnumber{1712.01164}

\begin{document} 
\maketitle
\flushbottom

\section{Introduction}

Recently, a series of papers \cite{Papadodimas1, Papadodimas2, Papadodimas3, Komargodski} initiated a systematic study on the correlation functions of operators in chiral rings of four-dimensional \ntwo\ superconformal field theories with exactly marginal couplings. In these papers, the so-called extremal correlators, 
$$\left\langle\mathcal{O}_{1}(x_{1})\cdots\mathcal{O}_{n}(x_{n})\,\mathcal{\overline{O}}_J(y)\right\rangle_{\mathbb{R}^4}\,$$
containing arbitrarily many chiral primary and one anti-chiral primary operators in the chiral rings, are exactly calculated. Because of the insertion of an anti-chiral operator, these correlators are \emph{non-holomorphic} functions of the marginal couplings, and thus hard to compute in comparison of the chiral correlators $\langle\mathcal{O}_{1}(x_{1})...\mathcal{O}_{n}(x_{n})\rangle_{\mathbb{R}^4}$ with topological twist. On the other hand, these correlators are known to satisfy the four-dimensional version of $tt^*$-equations \cite{Papadodimas4}. The equations are, nevertheless, insufficient to determine them all as in the two-dimensional situation. Therefore people in \cite{Papadodimas1, Papadodimas2, Papadodimas3} resorted to additional input data via supersymmetric localization \cite{Pestun} on \ntwo\ gauge theories. With the technique of superlocalization, one is able to compute exact partition functions $\mc Z[S^4]$ on $4$-sphere of the \ntwo\ SCFTs with Lagrangian descriptions, from which the extremal correlation functions $\langle\mathcal{O}_{1}(x_{1})\cdots\mathcal{O}_{n}(x_{n})\,\mathcal{\overline{O}}_J(y)\rangle_{S^4}\,$ on $S^4$ can be extracted. In the paper \cite{Komargodski}, an algorithm was further developed to successfully disentangle the operators mixing from $S^4$ to $\mathbb{R}^4$. Therefore they are able to find all $\langle\mathcal{O}_{1}(x_{1})\cdots\mathcal{O}_{n}(x_{n})\,\mathcal{\overline{O}}_J(y)\rangle_{\mathbb{R}^4}\,$ on $\mathbb{R}^4$, which also solve the $tt^*$-equations automatically.

In this paper, we consider the extremal correlators 
$$\left\langle\mathcal{\phi}_{1}(x_{1})\cdots\mathcal{\phi}_{n}(x_{n})\mathcal{\bar{\phi}}_J(y)\right\rangle_{\mathbb{R}^2}\,$$
in the (twisted) chiral rings of two-dimensional $\mc N=(2,\,2)$ SCFTs with exactly marginal coupling constants $\{\tau\,,\bar{\tau}\}$. The fields $\phi$'s and $\bar{\phi}$'s are primary (twisted) chiral operators and their Hermitian conjugates.
Some of these correlators, e.g.\,the Zamolodchikov metrics, as well as the $tt^*$-equations they satisfy have been intensively studied in \cite{Cecotti, Vafa}, where the input data mainly concerned about OPE coefficients computed from topological twisted theories.  We here in this note will instead provide an alternative, analogue to the method in \cite{Komargodski}, to apply the $2d$ supersymmetric localization as the input data to exactly compute these extremal correlators, both in perturbative and non-perturbative series, \emph{with no need} of knowledge on OPE coefficients. 

Compared to $4d$ \ntwo\ SCFTs, the nilpotent (twisted) chiral rings in $2d$ \ntt\ SCFTs are finite and not freely generated. Therefore the OPE of (twisted) chiral primaries are related to each others due to the specific equivalence relations in the (twisted) chiral rings, and the products of sufficiently many of them will eventually turn out to be zero, modulo non-singular superdescendants. This feature will impose many constraints on the $2d$ correlators and their $tt^*$-equations. Therefore, while the methodology in this paper is inspired by and similar to the work of \cite{Komargodski}, we are still motivated to establish the algorithm applicable to disentangle operators mixing from $S^2$ to $\mathbb R^2$ for $2d$ nilpotent (twisted) chiral rings, and develop the full details of exact determination of the extremal correlators. Furthermore, the two-dimensional SCFTs we consider beautifully interplay with geometries and topology. A given $2d$ $\mc N=(2,\,2)$ SCFT $\mc S$, with center charge $c\geq 3$, usually has geometric phases related to a Calabi-Yau manifold $\mc Y$. Their moduli spaces $\mc{M}(\mc{Y})$ and $\mc{M}(\mc{S})$ coincides with each other. Therefore the extremal correlators exactly encodes the information of the metrics of $\mc{M}(\mc{Y})$ and various vector bundles over it. One will see that, from the mathematical side, our algorithm developed in this paper admits a geometric interpretation as Griffiths transversality, and also reconstructs the $tt^*$-equations on Calabi-Yau complex moduli. Furthermore, via localization onto Higgs branch, we also relate the extremal correlators of a theory $\mc Y$ to the periods of its mirror $\widetilde{\mc Y}$ in the case of complete intersections in toric varieties. We wish that the exact computation of the extremal correlators would lead a detailed investigation on the structures of partition functions as well as the extremal correlators, integrability in $2d$ \ntt\ SCFTs, test of resurgence, and provide further implications to $2d/4d$ correspondence and so forth.

We plan the rest of the paper as follows. In section \ref{2} we review some basics on $\mc N=(2,\,2)$ SCFTs, their (twisted) chiral rings, and $tt^*$-equations the extremal correlators have to satisfy. In section \ref{3}, we review the method of supersymmetric localization on $S^2$ for SCFTs with irrelevant operator deformations, and establish the main algorithm to disentangle operators mixing from $S^2$ to $\mathbb R^2$. In section \ref{4} we explain how the algorithm could naturally arise as Griffiths transversality in Calabi-Yau complex moduli. We also use this observation to reconstruct $tt^*$-equations and constraints that the extremal correlators have to satisfy for chiral rings containing only marginal generators. At last, in section \ref{5}, we apply the method to several interesting Calabi-Yau manifolds and compute their extremal correlators in twisted chiral rings as well as chiral correlators in their A-twisted topological theories as a byproduct. We also provide a different formulation of these correlators via localization onto Higgs branch in the case of complete intersections in toric varieties.\\\\

\section{Preliminaries}
\label{2}
\subsection{Chiral rings in $\mc N=(2,\,2)$ SCFTs}
\indent We start from recalling some properties of \ntt\ superconformal algebra. Our notation follows the paper \cite{Boer}. In an Euclidean \ntt\ SCFT, we have left moving currents, $T(z),\,G^{\pm}(z)$ and $J(z)$, and right ones, $\bar{T}(\bar{z}),\,\bar{G}^{\pm}(\bar{z})$ and $\bar{J}(\bar{z})$, corresponding to the holomorphic and anti-holomorphic part of energy momentum tensor, supercurrents and $U(1)$ R-currents respectively. We from now on focus on the holomorphic part and NS sectors of the \ntt\ SCFTs. Among these operators, of particular importance is the anticommuting algebra of supercurrents:
\beq
\left\{G^-_r,\,G^+_s\right\}=2L_{r+s}-\left(r-s\right)J_{r+s}+\frac{c}{3}\left(r^2-\frac{1}{4}\right)\delta_{r+s,0}\,,
\label{superalgebraG}
\eeq
where $L_n$, $J_m$ and $G^{\pm}_{r}$ are the modes expansion of currents $T(z)$, $J(z)$ and $G^{\pm}(z)$. For any states $|\phi\rangle$, unitarity requires
\beq
\left\|G^{\pm}_{\pm 1/2}|\phi\rangle\right\|\geq 0\,.
\eeq
By superalgebra, the conformal weight $h_\phi$ is bounded by its R-charge $q_\phi$ 
\beq
h_\phi\geq\frac{1}{2}\left|q_\phi\right|
\eeq
The (anti)chiral primary states are those states saturating the above inequality. We define them as follows:
$$\emph{chiral primary states:}\ \ \ L_n|\phi_c\rangle=G^{\pm}_{n-1/2}|\phi_c\rangle=G^{+}_{-1/2}|\phi_c\rangle=0\ \ {\rm for}\ n\geq 1\,$$
\beq
\emph{antichiral primary states:}\ \ \ L_n|\phi_a\rangle=G^{\pm}_{n-1/2}|\phi_a\rangle=G^{-}_{-1/2}|\phi_a\rangle=0\ \ {\rm for}\ n\geq 1\,.
\eeq
With the aid of superconformal algebra, one can easily derive, from 
$$\left\{G^-_{1/2},\,G^+_{-1/2}\right\}|\phi_c\rangle=\left(2L_0-J_0\right)|\phi_c\rangle=0\,,$$
$$\left\{G^-_{-1/2},\,G^+_{1/2}\right\}|\phi_a\rangle=\left(2L_0+J_0\right)|\phi_a\rangle=0\,.$$
that the conformal dimension and $U(1)$ R-charge of any (anti)chiral primary states are related by
\beq
h_{\phi_c}=\frac{1}{2}q_{\phi_c},\ \ \ h_{\phi_a}=-\frac{1}{2}q_{\phi_a}\,.
\eeq
Besides, the unitarity requires further
\beq
\left\|G^+_{-3/2}|\phi_c\rangle\right\|\geq 0,\ \ \ \left\|G^-_{-3/2}|\phi_a\rangle\right\|\geq 0\,.\eeq
These two inequalities constrains the conformal dimension of (anti)chiral primary states
\beq
h\leq \frac{c}{6}\,.
\eeq
This bound fundamentally distinguishes the $2d$ chiral ring structure from that in $4d$, say the number of chiral ring operators is \emph{finitely} many in 2d \ntt\ SCFTs.

We next consider the OPE of chiral primary operators $\phi_i(z)$, which is associated to the chiral primary states $|\phi_i\rangle$ due to operator-state correspondence. In general OPE, one has to worry about the appearance of singularities when one operator $\phi_i(z)$ is approaching another $\phi_j(0)$,
\beq
\phi_i(z)\phi_j(0)\sim \sum_k\frac{C_{ij}^k}{z^{h_i+h_j-h_{k}}}\mc O_k(0)\,.\eeq
However, for the OPE of two chiral primary fields, their additive $U(1)$ R-charge guarantees that their OPE is actually non-singular and the leading constant coefficient terms must be also chiral primary fields \cite{Vafa2}, i.e.
\beq\phi_i(z)\phi_j(0)=\sum_k{C_{ij}^k}\phi_k(0)+{\rm nonsingular\ superdescendants}\,,\eeq
where $C_{ij}^k$ is the $z$-independent OPE coefficient. Therefore, modulo secondary fields, the chiral primary fields $\{\phi_i\}$ have a ring structure respect to their OPE coefficient, and form the so-called chiral ring $\mc R_z$. Since we have argued the number of chiral primary fields is finite, the chiral ring $\mc R_z$ is finitely generated but \emph{nilpotent.}. It is crucially different from the structure of $4d$ chiral ring, which is finitely and freely generated. This difference will be explicitly elucidated later when we compute the correlators of chiral primary fields.

One can also define antichiral rings $\overline{\mc R}_z$ in holomorphic sector in a similar fashion, as well as (anti)chiral rings $\mc R_{\bar z}$ ($\overline{\mc R}_{\bar z}$) in anti-holomorphic sector. For a non-chiral CFT, all states must be tensor products of holomorphic and anti-holomorphic sectors. We thus end up with four different combinations to define the (twisted) chiral primary fields and their hermitian conjugates, i.e.
$$\emph{(anti)chiral primary fields:}\ \  \phi_i\in\mc R_z\otimes \mc R_{\bar z},\ \ \bar{\phi}_i\in\overline{\mc R}_z\otimes \overline{\mc R}_{\bar z}$$
\beq\emph{twisted (anti)chiral primary fields:}\ \  \sigma_a\in\mc R_z\otimes \overline{\mc R}_{\bar z},\ \ \bar{\sigma}_a\in\overline{\mc R}_z\otimes \mc R_{\bar z}\,,\eeq
where we somewhat abuse the notation $\phi_i$ and the name ``chiral" that should not be confused with those defined in holomorphic sector.

Throughout the paper, theories we consider will only contain scalar operators in their (twisted) chiral rings. Therefore, the conformal weight $(h,\,\bar h)$ of $\phi$ and $(\tilde h,\,\bar{\tilde h})$ of $\sigma$ must obey
\beq
h=\bar h,\ \ \ \ \tilde h=\bar{\tilde h}\,.\eeq
On the other hand, for the $U(1)$ R-charge $(q,\,\bar q)$ of $\phi$ and $(\tilde q,\,\tilde{\bar q})$ of $\sigma$, we have
\beq
h=\frac{q}{2},\ \ \bar h=\frac{\bar q}{2};\ \ \ \tilde h=\frac{\tilde q}{2},\ \ \bar{\tilde h}=-\frac{\bar {\tilde q}}{2}\,.
\eeq
Therefore it is convenient to define the so-called $U(1)_V$ and $U(1)_A$ currents as linear combination of $J(z)$ and $\bar J(\bar z)$
$$J_V=J(z)+\bar J(\bar z)$$
\beq
J_A=J(z)-\bar J(\bar z)\,,\eeq
associated to the $U(1)_V$-charge $q_V$ and $U(1)_A$-charge $q_A$. Under the V-A notation, we see that the (twisted) chiral primary fields $\phi$ and $\tilde\phi$ have
$$\Delta_\phi=h+\bar h=\frac{q+\bar q}{2}=\frac{q_V}{2},\ \ \ q-\bar q=q_A=0$$
\beq\Delta_{\sigma}=\tilde h+\bar{\tilde h}=\frac{\tilde q-\bar{\tilde q}}{2}=\frac{q_A}{2},\ \ \ \tilde q+\bar{\tilde q}=q_V=0\,,\eeq
where $\Delta$ denotes the dimension of operators. In the language of field theories with Lagrangian description, we give the following important examples as (twisted) chiral primary fields: The \ntt\ chiral multiplet with dimension one, 
\beq\Phi=\left(\phi,\,\psi,\,\mc O\,\right)\,,\eeq
has its bottom component $\phi$ as a chiral primary field with charge $(q_V,q_A)=(2,0)$ and dimension $\Delta=1$. Its top component $\mc O$ has neutral R-charge and dimension 2, thus serves as a \emph{marginal} secondary field to perturb the CFT. Similarly for a \ntt\ twisted chiral multiplet, for example the field strength of $U(1)$ vector multiplet,
\beq\Sigma=\left(\sigma,\,\tilde\lambda,\,\widetilde{\mc O}\,\right)\,,\eeq
its bottom component $\sigma$ is a twisted chiral primary field with charge $(0,2)$ and dimension 1, and top component $\widetilde{\mc O}$ with neutral R-charge and dimension 2 is also marginal to perturb the CFT.\\

\subsection{Conformal manifolds of $\mc N=(2,\,2)$ SCFTs}
Now we turn to discuss how to perturb a CFT by its marginal operators. Suppose that for a $d$ dimensional CFT $\mc S_0$, there are exactly marginal operators $\mc O_i$. One can use these operators to deform the original theory $\mc S_0$ to
\beq\mc S_0\longrightarrow \mc S\equiv\mc S_0+\lambda^i\!\!\int{\rm d}^dx\,\mc O_i\,,\eeq
where $\{\lambda^i\}$ are exactly marginal couplings. Since the operators are exactly marginal, the coupling constants $\lambda^i$ are all dimensionless and their $\beta$-functions vanish,
\beq\beta_{\lambda^i}=0\,.\eeq
Therefore the deformed theory $\mc S$ is still conformal. We then in fact consider a family of CFTs, parametrized by the exactly marginal couplings $\{\lambda^i\}$. Put in other words, the conformal theory $\mc S$ has a moduli space, a.k.a the conformal manifold $\mc M(\mc S)$, whose local coordinates are $\{\lambda^i\}$. One can further define the Zamolodchikov metric $g_{ij}$ \cite{Boer} on $\mc M(\mc S)$ via the correlators of $\mc O_i$,
\beq\langle\mc O_i(x)\mc O_j(y)\rangle_{\mathbb{R}^2}=\frac{g_{ij}(\lambda)}{|x-y|^{2d}}\,,\eeq
where we evaluate the correlation function in the CFT with couplings $\{\lambda^i\}$. 

In the case of $\mc N=(2,\,2)$ SCFTs we consider, there are two types of exactly marginal operators (and their hermitian conjugates): the top components $\mc O_i$ of chiral primary multiplet $\Phi_i$ and $\mc{\widetilde O}_a$ of twisted chiral primary multiplet $\Sigma_a$. We formulate the marginal deformation in superspace,
\beqn
\mc S\equiv\mc S_0+\tau^i\!\!\int\!\!{\rm d}^2x\,{\rm d}^2\theta\,\,\Phi_i+\tilde\tau^a\!\!\int\!\!{\rm d}^2x\,{\rm d}^2\tilde\theta\,\,\Sigma_a+{\rm c.c.}\,
\label{SCFT}
\eeqn
where ${\rm d}^2\theta$ (${\rm d}^2\tilde\theta$) is the measure of (twisted) chiral sub-superspace. It is known \cite{Gomis1, Gomis2} that the moduli space of \ntt\ SCFTs is locally a direct product of two K\"{a}hler manifolds spanned by the chiral and twisted chiral descendants $\mc O_i$ and $\mc{\widetilde O}_a$,
\beq\mc M(\mc S)\simeq\mc M_{\rm c}(\tau, \bar\tau)\times\mc M_{\rm tc}(\tilde\tau, \bar{\tilde\tau})\,.\eeq
The corresponding Zamolodchikov metrics can be found by computing the correlators
\beq\left\langle\mc O_i(x)\mc{\overline{O}}_j(y)\right\rangle_{\mathbb{R}^2}=\frac{g_{i\bar{j}}(\tau, \bar\tau)}{|x-y|^{4}}\,,\,\,\,\,\,\left\langle\mc{\widetilde{O}}_a(x)\mc{\overline{\widetilde{O}}}_b(y)\right\rangle_{\mathbb{R}^2}=\frac{\tilde{g}_{a\bar{b}}(\tilde\tau, \bar{\tilde\tau})}{|x-y|^{4}}\,.\eeq
Or instead, noticing that
$$\mc O_i=\frac{1}{2}G^-_{-1/2}\bar{G}^-_{-1/2}\,\phi_i,\ \ \ \overline{\mc O}_j=\frac{1}{2}G^+_{-1/2}\bar{G}^+_{-1/2}\,\bar{\phi}_j\,;$$
\beq\widetilde{\mc O}_a=\frac{1}{2}G^-_{-1/2}\bar{G}^+_{-1/2}\,\sigma_a,\ \ \ \overline{\widetilde{\mc O}}_b=\frac{1}{2}G^+_{-1/2}\bar{G}^-_{-1/2}\,\bar{\sigma}_b\,;\eeq
by conformal Ward identities \cite{Boer}, we can directly evaluate the correlators of (twisted) chiral primary fields,
\beq
\left\langle\mc \phi_i(x)\mc{\bar{\phi}}_j(y)\right\rangle_{\mathbb{R}^2}=\frac{g_{i\bar{j}}(\tau, \bar\tau)}{|x-y|^{2}}\,,\,\,\,\,\,\left\langle\sigma_a(x)\bar{\sigma}_b(y)\right\rangle_{\mathbb{R}^2}=\frac{\tilde{g}_{a\bar{b}}(\tilde\tau, \bar{\tilde\tau})}{|x-y|^{2}}\,,
\eeq
where the ``$1/2$" is to normalize the superalgebra to avoid unwanted numerical factors.

Let us briefly remark that, by a simple dimension count, the operator $\phi_i$ ($\sigma_a$) has conformal weight $(\frac{1}{2},\,\frac{1}{2})$. The unitarity bound for chiral ring elements requires the center charge of our SCFTs, $c\geq 3$. Equivalently, only \ntt\ SCFTs with $c\geq 3$ have exactly marginal operators and thus moduli spaces. Throughout the paper, we only discuss theories subject to this condition, and require all chiral operators having \emph{integer} dimensions. Therefore the correlators $\left\langle\phi_i(x){\bar{\phi}}_j(y)\right\rangle_{\mathbb{R}^2}$ as well as $\left\langle\sigma_a(x)\bar\sigma_b(y)\right\rangle_{\mathbb{R}^2}$ are the first nontrivial extremal correlation functions to compute. The operators $\phi_i$'s ($\sigma_a$'s) are also the first non-trivial elements with lowest conformal weight in the (twisted) chiral rings of \ntt\ SCFTs. We will review more details of the ring structure right soon.\\

\subsection{Extremal correlators}
As we have seen that the Zamolodchikov metric is one of interesting objects to compute, we can in fact consider more general ``extremal correlators" in chiral rings (all discussions below equally work for twisted chiral rings).  These are correlators of the form
\beq\left\langle\phi_{1}(x_{1})...\phi_{n}(x_{n})\,\bar{\phi}_J(y)\right\rangle_{\mathbb{R}^2}\,\eeq
where $\phi_i$ are chiral primaries and $\bar{\phi}_J$ is antichiral. The selection rule respect to $U(1)_V$ symmetry requires the above correlator vanish unless 
\beq
q_J=-\sum_i q_i,\ \ \ {\rm or\ \ equivalently}\ \ \ \Delta_J=\sum_i \Delta_i\,.
\label{selection rule}
\eeq
In comparison of chiral correlators, which contain only chiral operators and holomorphically depend on marginal parameters $\{\tau\}$, the extremal correlators are in general non-holomorphic functions of marginal couplings $\{\tau, \bar{\tau}\}$. The main part of this paper is devoted to compute these extremal correlators both perturbatively and non-perturbatively.

To compute the extremal correlators, it is instrument to apply a standard conformal transformation, 
\beq
x'_i=\frac{x_i-y}{|x_i-y|^2}\,.\eeq
to hide the coordinates $``y"$ of the antichiral field to $``\infty"$,
\beq\left\langle\phi_{1}(x_{1})\cdots\phi_{n}(x_{n})\,\bar{\phi}_J(y)\right\rangle_{\mathbb{R}^2}=\frac{\left\langle\phi_{1}(x'_{1})\cdots\phi_{n}(x'_{n})\,\bar{\phi}_J(\infty)\right\rangle_{\mathbb{R}^2}}{|x_{1}-y|^{2\Delta_{1}}\cdots|x_{n}-y|^{2\Delta_{n}}}\,.\eeq
Next, one can show the numerator is actually spacetime independent. Notice that, by acting $\partial_{z^\prime_i}$ on $\left\langle\phi_{1}(x'_{1})\cdots\phi_{n}(x'_{n})\,\bar{\phi}_J(\infty)\right\rangle_{\mathbb{R}^2}$,
\beq\partial_{z^\prime_i}\phi_i(x^\prime_i)=G^+_{-1/2}G^-_{-1/2}\phi_i(x^\prime_i)\,.\eeq
By superconformal Ward identity, one can rewrite $G^+_{-1/2}$ acting on each of other  operators. $G^+_{-1/2}$ annihilates all other chiral primaries, while, acting on $\bar\phi_J(y)$, the correlator
\beq\left\langle\phi_{1}(x'_{1})\cdots\phi_{n}(x'_{n})\,G^+_{-1/2}\bar\phi_J(y)\right\rangle\eeq 
decays as $|y|^{-2\Delta_J-1}$. Therefore, when putting $``y"$ to infinity, this contribution will decay as $|y|^{-1}$ to zero as well. Overall we single out the spacetime dependent part from the extremal correlator, and show that
\beq\left\langle\phi_{1}(x_{1})\cdots\phi_{n}(x_{n})\,\bar{\phi}_J(\infty)\right\rangle_{\mathbb{R}^2}=\left\langle\phi_{1}(0)\cdots\phi_{n}(0)\,\bar{\phi}_J(\infty)\right\rangle_{\mathbb{R}^2}\,,\eeq
only depends on the marginal couplings $\{\tau, \bar{\tau}\}$.

Now one can apply OPE to these chiral primaries,
\beq\phi_{1}(0)\cdots\phi_{n}(0)=C^i_{12}C^j_{i3}\cdots C^k_{ln}\phi_k(0)\equiv\phi_I(0)\,\eeq 
modulo secondary operators which will not contribute to the correlator, and have
\beq\left\langle\phi_{1}(x_{1})\cdots\phi_{n}(x_{n})\,\bar{\phi}_J(y)\right\rangle_{\mathbb{R}^2}=\frac{\left\langle\phi_I(0)\,\bar{\phi}_J(\infty)\right\rangle_{\mathbb{R}^2}}{|x_{1}-y|^{2\Delta_{1}}\cdots|x_{n}-y|^{2\Delta_{n}}}\equiv\frac{g_{I\bar{J}}(\tau,\,\bar\tau)}{|x_{1}-y|^{2\Delta_{1}}\cdots|x_{n}-y|^{2\Delta_{n}}}\,.\eeq
Therefore, similar to Zamolodchikov metric, the computation of extremal correlators is to determine the $g_{I\bar J}$, which is referred as ``chiral ring data" of the SCFT \cite{Komargodski}.\\

\subsection{$tt^*$-geometries}
In this subsection, we will briefly review $tt^*$-equations of chiral ring data. For more details and derivations, we refer readers to \cite{Boer, Vafa}.

Given a chiral ring $\mc R$, we can grade it by the dimensions or R-charges of the operators in it,
\beq\mc R=\bigoplus^{N=c/3}_{\Delta_I=0}\mc R_I\,.\eeq
Since we work in NS sector, the vacuum state is unique, or say $\mc R_0$  contains only the unit operator $\mathds{1}$. As required before, the next level $\mc R_1$ contains chiral primaries with dimension $\Delta_1=1$, whose descendants gives the marginal operators to span the moduli space $\mc M(\mc S)$ of SCFT $\mc S$. $\mc R_I$ contains chiral primaries with dimension $\Delta_I=I$, and so on. At last the top sub-ring $\mc R_N$ also  contains only one operator with the highest dimension $c/3$ \cite{Vafa2}.

From the geometric perspective, one can interpret the (anti)chiral primaries $\phi_i$ and $\bar{\phi}_j$ in $\mc R_1$ and $\overline{\mc R}_1$ as sections on tangent bundle $\mc T\mc M(\mc S)$. Their correlator
\beq\left\langle\phi_i(0)\,\bar{\phi}_j(\infty)\right\rangle_{\mathbb{R}^2}=g_{i\bar{j}}(\tau,\bar{\tau})\eeq
designates a Hermitian metric on $\mc T\mc M(\mc S)$. Similarly operators $\phi_I$ and $\bar{\phi}_J$ living in $\mc R_I$ and $\overline{\mc R}_J$ can be also understood as sections living on certain vector bundles $\mc V_I$ and $\mc V_J$ over moduli space $\mc M(\mc S)$. The extremal correlators
\beq\left\langle\phi_I(0)\,\bar{\phi}_J(\infty)\right\rangle_{\mathbb{R}^2}=g_{I\bar{J}}(\tau,\bar{\tau})\,\delta_{\Delta_I\Delta_{\bar J}}\,,\eeq
analogously define Hermitian metrics on various bundle $\mc V_I$'s. Here the appearance of $\delta_{\Delta_I\Delta_{\bar J}}$ is imposed by the selection rule (\ref{selection rule}), which implies the total vector bundle 
\beq\mc V=\bigoplus^{N=c/3}_{\Delta_I=0}\mc V_I\eeq
is also graded by dimensions of the operators.

Now we are ready to discuss the $tt^*$-equations. Roughly speaking, $tt^*$-equations interpolate metrics $g_{I\bar J}$ defined on different bundles $\mc V_I$'s via the OPE coefficients of these (anti)chiral primaries. More specifically, let us consider the metric $g_{I\bar J}$ varied along certain direction $\tau^i$ (or $\bar{\tau}^{\bar j}$) of the moduli space $\MS$. It is equivalent, from the action (\ref{SCFT}), to compute 
\beq
\delta_{\tau}\,g_{I\bar J}\approx\delta\tau^i\partial_i g_{I\bar J}=\left\langle\delta\tau^i\int\!\!\dx\,\mc O_i(x)\phi_I(0)\bar{\phi}_J(\infty)\right\rangle\,.\eeq 
However the correlator is divergent when evaluating the integration. Therefore one has to renormalize it by adding counter terms. This process might lead the computation with ambiguities. It can be shown \cite{Zwiebach} that the renormalization process is equivalent to introduce connections $\nabla_\tau$ and $\nabla_{\bar{\tau}}$ on the vector bundle $\mc V$, and the variation of correlators along moduli space has to be modified as
\beq\delta_{\tau}\,g_{I\bar J}=\delta\tau^i\nabla_i g_{I\bar J}=\left\langle\delta\tau^i\int\!\!\dx\,\mc O_i(x)\phi_I(0)\bar{\phi}_J(\infty)\right\rangle_{\rm renormalized}\,.\eeq
In this sense, the physical process of renormalization in fact reflects non-trivial geometries of the moduli space $\MS$ and the bundle $\mc V$ over it.

The geometries are encoded in $tt^*$ equations which determine the curvature of the vector bundle $\mc V$ via the dynamics of the SCFT. In concrete, one can establish the equations
\beq
\left\lbrack\,\nabla_i,\, \nabla_{\bar j}\,\right\rbrack^K_{\ L}=g_{i\bar{j}}\delta^K_{\ L}\left(1-\frac{6}{c}\Delta_K\right)-\left[\,C_i,\, \bar{C_j}\,\right]^K_{\ L}\,.
\label{tt}
\eeq
The $C_i$ and $\bar{C}_j$ should be understood as OPE coefficient in matrix form, i.e.
\beq
\left(C_i\right)^K_{\ L}\equiv C^K_{iL}\,,\ \ \ \left(\bar{C}_j\right)_{\ L}^{K}\equiv g_{L\bar{N}}\bar{C}^{\bar N}_{\bar j \bar M}g^{\bar M K}\,,\eeq
where indexes $i,\,\bar j$ run for marginal operators, and $g^{\bar M L}$ stands for inverse of the metric. The $tt^*$-equations here is derived in NS sector \cite{Boer}, different from that in Ramond sector \cite{Vafa} by the first diagonal piece. We will come back in later section to comment more on this term as a matter of normalization, see section \ref{4.1}.

To see how the $tt^*$-equations relate metrics in various bundle $\mc V_I$, one can choose a holomorphic gauge as
\beq
\nabla_i=\partial_i-\left(\partial_i g_{I\bar J}\right)g^{\bar J K}\,,\ \ \ \nabla_{\bar{j}}=\partial_{\bar j}\,.\eeq
The holomorphic gauge can be always achieved and thus the metrics of the vector bundle $\mc V$ are constrained via $tt^*$-equations (\ref{tt}). The metrics, or say the chiral ring data, are  solutions to the $tt^*$-equations. Nevertheless in the paper we will not solve them from the equations. Instead, in next section, we will show that these chiral ring data can be directly computed via supersymmetric localization, and the results will automatically solve the $tt^*$-equations.

In the end of this subsection, for completeness, we would like to make some remarks. Above discussion on chiral rings can be identically repeated to twisted chiral rings. For \ntt\ SCFTs, the correlator of (anti)chiral and twisted (anti)chiral primaries always vanishes even when they have same dimensions, because their R-charges are different. It thereby implies the factorization of the moduli space. However this result breaks down in $\mc N=(4,4)$ SCFTs, whose moduli space is neither factorisable, nor even K\"{a}hlerian. More details on this issue are discussed in \cite{Seiberg2}.\\\\

\section{Chiral ring data via superlocalization on \boldmath $S^2$}
\label{3}
In this section, we will establish an algorithm to systematically compute (twisted) chiral ring data\footnote{In this section, we actually will consider twisted chiral primaries in details. All discussions on chiral primaries equally work for twisted chiral primaries, and vice versa.} $g_{I\bar J}$ of \ntt\ SCFTs with UV Lagrangian descriptions in the fashion of \cite{Komargodski}. The general idea is sketched as follows: For a chiral ring, $\mc R=\bigoplus_{\Delta_I=0}^{N=c/3}\mc R_I,$
every element in the ring can be uniquely represented by
\beq
\phi_{\{n_{\hat I_\alpha}\}}=\prod_{\alpha=1}\prod_{\hat I_\alpha}\phi_{\hat I_\alpha}^{n_{\hat I_\alpha}}\,\eeq
where the collection of $\{\hat I_1\}\equiv\{i\}$ labels the primitive generators\footnote{Here, by primitive generators, we mean the linearly independent chiral primary operators that spans the whole chiral ring, which later correspond to the generators of cohomology of Calabi-Yau manifolds, see section \ref{4} and also \cite{Morrison}.} in $\mc R_1$ corresponding to marginal deformation, $\{\hat I_\alpha|\alpha\geq 2\}$ enumerates the primitive generators in $\mc R_\alpha$ for a given dimension $\Delta=\alpha$, and $n_{\hat I_\alpha}$ specifies the power of a given generator. We can deform the SCFT by introducing not only marginal, but also irrelevant deformations respect to all chiral ring generators and their complex conjugates,
\beq
\mc S_0\rightarrow\mc S_{\rm deform}=\mc S_0+\frac{\tau^i}{2}\!\int\!\!{\rm d}^2x\,{\rm d}^2\theta\,\,\Phi_i+\frac{\tau^{\hat I_\alpha}}{2}\!\int\!\!{\rm d}^2x\,{\rm d}^2\theta\,\,\Phi_{\hat I_\alpha}+{\rm h.c.}\,,
\label{Sdeform}
\eeq
where the couplings are normalized with a factor of $4\pi$, and $\{\Phi_{\hat I_\alpha}\}$ denote the corresponding supermultiplets of chiral primaries  $\{\phi_{\hat I_\alpha}\}$. Such deformations surely break \ntt\ superconformal symmetries, while leaving a $\mathfrak{su}(2|1)$ sub-superalgebra intact. It is exactly the most general \ntt\ massive supersymmetries that can be preserved on $S^2$. Therefore we are able to place the deformed theory $S_{\rm deform}$ on $S^2$ and compute its partition function 
$$Z[S^2](\tau^i, \bar\tau^{\bar j}, \tau^{\hat I_\alpha}, \bar\tau^{\bar{\hat J}_\beta})$$ via localization techniques. Once we find $Z[S^2]$, by varying its parameters and utilizing supersymmetric Ward identities, one can obtain the extremal correlators of the chiral ring generators on $S^2$,
\beq
\left\langle\phi_{\hat I_\alpha}(N)\bar\phi_{\hat J_\beta}(S)\right\rangle_{S^2}=-\frac{1}{Z[S^2]}\partial_{\tau^{\hat I_\alpha}}\partial_{\bar\tau^{\bar{\hat J}_\beta}}Z[S^2]\Big|_{\tau^{\hat I_\gamma}=\bar\tau^{\bar{\hat  I}_\gamma}=0,\gamma\geq 2}\,,
\label{S2correlator}
\eeq
where ``N" and ``S" denote the north and south poles of the $S^2$. Finally, as the most important step, a Gram-Schmidt orthogonalization needs to be performed to extract extremal correlator, 
\beq
g_{\hat I_\alpha \bar{\hat J}_\beta}=\left\langle\phi_{\hat I_\alpha}(0)\bar\phi_{\hat J_\beta}(\infty)\right\rangle_{\mathbb{R}^2}\,,\eeq
on flat space from $\left\langle\phi_{\hat I_\alpha}(N)\bar\phi_{\hat J_\beta}(S)\right\rangle_{S^2}$.

Most of materials in this section can be regarded as a $2d$ version of discussion parallel to that for $4d$ \ntwo\ SCFTs in \cite{Komargodski}. As we will point out in section \ref{3.3}, the algorithm needs to be modified somewhat due to the nilpotency of the $2d$ chiral rings.\\

\subsection{Placing deformed theories on $S^2$}
The general methodology of putting theories on curved space supersymmetrically is developed in \cite{Seiberg}. Discussion specific to $2d$ \ntt\ theories is also explained in \cite{Gomis1, Benini, Gomis}, as well as in \cite{Gomis2} with an emphasis on spurious field analysis. We will follow \cite{Gomis1} with mild modification to place irrelevant operator deformations onto $S^2$ as well.

We have seen that a \ntt\ SCFT has $U(1)_V\times U(1)_A$ R-symmetries. Correspondingly elements in chiral ring $\mc R$ and twisted chiral ring $\widetilde{\mc R}$ take non-vanishing $U(1)_V$ and $U(1)_A$ charge respectively. The deformations (\ref{Sdeform}) are from F-terms and will break part of R-symmetries unless they are marginal. More explicitly, since superspace measure ${\rm d}^2\theta$ takes $(-2,0)$ R-charges, an irrelevant deformation
\beq
\frac{\tau^{\hat I_\alpha}}{2}\!\int\!\!{\rm d}^2x\,{\rm d}^2\theta\,\,\Phi_{\hat I_\alpha}
\eeq
inevitably breaks the $U(1)_V$ R-symmetry but keeps the $U(1)_A$ intact. The remaining massive superalgebra is labeled as $\mathfrak{su}(2|1)_B$. Similarly, an irrelevant deformation from a twisted chiral primary multiplet $\Sigma_{A_I}$, 
\beq\frac{\tilde\tau^{\hat I_\alpha}}{2}\!\int\!\!{\rm d}^2x\,{\rm d}^2\tilde\theta\,\,\Sigma_{\hat I_\alpha}\,,\eeq
will break $U(1)_A$ while preserving $U(1)_V$, whose remaining massive superalgebra is denoted as $\mathfrak{su}(2|1)_A$. The $\mathfrak{su}(2|1)_A$ and $\mathfrak{su}(2|1)_B$ are two \emph{inequivalent} sub-superalgebras in \ntt\ superconformal algebra. Interestingly they correspond to two inequivalent ways to place the deformed theories on $S^2$, that we will discuss in some details.

\subsubsection{ Deformations respect to $\mathfrak{su}(2|1)_A$}
The $\mathfrak{su}(2|1)_A$ type deformation allows us to use twisted chiral primaries $\{\Sigma_a, \Sigma_{\hat I_\alpha}\}$ to deform the original action while preserving $U(1)_V$ R-symmetry,
\beq
\mc S_{ A}=\mc S_0+\frac{\tilde\tau^a}{2}\!\int\!\!{\rm d}^2x\,{\rm d}^2\theta\,\,\Sigma_a+\frac{\tilde\tau^{\hat I_\alpha}}{2}\!\int\!\!{\rm d}^2x\,{\rm d}^2\theta\,\,\Sigma_{\hat I_\alpha}+{\rm h.c.}
\label{Adeform}
\eeq
For twisted chiral superfield with dimension $\Delta_\Sigma=\tilde\omega$,
\beq\Sigma=\left(\sigma,\, \tilde\lambda,\, \widetilde{\mc O}\,\right)\,,\eeq
its supersymmetric transformation on $S^2$ respect to $\mf{su}(2|1)_A$ is cast as 
\beqn
&&\delta\sigma=\zeta\cdot\tilde\lambda\,,\nonumber\\
&&\delta\tilde\lambda=i\gamma^\mu\tilde\zeta\,D_\mu\sigma+\zeta\,\widetilde{\mc O}-\frac{\tilde\omega}{R}\zeta\,\sigma\nonumber\\
&&\delta\widetilde{\mc O}=i\tilde\zeta\cdot\gamma^\mu D_\mu\tilde\lambda+\frac{\tilde\omega}{R}\zeta\cdot\tilde\lambda\,,
\label{sttc}
\eeqn
where $\zeta$ and $\tilde\zeta$ are Killing spinors parameterizing the $\mf{su}(2|1)_A$ superalgebra, $D_\mu$ is the covariant derivative on $S^2$ with radius $R$, and more about notations are summarized in appendix \ref{A}. Now placing $\Sigma$ from flat $\mathbb{R}^2$ to $S^2$,
\beq\int_{\mathbb{R}^2}\!\!{\rm d}^2x\,{\rm d}^2\theta\,\,\Sigma=\int_{\mathbb{R}^2}\!\!\dx\,\widetilde{\mc O}(x)\rightarrow\int_{S^2}\!\!\dx\sqrt{g}\,\widetilde{\mc O}(x)\,,\eeq
where $g$ is the determinant of metric on $S^2$. Apparently from eq.\,(\ref{sttc}), the above $F$-term is \emph{not} supersymmetric invariant,
\beq\delta\left(\int_{S^2}\!\!\dx\sqrt{g}\,\widetilde{\mc O}(x)\,\right)=\int_{S^2}\!\!\dx\sqrt{g}\,\frac{\tilde\omega-1}{R}\zeta\cdot\tilde\lambda\,,\eeq
unless $\tilde\omega=1$ corresponding to a marginal deformation. However, by compensating an additional piece proportional to $\sigma$, the modified F-term
\beq\int_{S^2}\!\!\dx\sqrt{g}\,\left(\widetilde{\mc O}(x)-\frac{\tilde\omega-1}{R}\sigma(x)\,\right)\eeq
is supersymmetric invariant on $S^2$. Therefore for a deformed theory (\ref{Adeform}), we can place it on $S^2$ with order of $1/R$ modifications as
\beq
\mc S_A[S^2]=\mc S_0[S^2]+\frac{\tilde\tau^{\hat I_\alpha}}{2}\!\int_{S^2}\!\!\dx\sqrt{g}\,\left(\widetilde{\mc O}_{\hat I_\alpha}(x)-\frac{\tilde\omega_{\hat I_\alpha}-1}{R}\sigma_{\hat I_\alpha}(x)\,\right)+{\rm h.c.}\,.
\label{SA}
\eeq
where $I_\alpha$ runs for all marginal and irrelevant couplings.

\subsubsection{ Deformations respect to $\mathfrak{su}(2|1)_B$}
Parallel to above discussion, the $\mathfrak{su}(2|1)_B$ superalgebra allows us to preserve $U(1)_A$ R-symmetries, which makes deformations by chiral primary multiplets $\{\Phi_i, \Phi_{\hat I_\alpha}\}$ feasible,
\beq
\mc S_{ B}=\mc S_0+\frac{\tau^i}{2}\!\int\!\!{\rm d}^2x\,{\rm d}^2\theta\,\,\Phi_i+\frac{\tau^{\hat I_\alpha}}{2}\!\int\!\!{\rm d}^2x\,{\rm d}^2\theta\,\,\Phi_{\hat I_\alpha}+{\rm h.c.}
\label{Bdeform}
\eeq
The supersymmetric transformation of a chiral superfield
\beq\Phi=\left(\phi,\,\psi,\,\mc O\,\right)\,\eeq
with dimension $\Delta_\Phi=\omega$, can be written down,
\beqn
&&\delta\phi=\tilde\epsilon\cdot\psi\,,\nonumber\\
&&\delta\psi=i\gamma^\mu\epsilon\,D_\mu\phi+\tilde\epsilon\,{\mc O}-\frac{\omega}{R}\,\tilde\epsilon\,\phi\nonumber\\
&&\delta{\mc O}=i\epsilon\cdot\gamma^\mu D_\mu\psi+\frac{\omega}{R}\,\tilde\epsilon\cdot\psi
\label{stc}
\eeqn
When placing the chiral primary on $S^2$, one can check,
\beq\delta\left(\int_{S^2}\!\!\dx\sqrt{g}\,\mc O(x)\,\right)=\int_{S^2}\!\!\dx\sqrt{g}\,\frac{\omega-1}{R}\tilde\epsilon\cdot\psi\,,\eeq
is not supersymmetric invariant, unless $\omega=1$ for marginal deformations. Therefore we modify the $F$-term by
\beq\int_{S^2}\!\!\dx\sqrt{g}\,\left(\mc O(x)-\frac{\omega-1}{R}\phi(x)\,\right)\,,\eeq
and thus the deformed action $(\ref{Bdeform})$, corrected as
\beq
\mc S_B[S^2]=\mc S_0[S^2]+\frac{\tau^{\hat I_\alpha}}{2}\!\int_{S^2}\!\!\dx\sqrt{g}\,\left({\mc O}_{\hat I_\alpha}(x)-\frac{\omega_{\hat I_\alpha}-1}{R}\phi_{\hat I_\alpha}(x)\,\right)+{\rm h.c.}\,,
\label{SB}
\eeq
is supersymmetric on $S^2$ with respect to $\mathfrak{su}(2|1)_B$.\\

\subsection{The $\mathfrak{su}(2|1)_A$ deformed partition functions on $S^2$}
\label{3.2}
Our discussion is focused on computing the partition function $Z_A$ on $S^2$ respect to $\mathfrak{su}(2|1)_A$ superalgebra, for one is always able to choose a ``twisted basis" \cite{Gomis3} to realize $\mathfrak{su}(2|1)_B$ deformation in terms of $\mathfrak{su}(2|1)_A$ superalgebra. Besides, under the assumption of mirror symmetry, for a theory $\mc S$ with $\mathfrak{su}(2|1)_B$ deformation, one can always find a mirror $\widetilde{\mc S}$, such that $Z_B(\mc S)=Z_A(\widetilde{\mc S})$. We will come back to this point in later sections.

The details of localization computation can be found in \cite{Benini, Gomis}. The partition function $Z_A(\mc S)$ captures the data of the twisted chiral ring $\widetilde{\mc R}$ of a theory $\mc S$.
We will adopt it by adding irrelevant deformations that correspond to all primitive generators in $\widetilde{\mc R}$ with dimension greater than one. 

An \ntt\ SCFT $\mc S$ considered here has a UV Lagrangian description, more concretely, realized as a gauged linear sigma model (GLSM) with gauge group $U(1)^s\times G$, where $G$ is a product of simple groups. The action of the theory,
\beq\mc S=\int\!\!\dx\,{\rm d}^4\theta\,\mc L_{\rm gauge}(V)+\mc L_{\rm matter}(\Phi, V)+\int\!\!\dx\,{\rm d}^2\theta\,\mc W(\Phi)+\int\!\!\dx\,{\rm d}^2\tilde\theta\,\widetilde{\mc W}(\Sigma)+{\rm h.c.}\eeq
contains gauge multiplets $V$, matter multiplets $\Phi$'s, superpotential $\mc W$ of matters, and twisted superpotential $\widetilde{\mc W}$ of field strength $\Sigma$, or say FI-terms, of $U(1)^s$ factors of gauge group. When placed on $S^2$, $\mc S$ will receive corrections in terms of $\mc O(1/R)$, 
$$\mc S\rightarrow \mc S_A=\mc S+\mc O\left(\frac{1}{R}\right)$$
as we have seen in eq.\,(\ref{SA}), and the modified action is invariant respect to supercharges $\mc Q\in \mathfrak{su}(2|1)_A$. We are allowed to add arbitrary $\mc Q$-exact terms $\mc Q\mc V$ with suitable asymptotic behaviors to evaluate the partition function
\beq
Z_A(\mc S)=\int\mc D\varphi\,{\rm e}^{-\mc S_A[\varphi]+t\mc Q\mc V}\eeq
without changing the final result. Therefore, by putting $t\rightarrow\infty$, we evaluate the above path integral on the locus $\mc M_{0}=\{\varphi_0\,|\,\mc Q\mc V(\varphi_0)=0\}$,
\beq
Z_A(\mc S)=\int\!\!{\rm d}\varphi_0\,Z_{\rm cl}(\varphi_0)\,Z_{\rm 1-loop}(\varphi_0),\eeq
where we have chosen to localize the theory onto the Coulomb branch, and the term $Z_{\rm 1-loop}$ corresponds to the one-loop fluctuation of $\mc Q\mc V$ around $\mc M_0$. We now spell out the detailed expression of $Z_A(\mc S)$ \cite{Benini, Gomis},  
\beq
Z_A(\mc S)=\frac{1}{\left|\mc W\right|}\sum_{\mathfrak{m}=\{\mathfrak{m_i},\mathfrak{{\tilde m}_l}\}}\int\left[\prod^{{\rm rank (G)}}_{i=1}\prod^{{s}}_{l=1}\frac{{\rm d}\sigma_i}{2\pi}\frac{{\rm d}\tilde\sigma_{l}}{2\pi}\right]Z_{\rm cl}(\sigma,\mf m)Z_{\rm gauge}\left(\{{\sigma_i}\},\{\mf m_i\}\right)\prod_{\Phi}Z_{\Phi}(\sigma,\mf m)\,,
\label{ZA}
\eeq
where we have scaled the radius of sphere $R=1$, $|\mc W|$ is the order of the Weyl group $G$ and $\sigma=\{\sigma_i,\tilde\sigma_l\}$. $\sigma_i\in\mathbb R^{{\rm rank}(G)}$ is in the Cartan subalgebra of $G$ and $\mf m_i\in\mathbb Z^{{\rm rank}(G)}$ is the GNO quantized magnetic charge of the Cartan part of $G$. Similarly $\tilde\sigma_l$ and $\tilde{\mf m}_l$ parametrize $\mathbb R^s$ and $\mathbb Z^s$ corresponding to the $U(1)^s$ factors of $G$. 

$Z_{\rm gauge}$ and $Z_{\Phi}$ are $1$-loop fluctuations of gauge and matter multiplets around $\mc Q\mc V$,
\beqn
&&Z_{\rm gauge}\left(\{{\sigma_i}\},\{\mf m_i\}\right)=\prod_{\alpha\in\Delta_+}\left(\frac{(\alpha,\mf m)^2}{4}+(\alpha,\sigma)^2\right)\,,\nonumber\\
&&Z_{\Phi}(\sigma,\mf m)=\prod_{\rho\in R_\Phi}\frac{\Gamma\left(\frac{1}{2}\mf q_\Phi-(\rho,i\sigma+\frac{1}{2}\mf m)-\sum_l Q_\Phi^l(i\tilde\sigma_l+\frac{1}{2}\tilde{\mf m}_l)\right)}{\Gamma\left(1-\frac{1}{2}\mf q_\Phi+(\rho,i\sigma-\frac{1}{2}\mf m)+\sum_l Q_\Phi^l(i\tilde\sigma_l-\frac{1}{2}\tilde{\mf m}_l)\right)}\,,
\eeqn
where $(\cdot,\cdot)$ is the standard inner product of Lie algebra $\mf g$ of $G$, $\alpha\in\Delta_+$ are positive roots over $\mf g$, $Q_\Phi^l$ is the gauge charge of $\Phi$ for $U(1)^s$ factors, $\rho$ is the weight of the representation $R_\Phi$ of $G$, and $\mf q_\Phi$ is the $U(1)_V$ R-charge of $\Phi$.

$Z_{\rm cl}(\sigma, \mf m)$ is the classical piece. For all gauge, matters and superpotential sectors are $\mc Q$-exact, so absorbed in $\mc V$, $Z_{\rm cl}$ interestingly only matters with twisted superpotentials $\widetilde{\mc W}$ and exactly encodes the information of twisted chiral ring of the theory $\mc S$. If $\mc S$ contains \emph{only} marginal deformations, the twisted potentials are FI-terms corresponding to the $U(1)^s$ factors,\footnote{The unusual sign in front of $\tilde\tau$ and $\bar{\tilde\tau}$ are due to analytical continuation of the theory from Minkowskian space.}
\beq
\widetilde{\mc W}(\Sigma)+{\rm h.c.}=\frac{1}{2}\sum_{l=1}^{s}\left(\tilde\tau^l\,\Sigma_l-\bar{\tilde\tau}^l\,\overline{\Sigma}_l\right)\,,\eeq
where we use $\Sigma_l\equiv\left(\tilde\sigma_l,\,\tilde\lambda_l,\,\widetilde{\mc O}_l\right)$ to denote the twisted super-field strength of the $U(1)^s$ gauge multiplets, and $\tilde\tau^l\equiv\frac{\tilde\theta^l}{2\pi}+i\,\tilde r^l$ are their complex FI-couplings. Evaluating it at locus $\mc M_0$ gives
\beq
Z_{\rm cl}\left(\tilde\sigma,\tilde{\mf m}\right)=\exp\left(-4\pi i\sum_{l=1}^s\tilde r^l\tilde\sigma_l-i\sum^s_{l=1}\tilde\theta^l\tilde{\mf m}_l\right)\,.\eeq


Now we introduce irrelevant deformations, see eq.\,(\ref{Adeform}), for generators with dimension greater than one in twisted chiral ring. Since all of them are twisted superpotentials, only the term $Z_{\rm cl}$ in eq.\,(\ref{ZA}) needs modifications. In case of gauge group $U(N)$, we spell out the deformed partition function. Following appendix \ref{B}, all generators are of the form 
$$\Tr\left(\Sigma\right),\,\Tr\left(\Sigma^2\right),...,\,\Tr\left(\Sigma^N\right)\,,$$ 
with $\Sigma=(\sigma,\, \lambda,\,iG-D)$ taking values on $\mf{u}(N)$ Lie algebra. We therefore introduce the deformations as the twisted superpotential
\beq
\widetilde{\mc W}(\Sigma)+{\rm h.c.}=\frac{1}{2}\Big(\tilde\tau_1\,\Tr\,\Sigma-\bar{\tilde\tau}_1\,\Tr\,\overline{\Sigma}\Big)+\frac{1}{2}\sum_{n=2}^N\Big(\tilde\tau_n\Tr\,\Sigma^n-\bar{\tilde\tau}_n\Tr\,\overline\Sigma^n\Big)\,,
\label{twisted superpotential}
\eeq
where $\tilde\tau_1$ is marginal and singled out, and $\tilde\tau_n\Tr\,\Sigma^n$ are irrelevant deformations\footnote{We here use notation ``$\tilde\tau_n$" with subscripts as the couplings of the deformations to avoid confusion with the powers of $\tilde\tau$.} with dimension $\Delta_n=n$. Their $F$-terms are 
\beq
\Tr\,\Sigma^n\Big|_{F-{\rm term}}=n\Tr\left\{(iG-D)\sigma^{n-1}\right\}+{\rm fermi.}
\eeq
When placed on $S^2$, following eq.\,(\ref{SA}), we correct them by
\beq
n\Tr\left\{(iG-D)\sigma^{n-1}\right\}+{\rm fermi.}-(n-1)\Tr\,\sigma^n\,.
\eeq
Localizing on Coulomb branch implies that we set the locus $\mc M_0$ at
\beqn
&&\sigma={\rm diag}\left\{\sigma_1+\frac{i}{2}m_1,\,\sigma_2+\frac{i}{2}m_2,...,\,\sigma_n+\frac{i}{2}m_n\right\}\equiv\sigma_c,\nonumber\\
&&D=-{\rm diag}\left\{\sigma_1,\,\sigma_2,...,\,\sigma_n\right\},\ \ G={\rm diag}\left\{\frac{m_1}{2},\,\frac{m_2}{2},...,\,\frac{m_n}{2}\right\}\nonumber\\
&&\bar\sigma=\sigma^\dagger_c\equiv\bar{\sigma}_c,\ \ \ \ \ \lambda=0
\label{locus}
\eeqn
Therefore we have 
\beq
Z_{\rm cl}(\sigma_c,\bar{\sigma}_c)=\exp\left\{-2\pi\sum_{n=1}^N\left(\tilde\tau_n\Tr\,\sigma_c^n-\bar{\tilde\tau}_n\Tr\,\overline\sigma_c^n\right)\right\}
\eeq
Overall our full deformed partition function for $U(N)$ gauge group is
\beq
Z_A(\tilde\tau,\bar{\tilde\tau})=\frac{1}{N!}\sum_{\{m_i\}\in\mathbb Z}\int_{-\infty}^\infty\prod^{{N}}_{i=1}\frac{{\rm d}\sigma_i}{2\pi}\,Z_{\rm cl}(\sigma_c,\bar\sigma_c)\,Z_{\rm gauge}(\sigma_c,\bar{\sigma}_c)\prod_{\Phi}Z_{\Phi}(\sigma_c,\bar{\sigma}_c)\,,
\label{main}
\eeq
with
\beqn
&&Z_{\rm cl}(\sigma_c,\bar\sigma_c)=\exp\left\{-2\pi\sum_{n=1}^N\left(\tilde\tau_n\Tr\,\sigma_c^n-\bar{\tilde\tau}_n\Tr\,\overline\sigma_c^n\right)\right\}\,,\nonumber\\
&&Z_{\rm gauge}(\sigma_c,\bar{\sigma}_c)=\prod_{i<j}\left(\frac{(m_i-m_j)^2}{4}+(\sigma_i-\sigma_j)^2\right)=\prod_{i<j}\left|\sigma_{c\,i}-\sigma_{c\,j}\right|^2\,,\nonumber\\
&&Z_{\Phi}(\sigma_c,\bar{\sigma}_c)=\prod_{\rho\in R_\Phi}\frac{\Gamma\left(\frac{1}{2}\mf q_\Phi-(\rho,i\bar\sigma_c)\right)}{\Gamma\left(1-\frac{1}{2}\mf q_\Phi+(\rho,i\sigma_c)\right)}\,.\nonumber
\eeqn
eq.\,(\ref{main}) serves as our main formula that will be used in Sec.\,\ref{5}. Different from $4d$ situation, where the Nekrasov's partition functions are not known yet for deformed theories \cite{Komargodski}, meanwhile our $2d$ deformed partition function here is \emph{exact}, because $2d$ localization onto Coulomb branch has no instanton correction! It would be very interesting to evaluate the $2d$ \emph{deformed} partition function through localization onto Higgs branch, which in principle could be written as discrete sum of vortex and anti-vortex partition functions as $2d$ version of Nekrasov partition function. We wish that it might shed light on how to compute the $4d$ deformed partition function exactly.\\

\subsection{Twisted chiral primary correlators from $S^2$ to $\mathbb R^2$}
\label{3.3} 
From action (\ref{SA}) and partition function (\ref{ZA}), one can extract exact correlation functions of the twisted chiral primaries,
\beqn
\frac{1}{Z_A[S^2]}\partial_{\tilde\tau^{\hat I_\alpha}}\partial_{\bar{\tilde\tau}^{\bar{\hat J}_\beta}}Z_A[S^2]\Bigg|_{\tilde\tau^{\hat I_\gamma}=\bar{\tilde\tau}^{\bar{\hat  I}_\gamma}=0,\gamma\geq 2}\=\Bigg\langle\left(\frac{1}{2}\!\int_{S^2}\!\!\!\dx\sqrt{g}\left(\mc O_{\hat I_\alpha}(x)-\left(\omega_{\hat I_\alpha}-1\right)\sigma_{\hat I_\alpha}(x)\right)\right)\nonumber\\
&&\left(\frac{1}{2}\!\int_{S^2}\!\!\!{\rm d}^2y\sqrt{g}\left(\overline{\mc O}_{\bar{\hat J}_\beta}(y)-\left(\omega_{\bar{\hat J}_\beta}-1\right)\bar\sigma_{\bar{\hat J}_\beta}(y)\right)\right)\Bigg\rangle_{S^2}\nonumber\\\nonumber\\
\=-4\pi^2\left\langle\sigma_{\hat I_\alpha}(N)\bar\sigma_{{\bar{\hat J}_\beta}}(S)\right\rangle_{S^2}
\label{ward}
\eeqn
where $\sigma_{\hat I}$ and $\mc O_{\hat I}$ are bottom and top terms of the twisted chiral primary multiplets $\Sigma_{\hat I}$ with dimension $\Delta_{\hat I}=\omega_{\hat I}$,
index $\hat I$, $\hat J$ labels all twisted chiral ring generators. The second equality is due to the $\mf{su}(2|1)_A$ supersymmetric Ward identity \cite{Gomis1}. In fact, taking derivative respect to $\tilde\tau_n$ on eq.\,(\ref{main}), we have
$$\partial_{\tilde\tau_n}Z_A=\frac{1}{N!}\sum_{\{m_i\}\in\mathbb Z}\int_{-\infty}^\infty\prod^{{N}}_{i=1}\frac{{\rm d}\sigma_i}{2\pi}\left(-2\pi\,\Tr\,\sigma^n_c\right)Z_{\rm cl}(\sigma_c,\bar\sigma_c)\,Z_{\rm gauge}(\sigma_c,\bar{\sigma}_c)\prod_{\Phi}Z_{\Phi}(\sigma_c,\bar{\sigma}_c)\,.$$
Noticing that $\sigma_c$ in eq.\,(\ref{locus}) is exactly the BPS solution evaluated at north pole \cite{Benini, Benini2}, we indeed have
\beq
\left\langle-2\pi\Tr\,\sigma^n\left(N\right)\right\rangle_{S^2}=\frac{1}{Z_A[S^2]}\partial_{\tilde\tau_n}Z_A[S^2]\Bigg|_{\tilde\tau_{n}=\bar{\tilde\tau}_n=0,n\geq2}
\label{1pt}
\eeq
Similarly, 
$$\left\langle2\pi\Tr\,\bar\sigma^n\left(S\right)\right\rangle_{S^2}=\frac{1}{Z_A[S^2]}\partial_{\bar{\tilde\tau}_n}Z_A[S^2]\Bigg|_{\tilde\tau_{n}=\bar{\tilde\tau}_n=0,n\geq2}\,,$$
for $\bar\sigma_c$ in eq.\,(\ref{locus}) is evaluated at south pole.

It has been throughout analyzed in \cite{Gomis1, Gomis2} that eq.\,(\ref{ward}) is the consequence of the unique regularization scheme respect to $\mf{su}(2|1)_A$ supersymmetries on $S^2$. Alternatively, one can understand it as that, to regularize the partition function $Z_A$ as well as correlators on $S^2$ respect to $\mf{su}(2|1)_A$, one has to introduce counter terms combined with the \ntt\ supergravity multiplet $\mf R$ preserving $\mf{su}(2|1)_A$,
\beq
\Gamma_{\rm c.t.}=\frac{1}{2}\!\int\dx{\rm d}^2\tilde\theta\,\mc E\,\mf R\,\mc F(\tilde\tau)+{\rm h.c.}\,,\eeq
where the supergravity multiplet $\mf R$ has dimension $\Delta_{\mf R}=1$ containing Ricci scalar curvature as the top component, $\mc E$ is the density in curved twisted superspace, and $\mc F(\tilde\tau)$ is a holomorphic function in terms of couplings $\{\tilde\tau^{\hat I}\}$.

More importantly, it is the multiplet $\mf R$ that mixes twisted chiral operator $\Sigma$ with other lower dimensional operators \cite{Komargodski}, i.e.
\beq
\Sigma_{\Delta}\longrightarrow\Sigma_{\Delta}+\alpha_1(\tilde\tau,\bar{\tilde\tau}; \Delta)\,\mf R\,\Sigma_{\Delta-1}+\alpha_2(\tilde\tau,\bar{\tilde\tau}; \Delta)\,\mf R^2\,\Sigma_{\Delta-2}+,\cdots,+\alpha_{\Delta}(\tilde\tau,\bar{\tilde\tau}; \Delta)\,\mf R^{\Delta}\,\mathds 1\,,
\label{mixing}
\eeq
where $\Delta$ denotes the dimension of $\Sigma_{\Delta}$, and $\alpha_i(\tilde\tau,\bar{\tilde\tau}; \Delta)$ are certain coefficients presenting mixing. Similar to the $4d$ case \cite{Komargodski}, the mixing only happens among twisted chiral primaries themselves, but not to twisted chiral mixing with twisted antichiral or (anti)chiral primaries. It is because only twisted chiral primaries are $\mf{su}(2|1)_A$ supersymmetric on the north pole of $S^2$, whereas twisted anti-chiral primaries are supersymmetric on the south pole of $S^2$, meanwhile (anti)chiral primaries respect to $\mf{su}(2|1)_B$ supersymmetries instead and thus are nowhere to be put on $S^2$ in $\mf{su}(2|1)_A$ regularization scheme. It explains the phenomenon that we observe \emph{nonzero} correlation functions between operators with different dimensions on $S^2$, see for example eq.\,(\ref{1pt}) as a result of $\Sigma_{\hat I_\alpha}$ mixing with identity operator. Therefore, to find correct correlation functions on $\mathbb R^2$, we need to perform Gram-Schmidt orthogonalization to disentangle twisted (anti)chiral operators with those of lower dimensions. We will see that the Gram-Schmidt procedure is adequate to disentangle operators mixing from $S^2$ to $\mathbb R^2$, as it admits a natural geometric interpretation in Calabi-Yau geometries we will discuss in section \ref{4}, and pass both perturbative and non-perturbative checks in the examples we give in section \ref{5}. However it would be interesting and important to investigate more detailed structures of the mixing coefficients $\alpha_i(\tilde\tau,\bar{\tilde\tau}; \Delta)$ in terms of conformal anomalies as analyzed in \cite{Gomis2} for $\alpha_1(\tilde\tau,\bar{\tilde\tau}; 1)$ corresponding to the mixing between marginal primaries and identity operator. We will leave the answer to this question in our subsequent work \cite{Chen}.

We now explain the algorithm to disentangle operators in somewhat detail. From eq.\,(\ref{mixing}), it is seen that the mixing only happens to twisted chiral operators themselves with dimension separation by multiples of one, therefore we disentangle them by induction on operators' dimensions. 

Since there will be many indexes appearing, we would like to summarize the notations we will use first. For a twisted chiral primary $\sigma_{a_\alpha}$, ``$\alpha$" labels the operator's dimension, and ``$a_\alpha$" enumerate the number of the operators with dimension $\Delta=\alpha$. We will see soon that, at each dimension $\Delta=\alpha$, operators $\sigma_{a_\alpha}$'s are in general not linear independent. We thus collect those linear independent operators and label them by $\sigma_{I_\alpha}$. At last $\sigma_{\hat I_\alpha}$'s still denote all twisted ring generators as before, with $\hat I_\alpha$ enumerating all primitive generators with dimension $\Delta=\alpha$. For $\sigma_{\hat I_\alpha}$'s must be linear independent for given $\alpha$, we have $\{\hat I_\alpha\}\subset\{I_\alpha\}$. With these notation, we start the orthogonalization from dimension zero.\\

\noindent$\Delta=0$: The unit operator $\mathds{1}$ is dimension zero and need no change.\\ 

\noindent$\Delta=1$: We have marginal twisted chiral primaries $\sigma_{a_1}\equiv\sigma_i$. For every primary specify a direction to marginally deform the SCFT, they are all linear independent, i.e. the index sets $\{I_1\}$, $\{a_1\}$ and $\{i\}$ identical. They are required to be orthogonal to unit operator $\mathds{1}$. We thus define
\beq
\hat{\sigma}_{a_1}\equiv\sigma_{a_1}-\frac{\left\langle\sigma_{a_1}(N)\,\bar{\mathds{1}}(S)\right\rangle_{S^2}}{\left\langle\mathds{1}(N)\,\bar{\mathds{1}}(S)\right\rangle_{S^2}}\,\mathds{1}\,,\ \ \ \ \hat{\bar\sigma}_{b_1}\equiv\bar\sigma_{b_1}-\frac{\left\langle\mathds{1}(N)\,\bar\sigma_{b_1}(S)\right\rangle_{S^2}}{\left\langle\mathds{1}(N)\,\bar{\mathds{1}}(S)\right\rangle_{S^2}}\,\bar{\mathds{1}}
\label{s1}
\eeq
One can check indeed 
\beq\left\langle\hat{\sigma}_{a_1}(N)\,\bar{\mathds{1}}(S)\right\rangle_{S^2}=\left\langle\mathds{1}(N)\,\hat{\bar\sigma}_{b_1}(S)\right\rangle_{S^2}=0\,,\eeq
and the twisted chiral ring data
\beqn
g^{(1)}_{a_1\bar b_1}\eqq\left\langle\sigma_{a_1}(0)\bar\sigma_{b_1}(\infty)\right\rangle_{\mathbb R^2}=\left\langle\hat{\sigma}_{a_1}(N)\,\hat{\bar\sigma}_{b_1}(S)\right\rangle_{S^2}\nonumber\\
\=\left\langle{\sigma}_{a_1}(N)\,{\bar\sigma}_{b_1}(S)\right\rangle_{S^2}-\frac{\big\langle\sigma_{a_1}(N)\,\bar{\mathds{1}}(S)\left\rangle_{S^2}\,\big\langle\mathds{1}(N)\,{\bar\sigma}_{b_1}(S)\right\rangle_{S^2}}{\left\langle\mathds{1}(N)\,\bar{\mathds{1}}(S)\right\rangle_{S^2}}\nonumber\\
\=-\partial_{i}\partial_{\,\bar j}\log\,Z_A[S^2]\equiv g^{(1)}_{I_1\bar J_1}\,,
\label{g1}
\eeqn
is exactly the Zamolodchikov metric of moduli space $\mc M$ of the SCFT.\\ 

\noindent$\Delta=2$: We define
\beq
\hat{\sigma}_{a_2}\equiv\sigma_{a_2}-\frac{\left\langle\sigma_{a_2}(N)\,\bar{\mathds{1}}(S)\right\rangle}{\left\langle\mathds{1}(N)\,\bar{\mathds{1}}(S)\right\rangle}\,\mathds{1}-\sum_{I_1,\bar J_1}g^{(1)\bar J_1 I_1}\left\langle\sigma_{a_2}(N)\,\hat{\bar\sigma}_{J_1}(S)\right\rangle\,\hat\sigma_{I_1}\,,
\label{s2}
\eeq
where $g^{(1)\bar J_1 I_1}$ is the inverse of metric $g^{(1)}_{I_1\bar J_1}$, and so can be defined for $\hat{\bar{\sigma}}_{b_2}$. One can firmly check that $\hat{\sigma}_{a_2}$ is orthogonal to all operators with dimension less than two, say $\{1,\,\hat\sigma_{I_1}\}$. The twisted chiral ring data on vector bundle $\mc V_2$ over $\mc M$ is computed by,
\beq
g^{(2)}_{a_2\bar{b}_2}\equiv\left\langle\sigma_{a_2}(0)\,\bar\sigma_{b_2}(\infty)\right\rangle_{\mathbb R^2}=\left\langle\hat{\sigma}_{a_2}(N)\,\hat{\bar\sigma}_{b_2}(S)\right\rangle_{S^2}\,.
\label{g2}
\eeq
The eq.\,(\ref{g1}) and (\ref{g2}) automatically satisfy the $tt^*$-equation (\ref{tt}) \cite{Komargodski}, where we choose the ``shift" basis\footnote{It is named ``diagonal" basis in \cite{Komargodski} for $4d$ chiral ring is freely generated. The nilpotency of $2d$ chiral rings actually realizes the OPE coefficients $C$ as shift matrices.} for OPE coefficient $C_{IJ}^K$, i.e.
\beq\sigma_{I_1}(x)\sigma_{J_1}(0)=\sigma_{I_1}\sigma_{J_1}(0)\equiv\sigma_{c_2}(0)\,.\eeq
We will stick to this basis for all operators with arbitrary dimensions. However as emphasized before, the $2d$ (twisted) chiral ring is finite and nilpotent. Therefore with this ``shift" basis, we will obtain too many operators which turns out to be \emph{linear dependent}. From level $\Delta=3$ we may encounter this problem. We thus need to collect those linear independent to continue orthogonalization.\\

\noindent$\Delta=3$: We want to continue the disentanglement as we have done at $\Delta=0,1,2$. However the metric (\ref{g2}) may be singular in general. In section \ref{4} and \ref{5} we will give such examples. The singular $g_{a_2\bar{b}_2}$ implies that not all $\hat\sigma_{a_2}$ of dimension two are linear independent. So we collect some of them to form a \emph{maximal} linear independent set, which includes all primitive generators of dimension two, and those generated by generators of dimension one. Assume we have picked such a set 
$$\mc A_2=\left\{\hat{\sigma}_{I_2}\right\}\,,$$ 
and computed the corresponding metric $g^{(2)}_{I_2\bar{J}_2}(\mc A_2)$. Now $g^{(2)}_{I_2\bar{J}_2}(\mc A_2)$ is non-singular and invertible. We can use its inverse $g^{(2)\bar{J}_2 I_2}(\mc A_2)$ to continue our orthogonalization for all operators $\sigma_{a_3}$ of $\Delta=3$,
\beqn
\hat{\sigma}_{a_3}\!\!&\equiv&\!\sigma_{a_3}-\frac{\left\langle\sigma_{a_3}(N)\,\bar{\mathds{1}}(S)\right\rangle_{S^2}}{\left\langle\mathds{1}(N)\,\bar{\mathds{1}}(S)\right\rangle_{S^2}}\,\mathds{1}-\sum_{I_1,\bar J_1}g^{(1)\bar J_1 I_1}\left\langle\sigma_{a_3}(N)\,\hat{\bar\sigma}_{J_1}(S)\right\rangle_{S^2}\,\hat\sigma_{I_1}\nonumber\\
\-\sum_{I_2,\bar{J}_2\in \mc A_2}g^{(2)\bar{J}_2 I_2}\left(\mc A_2\right)\left\langle\sigma_{a_3}(N)\,\hat{\bar\sigma}_{{J}_2}(S)\right\rangle_{S^2}\,\hat\sigma_{I_2}\,.
\label{s3}
\eeqn
Now $\hat\sigma_{a_3}$ is orthogonal to all lower dimensional operators, $\{1,\sigma_{a_1},\sigma_{a_2}\}$.

At last we need show such construction does \emph{not} depend on the choice of $\mc A_2$. If we choose another maximal linear independent set $\mc A_2^\prime=\{\hat{\sigma}_{I_2^\prime}\}$, one can always find the linear transformation $\mc T$ relating them, and their hermitian conjugates,
\beq\hat{\sigma}_{I_2^\prime}=\mc T_{I_2^\prime}^{\ I_2}\,\hat{\sigma}_{I_2}\,,\ \ \ \hat{\bar\sigma}_{J_2^\prime}=\overline{\mc T_{J_2^\prime}^{\ J_2}}\,\hat{\bar\sigma}_{J_2}\,,\eeq 
where $\overline{\mc T_{J_2^\prime}^{\ J_2}}$ is the complex conjugate of $\mc T_{J_2^\prime}^{\ J_2}$. Correspondingly the inverse of metric transforms as
\beq
g^{(2)\bar{J}^\prime_2 I^\prime_2}(\mc A^\prime_2)=\left(\mc T^{-1}\right)^{\ I^\prime_2}_{I_2}\,\overline{\left(\mc T^{-1}\right)^{\ J^\prime_2}_{J_2}}\,g^{(2)\bar{J}_2 I_2}(\mc A_2)\,.\eeq 
Therefore we show that eq.\,(\ref{s3}) is indeed independent of the choice of $\mc A$. Based on the new set of $\{\hat\sigma_{a_3}\}$, we can compute the twisted chiral ring data on $\mc V_3$ over $\mc M$,
\beq
g^{(3)}_{a_3\bar{b}_3}\equiv\left\langle\sigma_{a_3}(0)\,\bar\sigma_{b_3}(\infty)\right\rangle_{\mathbb R^2}=\left\langle\hat{\sigma}_{a_3}(N)\,\hat{\bar\sigma}_{b_3}(S)\right\rangle_{S^2}\,,
\label{g3}
\eeq
which can be shown again satisfying the $tt^*$-equation (\ref{tt}) up to $\Delta=3$.\\ 

\noindent$\Delta=n$: For generic $n$, by induction, we have orthogonalized operators up to $\Delta=n-1$. Among operators $\hat\sigma_{a_{\alpha}}$ of $\Delta=\alpha$, we can collect a maximal linear independent set $\mc A_{\alpha}$ for $\alpha=0,1,...,n-1$, where $\mc A_0=\{\mathds 1\}$ and $\mc A_1=\{\hat\sigma_{I_1}\}$. We compute their metrics $g_{I_\alpha\bar{J}_\alpha}$ and inverse $g^{\bar{J}_\alpha I_\alpha}$, and define
\beq
\hat\sigma_{a_n}\equiv\sigma_{a_n}-\sum_{\alpha=0}^{n-1}\,\sum_{I_\alpha,\bar{J}_\alpha\in \mc A_\alpha}g^{(\alpha)\bar{J}_\alpha I_\alpha}(\mc A_\alpha)\,\left\langle\sigma_{a_n}(N)\,\hat{\bar\sigma}_{{J}_\alpha}(S)\right\rangle_{S^2}\,\hat\sigma_{I_\alpha}\,,
\label{sn}
\eeq
where $g^{(0)\bar{J}_0 I_0}\equiv\left(\left\langle\mathds{1}(N)\,\bar{\mathds{1}}(S)\right\rangle_{S^2}\right)^{-1}$\,.
It allows us to compute the twisted chiral ring data at level $\Delta=n$ for bundle $\mc V_n$ over $\mc M$,
\beqn
g^{(n)}_{a_n\bar{b}_n}\!\!&\equiv&\!\left\langle\sigma_{a_n}(0)\,\bar\sigma_{b_n}(\infty)\right\rangle_{\mathbb R^2}=\left\langle\hat{\sigma}_{a_n}(N)\,\hat{\bar\sigma}_{b_n}(S)\right\rangle_{S^2}\nonumber\\
\=\left\langle{\sigma}_{a_1}(N)\,{\bar\sigma}_{b_1}(S)\right\rangle_{S^2}-\sum_{\alpha=0}^{n-1}\sum_{I_\alpha,\bar{J}_\alpha\in \mc A_\alpha}\!\!\!\!\!\!g^{(\alpha)\bar{J}_\alpha I_\alpha}(\mc A_\alpha)\left\langle\sigma_{a_n}(N)\,\hat{\bar\sigma}_{{J}_\alpha}(S)\right\rangle_{S^2}\,\left\langle\hat{\sigma}_{I_\alpha}(N)\,\bar{\sigma}_{{b}_n}(S)\right\rangle_{S^2}\nonumber\\
\label{gn}
\eeqn
In practice, since we have showed that eq.\,(\ref{sn}) and (\ref{gn}) are independent of the choice of $\mc A_\alpha$, one can freely choose any convenient set of $\{\mc A_\alpha\}$ to perform the orthogonalization. We therefore for convenience pick up the original operators $\{\sigma_{I_\alpha}\}$, instead of $\{\hat\sigma_{I_\alpha}\}$ in all sets of $\{\mc A_{\alpha}\}$, to perform the orthogonalization.\\ 

Now we summarize the algorithm. For a given set of primitive generators in twisted chiral ring, $\mc G=\{\mathds 1,\sigma_{\hat I_1},\sigma_{\hat I_2},...,\sigma_{\hat I_A}\}$, where $A$ labels the maximal dimension of the generators, we choose the ``shift" OPE basis as before
\beq
\sigma_{\hat I_\alpha}(x)\sigma_{\hat J_{\beta}}(0)=\sigma_{\hat I_\alpha}\sigma_{\hat J_{\beta}}(0)\,,
\label{OPEb}
\eeq
where on RHS, $\sigma_{\hat I_\alpha}\sigma_{\hat J_{\beta}}$ stands for an element with dimension $\Delta={\alpha+\beta}$ in the ring. Under this basis, we collect all possible elements and arrange them by dimensions,
\beq
\left\{\mathds{1}\,;\,\,\left\{\sigma_{\hat I_1}\right\}\,;\,\,\left\{\sigma_{\hat I_1}\sigma_{\hat J_1},\,\sigma_{\hat I_2}\right\}\,;\,\,\left\{\sigma_{\hat I_1}\sigma_{\hat J_1}\sigma_{\hat K_1},\,\sigma_{\hat I_2}\sigma_{\hat J_1},\,\sigma_{\hat I_3}\right\}\,;...\right\}\,,
\eeq
which can be uniquely expressed by the primitive generators in $\mc G$. In above we do not take account of any equivalent relations among the generators, and treat all elements as freely generated, which is the reason that many of them are linear dependent. For any two of them,
\beq
\sigma_{a_\alpha}\equiv\sigma_{\{n_{\hat I_\gamma}\}}=\prod^{A}_{\gamma=1}\prod_{\hat I_\gamma}\sigma_{\hat I_\gamma}^{n_{\hat I_\gamma}}\,,\ \ \ \ \ \sigma_{b_\beta}\equiv\sigma_{\{n^\prime_{\hat I_\gamma}\}}=\prod^{A}_{\gamma=1}\prod_{\hat I_\gamma}\sigma_{\hat I_\gamma}^{n^\prime_{\hat I_\gamma}}\,,\eeq
we can compute their correlator from $Z_A$, the partition function (\ref{ZA}),
\beqn
M_{a_\alpha\bar b_\beta}\!\!&\equiv&\!\left\langle\sigma_{a_\alpha}(N)\,\bar\sigma_{b_\beta}(S)\right\rangle_{S^2}\nonumber\\
\=\frac{1}{Z_A}\prod_{\gamma=0}^A\prod_{\hat I_\gamma,\bar{\hat J}_\gamma}\left(-\frac{1}{2\pi}\frac{\partial}{\partial\tau^{\hat I_\gamma}}\right)^{n_{\hat I_\gamma}}\left(\frac{1}{2\pi}\frac{\partial}{\partial\bar\tau^{\bar{\hat J}_\gamma}}\right)^{n^\prime_{\bar{\hat J}_\gamma}}Z_A\Bigg|_{\tau^{\hat K_\delta}=\bar\tau^{\bar{\hat K}_\delta}=0,\delta\geq 2}
\label{bigM}
\eeqn
where we use matrix $M_{a_\alpha\bar b_\beta}$ to relabel all these correlators, where as before $\alpha$, $\beta$ denote the dimension of operators, and $a_\alpha$, $b_\beta$ enumerate all operators of dimension $\alpha$ and $\beta$. Since we have argued that, in general, $M$ is a singular matrix and not invertible, we have to \emph{remove all but one} of the rows and columns corresponding to those linear dependent operators. One can perform this operation level by level respect to the dimensions, $\Delta=n-1$, of the operators, and finally obtain a matrix 
\beq\widetilde M_{n-1,\,I_\alpha \bar{J}_\beta}=\left\langle\sigma_{I_\alpha}(N)\,\bar\sigma_{J_\beta}(S)\right\rangle_{S^2}\,,
\label{tildeM}
\eeq
where $I_\alpha$, $\bar{J}_\beta$ denote only those linear independent operators up to dimension $\Delta$. Since we also showed that orthogonalization does not depend on the choice of $\{\mc A_\alpha\}$, we can use $\widetilde M_{n-1}$, instead of $g_{I_\alpha \bar{J}_\beta}$, in eq.\,(\ref{sn}), i.e.
\beq
\hat\sigma_{a_n}=\sigma_{a_n}-\sum_{\alpha,\beta=0}^{n-1}\,\sum_{I_\alpha,{\bar J}_\beta}\widetilde M_{n-1}^{\bar{J}_\beta I_\alpha}\,\big\langle\sigma_{a_n}(N)\,{\bar\sigma}_{{J}_\beta}(S)\big\rangle_{S^2}\,\sigma_{I_\alpha}\,,
\label{snm}
\eeq
where $\widetilde M_{n-1}^{\bar {J}_\beta I_\alpha}$ is the inverse of $\widetilde M_{n-1}$. Similarly, instead of using eq.\,(\ref{gn}), for any two elements $\sigma_{c_n}$ and $\bar\sigma_{d_m}$ with dimension $n$ and $m$, their correlator can be expressed in terms of $\widetilde M_{n-1}^{\bar {J}_\beta I_\alpha}$ from eq.\,(\ref{snm}) as well. Finally we have
\beqn
g^{(n)}_{c_n\bar{d}_m}\!\!&\equiv&\!\delta_{nm}\,\left\langle\sigma_{c_n}(0)\,\bar\sigma_{d_n}(\infty)\right\rangle_{\mathbb R^2}=\delta_{nm}\,\big\langle\hat{\sigma}_{c_n}(N)\,\hat{\bar\sigma}_{d_n}(S)\big\rangle_{S^2}\nonumber\\
\=\delta_{nm}\,\bigg(\left\langle\sigma_{c_n}(N)\,\bar\sigma_{d_n}(S)\right\rangle_{S^2}\nonumber\\
\-\sum_{\alpha,\beta=0}^{n-1}\,\sum_{I_\alpha,{\bar J}_\beta}\left\langle\sigma_{c_n}(N)\,{\bar\sigma}_{{J}_\beta}(S)\right\rangle_{S^2}\,\widetilde M_{n-1}^{\bar {J}_\beta I_\alpha}\,\left\langle\sigma_{I_\alpha}(N)\,{\bar\sigma}_{d_n}(S)\right\rangle_{S^2}\bigg)\,.
\label{gnm}
\eeqn
eq.\,(\ref{gnm}) is the main formula that will be applied later. It automatically satisfies the $tt^\ast$-equations as eq.\,(\ref{g2}) and (\ref{g3}). However since we choose the OPE basis eq.\,(\ref{OPEb}) regardless of the equivalence relations in the rings, there would be additional constraints imposed on these metrics $g_{c_n\bar{d}_m}$. We will discuss these constraints next section in the context of Calabi-Yau manifolds.\\\\

\section{Chiral ring data in Calabi-Yau manifolds}
\label{4}
In this section, we will consider $d$-(complex) dimensional compact Calabi-Yau manifolds as our important examples of \ntt\ SCFTs with center charge $c=3d$. Discussion in the first two subsections is focused on the complex moduli space of CY-manifolds with one complex dimension, or say the chiral ring generated by a single marginal chiral primary, then  generalized to complex moduli of higher dimensions. We will reconstruct $tt^*$-equations and their solutions, the chiral ring data, via variation of the Hodge structure of the horizontal cohomology classes in Calabi-Yau manifolds. In fact the equivalence between geometries of Hodge structure and $tt^*$-equations has been long time known \cite{Vafa}. The chiral ring data are uniquely determined by $tt^*$-equations if we have full OPE data $C_{IJ}^{K}$ which can be obtained from topological twist \cite{Warner, Dijkraaf, Witten2, Witten3, Witten4}. 

On the other hand, in this note, we resort to an alternative route to find chiral ring data, i.e. extracting them directly from partition functions with deformation eq.\,(\ref{main}). In this scenario, it is \emph{not} necessary to know any OPE data, which allows us to simply work under the ``shift" OPE basis (\ref{OPEb}). However the price we pay is that we are blind for equivalence relations in the chiral rings, and we have to collect linear independent operators out of excessively many linear dependent ones. The algorithm developed in section \ref{3} resolves this problem. It therefore must be compatible with $tt^*$-equations as well as the geometries of Hodge structure in the context of CY-manifolds.

Indeed, in this section, we will show that the Gram-Schmidt orthogonalization in the algorithm admits a natural geometric interpretation as Griffith transversality with projections \cite{Griffiths, Morrison, Alim} on the Hodge structure of CY-manifolds, and there will be more constraints imposed on the chiral ring data if we use the ``shift" OPE basis ($\ref{OPEb}$). On the contrary, Griffiths transversality with projection on generic complex moduli of higher dimensions can be reciprocally defined with help of the algorithm as a spinoff in section \ref{4.3}. \\

\subsection{Variation of Hodge structure}
\label{4.1}
For a given $d$-dimensional Calabi-Yau manifold $\mc Y$, its metric deformations fall into two families: K\"{a}hler class and complex structure deformations, both of which form the moduli spaces of $\mc Y$. The K\"{a}hler and complex moduli, labeled as $\mc M_K(\mc Y)$ and $\mc M_C(\mc Y)$, themselves are K\"{a}hler manifolds\footnote{We will not consider CY-manifolds like $K3$ surface which corresponds to a $\mc N=(4,4)$ SCFT with a homogenous moduli space.} with dimensions $h^{1,1}(\mc Y)$ and $h^{d-1,1}(\mc Y)$, the non-trivial Hodge numbers of $\mc Y$. On the other hand, a supersymmetric sigma model $\mc S$ with the target space of $\mc Y$ is a \ntt\ SCFT with center charge $c=3d$. The moduli space $\mc M_{\rm c}(\mc S)$ spanned by marginal primaries coincides with $\mc M_C(\mc Y)$, and $\mc M_{\rm tc}(\mc S)$ with $\mc M_K(\mc Y)$. 

In fact the identification between $\mc M_{\rm c}$ and $\mc M_C$ can be extended to the chiral ring bundle $\mc V$ over $\mc M_{\rm c}$ with the fiber of chiral ring $\mc R$, and the Hodge bundle $\mc H$ over $\mc M_C$, via a bundle map $\varphi$,
\beq\varphi:\ \mc V\simeq\mc H\,,\eeq
where $\mc H$ has fibers $H^d(\mc Y, \mathbb C)$, i.e. the horizontal cohomology classes of $\mc Y$. We will show, in the case of $\mc M_C$ being one-dimensional, that the chiral ring data in $\mc V$ can be computed via the canonical non-degenerate bilinear form on the fiber $H^d(\mc Y, \mathbb C)$ once after we specify the isomorphism $\varphi$. Below we will explain in concrete how the $\varphi$ can be introduced by identifying elements in the chiral ring $\mc R$ and $H^d(\mc Y, \mathbb C)$ on a fixed fiber of $\mc V$ and $\mc H$. 

A given point $\tau$ on $\mc M_C$ specifies a complex structure of $\mc Y$, we therefore have a Hodge decomposition of the fiber
\beq
H^d(\mc Y_\tau, \mathbb C)\simeq \bigoplus_{\alpha+\beta=d}H^{\beta,\alpha}(\mc Y_\tau)\,,\eeq
respect to the complex structure. A holomorphic $d$-form $\Omega(\tau)$ spanning $H^{d,0}(\mc Y_\tau)$ is a natural section over a line bundle $\mc L\subset\mc H$. We want to consider how $\Omega(\tau)$ varies respect to $\tau$. It turns out that moving on the moduli space $\mc M_C$ will change complex structure, so that $\partial_\tau\Omega$ will not be in $H^{d,0}(\mc Y_\tau)$. In general, for any element in $H^{\beta,\alpha}(\mc Y_\tau)$, its variation respect to $\tau$ will not be an element in $H^{\beta,\alpha}(\mc Y_\tau)$ anymore. The inherent conflict between the basis of $H^{\beta,\alpha}(\mc Y_\tau)$ and that varying holomorphically on $\tau$ has been recognized since the work of Griffiths \cite{Griffiths}. To circumvent this conflict, instead of working on $H^{\beta,\alpha}(\mc Y_\tau)$, one considers the Hodge filtration $\mc F^{\bullet}(\mc Y)=\{\mc F^\beta(\mc Y)\}^d_{\beta=0}$ as
\beq
\mc F^\beta=\bigoplus_{a\geq \beta}H^{a,d-a}(\mc Y)\,,\ \ \ H^d(\mc Y,\mathbb C)=\mc F^0\supset\mc F^1\supset...\supset\mc F^{d+1}=0	\,.\eeq
Now the variation $\partial_\tau$ equipping on $\mc F^\beta$ is a flat connection, or say the Gauss-Manin connection, with the important property, see also \cite{Alim},
\beq
\partial_\tau:\ \mc F^\beta\longrightarrow\mc F^{\beta-1}\,,
\label{GT}
\eeq
which is called Griffiths transversality. The $d$-form $\Omega(\tau)\in\mc F^d$ varies holomorphically on the pages $\{\mc F^\beta\}$. To project elements in $\mc F^\beta$ back to states in $H^{\beta,\alpha}(\mc Y_\tau)$, we introduce the anti-holomorphic filtration, $\overline{\mc F^{\bullet}(\mc Y)}=\left\{ \overline{\mc F^\alpha(\mc Y)}\right\}^d_{\alpha=0}$ by the virtue of $\overline{H^{\beta,\alpha}(\mc Y_\tau)}=H^{\alpha,\beta}(\mc Y_\tau)$, and we thus have
\beq
H^{\beta,\alpha}(\mc Y_\tau)=\mc F^\beta(\mc Y)\cap\overline{\mc F^\alpha(\mc Y)}\,.
\label{FF}
\eeq
Eq.\,(\ref{FF}) accompanied by the canonical non-degenerate pair $\langle\cdot\,,\cdot\rangle$ on compact $\mc Y$ will project $\partial_\tau^n\Omega$ to various $(\beta,\alpha)$-pure states in $H^{\beta,\alpha}(\mc Y_\tau)$, where
\beq\left\langle\cdot\,,\cdot\right\rangle:\ H^{\beta,\alpha}(\mc Y_\tau)\times H^{d-\beta,d-\alpha}(\mc Y_\tau)\longrightarrow\mathbb R\,.\eeq
Especially for the $d$-form $\Omega(\tau)$ and its complex conjugate $\overline{\Omega}(\bar\tau)$, we have the following result \cite{Tian},
\beq
{\rm e}^{-K(\tau,\bar\tau)}=\left\langle\Omega(\tau)\,,\overline{\Omega}(\bar\tau)\right\rangle\equiv i^{d^2}c_d\int_{\mc Y}\Omega(\tau)\wedge\overline{\Omega}(\bar\tau)\,,
\label{K}
\eeq
where $K(\tau,\bar\tau)$ is the K\"{a}hler potential of moduli space $\mc M_C(\mc Y)$, and $c_d$ is a specific constant, the degree of $\mc M_C$. With eq.\,(\ref{GT}), (\ref{FF}) and (\ref{K}), we are able to project, for example, 
$$\partial_\tau\Omega\in H^{d,0}\oplus H^{d-1,1}\,,$$
onto $H^{d-1,1}$. Explicitly, we decompose
$$\partial_\tau\Omega=\omega\,\Omega+X^{(d-1,1)}\,,$$
where $X^{(d-1,1)}\in H^{d-1,1}$ and $\omega$ is about to be determined. Wedging $\overline\Omega(\bar\tau)$ and by the virtue of its anti-holomorphy, we find 
\beq
X^{(d-1,1)}=\partial_\tau\Omega-\frac{\big\langle\partial_\tau\Omega\,,\overline{\Omega}\big\rangle}{\big\langle\Omega\,,\overline{\Omega}\big\rangle}\,\Omega=(\partial_\tau+\partial_\tau K)\,\Omega\equiv D_\tau\Omega\in H^{d-1,1}(\mc Y_\tau)\,,
\label{D1}
\eeq
where 
$$D_\tau:\,\mc L\rightarrow\mc L\otimes\mc T^*\mc M_C$$ 
is defined as a covariant derivative from bundle $\mc L$ to $\mc L\otimes\mc T^*\mc M_C$ with fiber $H^{d-1,1}(\mc Y_\tau)$. One can further express the metric of $\mc M_C$ as
\beq
g_{\tau\bar\tau}=\partial_\tau\partial_{\bar\tau}K=-\frac{\left\langle D_\tau\Omega\,,\overline{D_\tau\Omega}\right\rangle}{\left\langle \Omega\,,\overline{\Omega}\right\rangle}=-i^{d^2}c_d\,{\rm e}^{K}\int_{\mc Y}D_\tau\Omega\wedge\overline{D_\tau\Omega}\,.
\label{metric}
\eeq
Eq.\,(\ref{D1}) and (\ref{metric}) imply that the bundle map $\varphi$, on a fixed fiber, is specified as
\beq
\varphi:\ \mathds 1\mapsto\Omega\,,\ \ \ \phi\mapsto D_\tau\Omega\,,
\label{bm}
\eeq
up to a scaling factor, where $\phi$ is the marginal primary in the chiral ring $\mc R$, meanwhile the OPE in $\mc R$ should be identified to the covariant derivative $D_\tau$, see also \cite{Morrison}.

Acting an additional $\partial_\tau$ onto $\partial_\tau\Omega$, by Griffith transversality we have
$$\partial_\tau^2\,\Omega=\omega_1\,\Omega+\omega_2\,D_\tau\Omega+X^{(d-2,2)}\,,$$
such that $X^{(d-2,2)}\in H^{d-2,2}$\,. Successively wedging $\overline\Omega$ and $\overline{D_\tau\Omega}$, after some algebra,  one arrives
\beqn
X^{(d-2,2)}&=&\partial^2_\tau\Omega-\frac{\left\langle\partial^2_\tau\Omega\,,\overline{\Omega}\right\rangle}{\left\langle\Omega\,,\overline{\Omega}\right\rangle}\,\Omega-\frac{\left\langle\partial^2_\tau\Omega\,,\overline{D_\tau\Omega}\right\rangle}{\left\langle D_\tau\Omega\,,\overline{D_\tau\Omega}\right\rangle}\,D_\tau\Omega\nonumber\\\nonumber\\
&=&(\partial_\tau+\partial_\tau K-\Gamma^{\tau}_{\tau\tau})D_\tau\Omega\equiv D_\tau(D_\tau\Omega)\in H^{d-2,2}(\mc Y_\tau)\,,
\label{D2}
\eeqn
where $\Gamma^{\tau}_{\tau\tau}$ is the Christoffel connection respect to metric $g_{\tau\bar\tau}$, and 
$$D_\tau:\,\mc L\otimes\mc T^*\mc M_C\rightarrow\mc L\otimes{\rm Sym}_2\mc T^*\mc M_C\ {\rm with\ fiber}\ H^{d-2,2}(\mc Y_\tau)\,.$$ 
The state $D^2_\tau\,\Omega$ in $H^{d-2,2}(\mc Y_\tau)$ is identified to $\phi\cdot\phi\equiv\phi^2$ according to our OPE basis.

One can repeat the process by successively acting $\partial_\tau$ and projection to find more pure states in $H^{d-\alpha,\alpha}(\mc Y_\tau)$ with $\alpha=0,1,...,d$,
\beq
X^{(d-\alpha,\alpha)}=\partial^\alpha_\tau\Omega-\sum_{\beta=0}^{\alpha-1}\frac{\left\langle\partial^\alpha_\tau\Omega\,,\overline{D^\beta_\tau\Omega}\right\rangle}{\left\langle D^\beta_\tau\Omega\,,\overline{D^\beta_\tau\Omega}\right\rangle}\,D^\beta_\tau\Omega\equiv D^\alpha_\tau\Omega\in H^{d-\alpha,\alpha}(\mc Y_\tau)\,\,,
\label{Dq}
\eeq
which hence specifies the bundle map 
\beq\varphi|_{\mc R}:\ \ \left\{\mathds 1,\,\phi,\,\phi^2...,\,\phi^d\right\}\mapsto\left\{\Omega,\,D_\tau\Omega,\,D^2_\tau\Omega,...,\,D^d_\tau\Omega\right\}\,.\eeq

Comparing eq.\,(\ref{D1}), (\ref{D2}) as well as eq.\,(\ref{Dq}) to eq.\,(\ref{s1}), (\ref{s2}) and (\ref{sn}),
we find that the Griffiths transversality with projection naturally gives rise to the Gram-Schmidt orthogonalization we developed in last section. The reason behind is actually simple: First, from the geometric perspective, the bundle map $\varphi$ gives correct correspondence between chiral ring operators $\phi^\alpha$ and cohomology states $D_\tau^\alpha\Omega$ in $H^{d-\alpha,\alpha}(\mc Y_\tau)$, both of which are graded in dimension and degree $\alpha$. Therefore the correct chiral ring data will be expressed in terms of these states. On the other hand, the partition function $Z_B$ computed via localization respect to $\mf{su}(2|1)_B$ is exactly eq.\,(\ref{K}) \cite{Gomis3, Gomis4},
\beq
Z_B[S^2]={\rm e}^{-K(\tau,\bar\tau)}=\left\langle\Omega(\tau)\,,\overline{\Omega}(\bar\tau)\right\rangle\equiv i^{d^2}c_d\int_{\mc Y}\Omega(\tau)\wedge\overline{\Omega}(\bar\tau)\,.\eeq
Overall, to produce correct correlators from $Z_B[S^2]$, one has to apply Griffiths transversality with projection respect to the states' degree $\alpha$, which is nothing more than the Gram-Schmidt orthogonalization respect to the operators' dimension $\alpha$.\\

\subsection{$tt^*$-equations of chiral ring data on complex moduli}
Let us now work out in detail the $tt^*$-equations of chiral ring data on one-dimensional complex moduli. It turns out to be the Toda chain equations with constraints. The derivation is only based on the orthogonality of the pure states $D^\alpha_\tau\Omega\in H^{d-\alpha,\alpha}(\mc Y_\tau)$, or say the chiral primaries. In this sense, the $tt^*$-equations that the chiral ring data need to satisfy are universal for both $2d$ and $4d$ cases \cite{Komargodski}. However with the help of geometry, we will see that there are more constraints that the $2d$ ones must obey, which as we emphasized is due to the nilpotency of the $2d$ chiral ring. 

For simplicity, we label the chiral ring data as
\beq
g^{(\alpha)}\equiv(-1)^\alpha\,\frac{\left\langle \phi_\alpha\,,\overline{\phi_\alpha}\right\rangle}{\left\langle \phi_0\,,\overline{\phi_0}\right\rangle}\,,\ \ {\rm with}\ \ \phi_\alpha\equiv D^\alpha_\tau\Omega\,,\ \ \alpha=0,1,...,d\,.
\label{gq}
\eeq
Before establishing equations on $g_\alpha$'s, we first prove a useful lemma that 
\beq
lemma:\ \ \ \partial_{\bar\tau}\phi_\alpha\in\bigoplus_{\beta=1}^\alpha H^{d-\alpha+\beta,\alpha-\beta}(\mc Y_\tau)\,,\ \ {\rm for}\ \ \phi_\alpha\in H^{d-\alpha,\alpha}(\mc Y_\tau)\,.
\label{lemma}
\eeq
It can be shown from eq.\,(\ref{Dq}) by the holomorphicity of $\partial^q_\tau\Omega$ and induction on $q$. Next we show that 
\beq
\partial_\tau\phi_\alpha\in H^{d-\alpha,\alpha}(\mc Y_\tau)\oplus H^{d-\alpha-1,\alpha+1}(\mc Y_\tau)\,.\eeq
From eq.\,(\ref{Dq}) and Griffiths transversality, $\partial_\tau\phi_\alpha\in\bigoplus_{\beta=0}^{\alpha+1}H^{d-\beta,\beta}$. Further by wedging $\overline{\phi_\beta}$ for $\beta=0,1,...,\alpha-1$, we have
$$\big\langle \partial_\tau\phi_\alpha\,,\overline{\phi_\beta}\big\rangle=-\big\langle \phi_\alpha\,,\partial_\tau\overline{\phi_\beta}\big\rangle=0\,\ \ {\rm for}\ \ \beta=0,1,...,\alpha-1\,,$$
where the second equality is due to lemma (\ref{lemma}). Therefore we can express $\phi_{\alpha+1}$ in eq.\,(\ref{Dq}) in terms of $\phi_\alpha$,
\beq\partial_\tau\phi_\alpha=\phi_{\alpha+1}+\Gamma_\alpha\,\phi_\alpha\,,\eeq
and further determine $\Gamma_\alpha$ as
\beqn 
\Gamma_\alpha=\frac{\big\langle \partial_\tau\phi_\alpha\,,\overline{\phi_\alpha}\big\rangle}{\big\langle \phi_\alpha\,,\overline{\phi_\alpha}\big\rangle}=\partial_\tau\log\big\langle \phi_\alpha\,,\overline{\phi_\alpha}\big\rangle=-\partial_\tau\,K+g^{{(\alpha)}-1}\partial_\tau g^{(\alpha)}\,,
\label{connection}
\eeqn
where lemma (\ref{lemma}) is used in the second equality. For $\alpha=1$, $\Gamma_1$ is the standard connection on $\mc L\otimes\mc T^*\mc M_C$, see eq.\,(\ref{D2}), and for arbitrary $\alpha$, it serves as the connection on subbundle $\mc L\otimes{\rm Sym}_\alpha\mc T^*\mc M_C$. It will be seen more explicitly when treating higher dimensional moduli. We thus define the covariant derivative $D_\tau:\,\mc L\otimes{\rm Sym}_\alpha\mc T^*\mc M_C\rightarrow \mc L\otimes{\rm Sym}_{\alpha+1}\mc T^*\mc M_C$
\beq
\phi_{\alpha+1}=D_\tau\phi_\alpha\equiv\left(\partial_\tau-\Gamma_\alpha\right)\phi_\alpha\,.
\label{q+1q}
\eeq
With eq.\,(\ref{q+1q}), we have
\beqn
\left\langle \phi_{\alpha+1}\,,\overline{\phi_{\alpha+1}}\right\rangle=-\left\langle \partial_{\bar{\tau}}D_\tau\phi_{\alpha}\,,\overline{\phi_{\alpha}}\right\rangle=\left(\sum_{\beta=0}^\alpha\partial_{\bar\tau}\Gamma_\beta\right)\left\langle \phi_{\alpha}\,,\overline{\phi_{\alpha}}\right\rangle\,,
\label{iteration}
\eeqn
where the last equality is obtained by repeatedly computing the commutator $\left[\partial_{\bar\tau}, D_\tau\right]$ and applying lemma (\ref{lemma}). One can further rewrite eq.\,(\ref{iteration}) as
\beq
\partial_{\bar\tau}\Gamma_\alpha=\partial_{\bar\tau}\partial_\tau\log\left\langle \phi_{\alpha}\,,\overline{\phi_\alpha}\right\rangle=\frac{\left\langle \phi_{\alpha+1}\,,\overline{\phi_{\alpha+1}}\right\rangle}{\left\langle \phi_{\alpha}\,,\overline{\phi_{\alpha}}\right\rangle}-\frac{\left\langle \phi_{\alpha}\,,\overline{\phi_{\alpha}}\right\rangle}{\left\langle \phi_{\alpha-1}\,,\overline{\phi_{\alpha-1}}\right\rangle}\,,
\label{Toda1}
\eeq 
or in terms of $g^{(\alpha)}$'s by eq.\,(\ref{gq})
\beqn
&&\partial_{\bar\tau}\partial_\tau\log Z=-g^{(1)}\,,\nonumber\\
&&\partial_{\bar\tau}\partial_\tau\log g^{(\alpha)}=\frac{g^{(\alpha)}}{g^{(\alpha-1)}}-\frac{g^{(\alpha+1)}}{g^{(\alpha)}}+g^{(1)}\,,\ \ {\rm for}\ \ 1\leq \alpha\leq d-1\,,\nonumber\\
&&\partial_{\bar\tau}\partial_\tau\log g^{(d)}=\frac{g^{(d)}}{g^{(d-1)}}+g^{(1)}\,,\ \ {\rm with}\ \ g^{(0)}=1
\label{Toda2}
\eeqn
i.e. the celebrated Toda chain equations as one-dimensional $tt^*$-equations. 

Now let us figure out the constraints imposed on $g_\alpha$'s. First noticing that $\phi_d\in H^{0,d}(\mc Y_\tau)$ is linear dependent on the anti-holomorphic $d$-form $\overline\Omega=\overline{\phi_0}$, we thus write
\beq\phi_d=\mc C^{(d)}\,{\rm e}^{K}\,\overline{\phi_0}\,,\eeq
where ${\rm e}^{K}$ is for convenient normalization. $\mc C^{(d)}$ is determined by wedging $\phi_0$ as
\beq
\mc C^{(d)}(\tau)=\left\langle \phi_{0}\,,{\phi_{d}}\right\rangle=\left\langle\phi\cdot\phi\cdot...\cdot\phi\right\rangle_{S^2}
\label{dpt}
\eeq
$\mc C^{(d)}$ is actually the $d$-point chiral correlation function computed via B-twist on $S^2$ \cite{Morrison}. Its holomorphicity on $\tau$ can be shown by acting $\partial_{\bar\tau}$ and use lemma (\ref{lemma}). In terms of $\mc C^{(d)}$, one can relate $\phi_\alpha$ with $\overline{\phi_{d-\alpha}}$ as
\beq\phi_\alpha=\mc C^{(d)}\,{\rm e}^{K}\,\left(g^{(d-\alpha)}\right)^{-1}\,\overline{\phi_{d-\alpha}}\,,\ \ \ \overline{\phi_\alpha}=\overline{\mc C^{(d)}}\,{\rm e}^{K}\,\left(g^{(d-\alpha)}\right)^{-1}\,\phi_{d-\alpha}\eeq
Therefore we have the additional constraints imposed on $g^{(\alpha)}$'s
\beq
g^{(\alpha)}\,g^{(d-\alpha)}={\rm e}^{2K}\,\overline{\mc C^{(d)}}\,\mc C^{(d)}\,\ \ {\rm for}\ \ \alpha=1,2...,d\,.
\label{constraints}
\eeq 
Eq.\,(\ref{Toda2}) together with constraints (\ref{constraints}) will completely determine the \emph{full} chiral ring data of one-dimensional complex moduli under the ``shift" OPE basis (\ref{OPEb}). The constraints (\ref{constraints}) in turn will give consistency check of our computation in next section.

In the end of this subsection, we make some remarks on the $tt^*$-equations for states in NS and Ramond sectors. If we do not normalize the vacuum state $\mathds 1$ with $\langle\mathds 1\,,\overline{\mathds 1}\rangle=1$, the $tt^*$-equations (\ref{Toda1}) is actually derived in Ramond sector \cite{Vafa}. Meanwhile the $tt^*$-equations in NS sector need to be modified by an additional piece in Eq.\,(\ref{tt}), see also \cite{Boer}, where they derived the equations from the OPE of SCFTs without twisting the theories. 

On the other hand, it has been shown in the work of \cite{Gomis4} that, if one places the states $\phi_\alpha$ and its complex conjugate $\overline{\phi_\alpha}$ on the north and south poles of the sphere and drags the sphere to an infinitely long cigar, the deformed partition function $Z_B[{\rm cigar}]$ will exactly realize Cecotti and Vafa's topological-antitopological twist construction. Therefore the unnormalized correlators $\left\langle \phi_{\alpha}\,,\overline{\phi_{\alpha}}\right\rangle$ are computed in Ramond sector and thus satisfy eq.\,(\ref{Toda1}). Furthermore, it is also known that partition function $Z_B[{S^2}]=Z_B[{\rm cigar}]$ on round sphere. And computing $Z_B[{S^2}]$ treats fermionic states in NS sector. So we should expect that the $tt^*$-equations (\ref{tt}) in NS sector would be obtained with appropriate normalizations of states $\phi_\alpha$. 

Indeed, the additional diagonal piece in eq.\,(\ref{tt}) only matters with normalization. The ambiguous normalization is encoded in the non-trivial curvature of line bundle $\mc L$ \cite{Boer, Komargodski, Morrison}. If the states are normalized as eq.\,(\ref{gq}), we will reach $tt^*$-equations (\ref{Toda2}), where the additional piece $g_1$ is the curvature of $\mc L$. To normalize the states in standard NS sector, we require the unique vacuum state $\phi_0$ and highest chiral state $\phi_d$ normalized as
$$\left|\mathds 1\right\rangle\equiv{\rm e}^{\frac{1}{2}K}\phi_0\,,\ \ {\rm and}\ \ \ \left|\,d\,\right\rangle\equiv{\rm e}^{-\frac{1}{2}K}\phi_d\,,$$ 
so that \cite{Boer}
$$\langle\, \bar{\mathds 1}\,|\mathds 1\,\rangle=1\,,\ \ {\rm and}\ \ \ \langle \bar d\,|\,d\,\rangle=\overline{\mc C^{(d)}}\,\mc C^{(d)}\,.$$
All other states $\phi_\alpha$ are uniformly placed between $\left[-\frac{K}{2},\,\frac{K}{2}\right]$ respect to their degree $q$, 
$$|\,\alpha\,\rangle\equiv{\rm e}^{\frac{1}{2}\left(1-\frac{2}{d}\alpha\right)K}\,\phi_\alpha\,.$$
With these normalizations, we restore the $tt^*$-equations (\ref{tt}) in NS sector.\\

\subsection{Chiral ring data in complex moduli of higher dimensions}
\label{4.3} 
Now we generalize the previous results to the case of complex moduli of higher dimensions. The equations we will construct for chiral ring data are essentially the $tt^*$-equation in ``shift" OPE basis (\ref{OPEb}). We assume that \emph{all} primitive generators belong to $H^{d-1,1}(\mc Y_\tau)$. Similar to one-dimensional situation, we start from a holomorphic $d$-form
\beq\phi_0\equiv\Omega(\tau^i)\,,\eeq 
parametrized by moduli coordinates $\{\tau^i\}$ and spanning $H^{d,0}(\mc Y_\tau)$. States in $H^{d-\alpha,\alpha}(\mc Y_\tau)$ can be further build out via Griffiths transversality with projection. For simplicity, we consider \emph{unnormalized} chiral ring data and construct their equations in Ramond sector by degree $\alpha$.\\


\noindent$\alpha=1$: Obviously, for states $\phi_{i}$ with degree one, they span the cotangent space of $\mc M_C$, where
\beq
\phi_{i}\equiv D_i\phi_0=\partial_i\phi_0-\frac{\left\langle\partial_i\phi_0\,,\overline{\phi_0}\right\rangle}{\left\langle\phi_0\,,\overline{\phi_0}\right\rangle}\,\phi=(\partial_i+\partial_i K)\,\phi_0\,,
\label{phi1}
\eeq
analogue to eq.\,(\ref{D1}). The unnormalized chiral ring data 
$$G^{(1)}_{i\bar j}\equiv-\left\langle \phi_i\,,\overline{\phi_j}\right\rangle=\left\langle \partial_{\bar j}D_i\phi_0\,,\overline{\phi_0}\right\rangle=g^{(1)}_{i\bar j}\cdot\left\langle \phi_0\,,\overline{\phi_0}\right\rangle\equiv g^{(1)}_{i\bar j}\,G^{(0)}_{0\bar 0}\,,$$
where $g^{(1)}_{i\bar j}$ is the metric of $\mc M_C$. Instead, we can rewrite it as
\beq
\partial_{\bar j}\left(\partial_iG^{(0)}_{0\bar 0}\,G^{(0)\bar 0 0}\right)=-G^{(1)}_{i\bar j}\,{G^{(0)\bar 00}}\,,
\label{tt0}
\eeq
where $G^{(0)\bar 0 0}$ is the inverse of $G^{(0)}_{0\bar 0}$. eq.\,(\ref{tt1}) is the $tt^*$-equation at degree $\alpha=0$ in Ramond sector, see also eq.\,(\ref{tt}).\\

\noindent$\alpha=2$: Because of the assumption that there is no primitive generators in $H^{d-2,2}(\mc Y_\tau)$, all states herein are generated by those with degree one. Similar to eq.\,(\ref{D2}) and eq.\,(\ref{q+1q}), let us spell them out,
\beqn
\phi_{ij}&\equiv&D_iD_j\phi_0=\partial_i\partial_j\phi_0-\frac{\left\langle\partial_i\partial_j\phi_0\,,\overline{\phi_0}\right\rangle}{\left\langle\phi_0\,,\overline{\phi_0}\right\rangle}\,\phi_0+\sum_{k,\bar l}G^{(1)\bar l k}{\left\langle\partial_i\partial_j\phi_0\,,\overline{\phi_l}\right\rangle}\,\phi_k\nonumber\\
&=&\partial_i\phi_j-\partial_iG^{(1)}_{j\bar l}\,G^{(1)\bar l k}\,\phi_k\,,
\label{phi2}
\eeqn
where $G^{(1)\bar l k}$ is the inverse of $G^{(1)}_{k\bar l}$ and the second line is obtained by lemma (\ref{lemma}) similar to eq.\,(\ref{q+1q}). Apparently $\phi_{ij}$ is symmetric respect to $i\,,j$, so the covariant derivative $D_i$ defines a map
$$D_i:\,\mc L\otimes\mc T^*\mc M_C\rightarrow\mc L\otimes{\rm Sym}_2\mc T^*\mc M_C\ {\rm with\ fiber}\ \ H^{d-2,2}(\mc Y_\tau)\,.$$ 
The unnormalized chiral ring data
\beq
G^{(2)}_{ik,\bar j\bar l}\equiv\left\langle \phi_{ik}\,,\overline{\phi_{jl}}\right\rangle=-\left\langle \partial_{\bar j}D_i\phi_k\,,\overline{\phi_l}\right\rangle=-\partial_{\bar j}\left(\partial_iG^{(1)}_{k\bar n}G^{(1)\bar nm}\right)G^{(1)}_{m\bar l}-\partial_{\bar j}\left(\partial_kG^{(0)}_{0\bar 0}\,G^{(0)\bar 0 0}\right)G^{(1)}_{i\bar l}\,,
\label{G2}
\eeq
where the computation is similar to eq.\,(\ref{iteration}). Using eq.\,(\ref{tt1}), we obtain the $tt^*$-equations at degree $\alpha=1$ in Ramond sector,
\beq
\partial_{\bar j}\left(\partial_iG^{(1)}_{k\bar n}G^{(1)\bar nm}\right)=-G^{(2)}_{ik,\bar j\bar l}G^{(1)\bar l m}+G^{(1)}_{k\bar j}\,{G^{(0)\bar 00}}\delta_i^{\,m}\,.
\label{tt1}
\eeq\\

\noindent$\alpha=3$: When constructing states $\phi_{ijk}\in H^{d-3,3}(\mc Y_\tau)$, we encounter the problem that one needs to figure out the ``inverse" of $G^{(2)}_{ik,\bar j\bar l}$, see also \cite{Morrison}, which in terms of the geometric data on $\mc M_C$ is
\beq
G^{(2)}_{ik,\bar j\bar l}\,G^{(0)\bar 00}=g^{(1)}_{i\bar j}g^{(1)}_{k\bar l}+g^{(1)}_{i\bar l}g^{(1)}_{k\bar j}-R^{(2)}_{i\bar jk\bar l}\,.
\label{2g-R}
\eeq
However, analogue to the discussion in section \ref{3.3}, $\left\{\phi_{ij}\right\}$ are not necessarily linearly independent. Therefore $G_{ik,\bar j\bar l}$ might be singular and not invertible. The resolution is still to pick up a maximal set of linearly independent states, denoted as $\{\phi_{I_2}\}\subset\{\phi_{ij}\}$. Therefore the unnormalized metric on $\mc L\otimes{\rm Sym}_2\mc T^*\mc M_C$
$$G^{(2)}_{I_2\bar J_2}\equiv\left\langle \phi_{I_2}\,,\overline{\phi_{J_2}}\,\right\rangle$$
can be defined, and its inverse $G^{(2)\bar J_2 I_2}$ exists. With the aid of $G^{(2)\bar J_2 I_2}$, we are able to obtain states projected onto $H^{d-3,3}(\mc Y_\tau)$,
\beqn
\phi_{ijk}&\equiv& D_iD_jD_k\phi_0\equiv\partial_{i}\partial_{j}\partial_{k}\phi_0-\sum_{\alpha=0}^2\sum_{I_\alpha,\bar J_\alpha}G^{(2)\bar J_\alpha I_\alpha}\left\langle \partial_{i}\partial_{j}\partial_{k}\phi_0\,,\overline{\phi_{J_\alpha}}\,\right\rangle\,\phi_{I_\alpha}\nonumber\\
&=&\partial_i\phi_{jk}-\partial_iG^{(2)}_{jk,\bar J_2}G^{(2)\bar J_2 I_2}\,\phi_{I_2}\,,
\eeqn    
where $\{I_0\}\equiv\{0\}$, $\{I_1\}\equiv\{i\}$, and $G^{(2)}_{jk,\bar J_2}=\left\langle \phi_{jk}\,,\overline{\phi_{J_2}}\,\right\rangle$\,. It can be shown similar to the argument in section \ref{3.3} that $\phi_{ijk}$ is well-defined respect to different choice of the maximal set $\{\phi_{I_2}\}$. $\phi_{ijk}$ is symmetric respect to $i$, $j$ and $k$, and $D_i$ defines a map
$$D_i:\,\mc L\otimes{\rm Sym}_2\mc T^*\mc M_C\rightarrow\mc L\otimes{\rm Sym}_3\mc T^*\mc M_C\ {\rm with\ fiber}\ \ H^{d-3,3}(\mc Y_\tau)\,.$$ 
The unnormalized chiral ring data
\beqn
G^{(3)}_{ikm,\,\bar j\bar J_2}\!\!&\equiv&\!-\left\langle \phi_{ikm}\,,\overline{\phi_{jJ_2}}\right\rangle=\left\langle \partial_{\bar j}D_i\phi_{km}\,,\overline{\phi_{J_2}}\right\rangle\nonumber\\
\=-\partial_{\bar j}\left(\partial_iG^{(2)}_{km,\bar L_2}G^{(2)\bar L_2K_2}\right)G^{(2)}_{K_2\bar J_2}-\partial_{\bar j}\left(\partial_kG^{(1)}_{m\bar n}G^{(1)\bar np}\right)G^{(2)}_{ip,\bar J_2}\nonumber\\
\-\partial_{\bar j}\left(\partial_mG^{(0)}_{0\bar 0}\,G^{(0)\bar 0 0}\right)G^{(2)}_{ik,\bar J_2}\,.\nonumber
\eeqn
Applying eq.\,(\ref{tt0}) and (\ref{tt1}), we obtain
\beq
\partial_{\bar j}\left(\partial_iG^{(2)}_{km,\bar J_2}G^{(2)\bar J_2K_2}\right)=-G^{(3)}_{ikm,\bar j\bar J_2}G^{(2)\bar J_2 K_2}+G^{(2)}_{km,\bar j\bar l}G^{(1)\bar lp}\delta_{ip}^{\,K_2}+G^{(1)}_{m\bar j}G^{(0)\bar 00}\delta_{ik}^{\,K_2}
\label{tt2}
\eeq
Choosing index $\{km\}\subset\{I_2\}$ and ``shift" OPE basis $(\ref{OPEb})$, we reconstruct eq.\,(\ref{tt}) up to degree $\alpha=2$. One can continue this procedure and reconstruct the $tt^*$-equations to all degrees. We will not go through the details. 

Now we turn to study the constraints imposed on the chiral ring data $G_{a_\alpha\bar b_\alpha}$, where $\{a_\alpha\}$ and $\{\bar b_\alpha\}$ enumerate all states with degree $\alpha$. Resembling one-dimensional case, first we have
$$\phi_{a_d}=\mc C^{(d)}_{a_d}\,{\rm e}^K\,\overline{\phi_0}\,,$$
from which we compute
\beq
\mc C^{(d)}_{a_d}(\tau)=\left\langle \phi_{0}\,,\phi_{a_d}\right\rangle\,.
\label{dpt2}
\eeq
It contains various $d$-point chiral correlators computed in B-twisted models. The holomorphicity is guaranteed by lemma (\ref{lemma}) as well. Further because of the symmetry of horizontal cohomology classes $H^d(\mc Y_\tau)$, we have
$$\phi_{a_\alpha}=\mc C_{a_\alpha I_{d-\alpha}}^{(d)}\,G^{(d-\alpha)\bar J_{d-\alpha}I_{d-\alpha}}\,\overline{\phi_{J_{d-\alpha}}}\,,\ \ \ \overline{\phi_{b_\alpha}}=\overline{\mc C^{(d)}_{b_\alpha J_{d-\alpha}}}\,G^{(d-\alpha)\bar J_{d-\alpha}I_{d-\alpha}}\,\phi_{I{d-\alpha}}\,,$$
where the index $a_\alpha I_{d-\alpha}$ specifies an element in $\{a_d\}$. Overall we have the constraints
\beq
G^{(\alpha)}_{a_\alpha\bar b_\alpha}=\mc C_{a_\alpha I_{d-\alpha}}^{(d)}\overline{\mc C^{(d)}_{b_\alpha J_{d-\alpha}}}\,G^{(d-\alpha)\bar J_{d-\alpha}I_{d-\alpha}}\nonumber
\eeq
or
\beq
g^{(\alpha)}_{a_\alpha\bar b_\alpha}=\mc C_{a_\alpha I_{d-\alpha}}^{(d)}\overline{\mc C^{(d)}_{b_\alpha J_{d-\alpha}}}\,{\rm e}^{2K}\,g^{(d-\alpha)\bar J_{d-\alpha}I_{d-\alpha}}
\label{constraints2}
\eeq
For example, in $d=3$ the CY-threefold case, putting $\alpha=2$ we obtain the constraint on eq.\,(\ref{2g-R}), see also \cite{Alim},
\beq
g^{(2)}_{ik,\bar j\bar l}=g^{(1)}_{i\bar j}g^{(1)}_{k\bar l}+g^{(1)}_{i\bar l}g^{(1)}_{k\bar j}-R^{(2)}_{i\bar jk\bar l}=\mc C^{(3)}_{ikm}\overline{\mc C^{(3)}_{jln}}\,{\rm e}^{2K}\,g^{(2)\bar n m}\,.
\label{constraints3}
\eeq
eq.\,(\ref{dpt2}) and (\ref{constraints}) will serve as consistency checks of computation in next section.

At last, we comment on when there are primitive generators $\phi_{\hat I_\alpha}$ of degree $\alpha\geq 2$ in $H^d(\mc Y_\tau)$. In this situation, we are \emph{unable} to establish the $tt^*$-equations of chiral ring data including $\phi_{\hat I_\alpha}$ only from Griffiths transversality, because we have insufficient \emph{input} data. Recall, from eq.\,(\ref{bm}), that Griffiths transversality establishes the relation between the OPE $\phi\cdot\mathds 1$ and $D_\tau\Omega$, and so forth. Therefore only when bridging the OPE $\phi_{\hat I_\alpha}\cdot\mathds 1$ and some operations acting on $\Omega$, can we establish corresponding equations on $\phi_{\hat I_\alpha}$. Fortunately the localization method discussed in section \ref{3} indeed provides enough input data to compute all chiral ring data, where the partition function with irrelevant deformation eq.\,(\ref{main}) can be regarded as the generating function of all (twisted) chiral ring data. Explicit examples and computation will be given in next section.\\\\

\section{Examples}
\label{5}
We will compute the twisted chiral ring data of compact Calabi-Yau manifolds. All of the examples, collected from \cite{Morrison2, Honma}, have GLSM descriptions at UV regime and flow to the CY geometric phase in deep infrared. The twisted chiral ring data encode geometries of K\"{a}hler moduli $\mc M_K(\mc Y)$ as well as the data of vertical cohomology classes of $\mc Y$. Our algorithm in section \ref{3} is designed for twisted chiral operator deformations and thus can be directly applied to these examples. On the other hand, the constraints imposed on chiral ring data are derived for complex moduli $\mc M_C(\mc Y)$ and horizontal cohomology classes of $\mc Y$ in section \ref{4}. With the property of mirror symmetry, they equally work for twisted chiral ring data as well.

The K\"{a}hler moduli $\mc M_K(\mc Y)$ are parameterized by marginal FI parameters $\{\tilde\tau,\,\bar{\tilde\tau}\}$,
$$\tilde\tau=\frac{\theta}{2\pi}+i\,r\,,$$
of the given GLSM at UV regime. The twisted chiral ring data, $\widetilde g_{\alpha}(\tilde\tau,\,\bar{\tilde\tau})$, as well as partition function $Z_A(\tilde\tau,\,\bar{\tilde\tau})$, are non-holomorphic functions of $\{\tilde\tau,\,\bar{\tilde\tau}\}$. Sometimes, it is convenient to perform all computatoin in the large volume limit of $\mc M_K(\mc Y)$, 
$$r\gg 0\,.$$ 
In this region, one can instead expand $Z_A$ and $\widetilde g^{(\alpha)}$ in terms of the flat coordinates $\{t,\,\bar t\,\}$ of $\mc M_K(\mc Y)$ \cite{Morrison2}, where the expression will be greatly simplified. The charts $\{t,\,\bar t\,\}$ is related to $\{\tilde\tau,\,\bar{\tilde\tau}\}$ via the ``mirror map",
\beq
t=f(\tilde\tau)\,,\ \ \ \ \bar t=\bar f(\bar{\tilde\tau})\,.
\label{mirror map}
\eeq
Therefore one express 
\beq
\widetilde g^{(\alpha)}(\tilde\tau,\,\bar{\tilde\tau})=\left(\frac{\partial f}{\partial \tilde\tau}\right)^{\alpha}\left(\frac{\partial \bar f}{\partial \bar{\tilde\tau}}\right)^{\alpha}g^{(\alpha)}(t,\,\bar t)\,,
\label{GGtilde}
\eeq
where $\alpha$, as before, labels the degree of twisted chiral ring data, and we have omitted other indexes for brevity. We will compute 
\beq g^{(\alpha)}(t,\,\bar t)=\mc O_{\rm pert.}\left(\frac{1}{{\rm Im}\,t}\right)+\mc O_{\rm inst.}\left({\rm e}^{2\pi i t}+{\rm c.c.}\right)\,,\eeq
where $\mc O_{\rm pert.}$ and $\mc O_{\rm inst.}$ respectively collect the perturbative series and non-perturbative corrections of $ g_{\alpha}(t,\,\bar t)$.\\

\subsection{The sextic fourfold: $X_6\subset \mathds P^5$}
\begin{table}[t]
\centering
\begin{tabular}{|c||c|c|c|}
	\hline Field & $U(1)$ & $U(1)_V$ & $U(1)_A$\\
	\hline $\Phi_i$ & $+1$ & $2\mf q$ & $0$\\
	\hline $P$ & $-6$ & $2-12\mf q$ & $0$\\
	\hline
\end{tabular}
\caption{The $U(1)$ gauge charge, $U(1)_V$ and $U(1)_A$ R-charge of matter fields $P$, $\Phi_i$ for $i=1,2...,6$.}
\label{T1}
\end{table}
The first example we consider is the Fermat sextic fourfold $X_6\subset \mathds P^5$ \cite{Honma}, defined by a degree six hypersurface in $\mathds P^5$. It can be realized as an $U(1)$ Abelian \ntt\ GLSM with matter content summarized in Table \ref{T1}. The model has one-dimensional K\"{a}hler moduli $\mc M_K(X_6)$ spanned by the twisted chiral primary $\Sigma$, as the field strength of $U(1)$ vector multiplet $V$, see also in appendix \ref{B}, associated to the marginal FI-coupling
$$\tilde\tau=\frac{\theta}{2\pi}+i\,r\,.$$
The model also has a superpotenial $W=PW_6(\Phi)$ where $W_6(\Phi)$ is a homogeneous degree six polynomial of $\Phi_i$. Although $W$ is supersymmetric exact respect to $\mf{su}(2|1)_A$, it restricts the $U(1)_V$ R-charge of the matter contents up to an arbitrary number $\mf q$. For convergent reason, we require $0<\mf q<\frac{1}{6}$ to compute the partition function in the same way of \cite{Morrison2}, using eq.\,(\ref{ZA}) and (\ref{main}),
\beq
Z_{X_6}={\rm e}^{-K(\tilde\tau,\bar{\tilde\tau})}=\sum_{m\in\mathds Z}{\rm e}^{-i\theta m}\int_{-\infty}^{+\infty}\frac{\rm d\sigma}{2\pi}{\rm e}^{-4\pi i\,r\sigma}\frac{\Gamma(\mf q-i\sigma-\frac{1}{2}m)^6\,\Gamma(1-6\mf q+6i\sigma+3m)}{\Gamma(1-\mf q+i\sigma-\frac{1}{2}m)^6\,\Gamma(6\mf q-6i\sigma+3m)}\,,
\label{ZX6}
\eeq
where $K(\tilde\tau,\bar{\tilde\tau})$ is the K\"{a}hler potential of $\mc M_K(X_6)$.
\subsubsection{Calabi-Yau phase}
In the Calabi-Yau phase, $r\gg 0$, the integral is evaluated as,
\beqn
Z^{\rm CY}_{X_6}=(\bar zz)^{\mf q}\oint\frac{\rm d\epsilon}{2\pi i}(\bar zz)^{-\epsilon}\,\frac{\pi^5\sin(6\pi\epsilon)}{\sin^6(\pi\epsilon)}\bigg|\sum_{k=0}^{\infty}z^k\frac{\Gamma(1+6k-6\epsilon)}{\Gamma(1+k-\epsilon)^6}\bigg|^2\,
\label{ZX6CY}
\eeqn
where $z={\rm e}^{2\pi i\tilde\tau}$ and the complex conjugate does not act on $\epsilon$. We will expand $Z_A$ respect to flat coordinates as mentioned before. Therefore, following \cite{Morrison2, Honma}, we perform a K\"{a}hler transformation,
\beq
K(\tilde\tau,\bar{\tilde\tau})\longrightarrow K(\tilde\tau,\bar{\tilde\tau})+\log T(z)+\log \overline{T(z)}\,,\ \ \ {\rm with}\ \ \ T(z)=z^{\mf q}\sum_{k=0}^\infty z^k\frac{\Gamma(1+6k)}{\Gamma(1+k)^6}
\label{KT}
\eeq
and read off the coefficient of $\log^3\bar z$, which defines the mirror map (\ref{mirror map}), 
$$2\pi i\,t=2\pi if(\tilde\tau)=\log z+6264z + 67484340z^2+1272752107200z^3+\cdots \,.$$
Inverse the map and expand $Z_A$ in terms of $\{t\,,\bar t\,\}$, we have
\beqn
Z^{\rm CY}_{X_6}(t,\bar t\,)\=\frac{1}{4}\,\xi^{-4}+840\,\zeta(3)\,{\xi}^{-1}\nonumber\\\nonumber\\
\+30248\,(\bar q+q\,)\left(\xi^{-2}+2\xi^{-1}\right)+609638400\,\bar qq+\mc{O}(q^2)+{\rm c.c.}\,\,,
\label{ZtX6}
\eeqn
where we have set
$$\xi\equiv\frac{1}{4\pi\,{\rm Im}\,t}\,,\ \ \ {\rm and}\ \ \ q\equiv{\rm e}^{2\pi i\,t}\,.$$ 
The first line of eq.\,(\ref{ZtX6}) includes all perturbative contributions, and the second line collects the non-perturbative ones starting from one-instanton correction to the K\"{a}hler potential. In the case of $X_6$ as a CY-fourfold with center charge $c=12$, we have five twisted chiral ring data, see eq.\,(\ref{gq}),
\beq
g^{(\alpha)}\left(t,\bar t\,\right)\equiv\big\langle\sigma^\alpha(0)\,\bar\sigma^\alpha(\infty)\big\rangle_{\mathbb R^2}\,,\ \ {\rm for}\ \ \alpha=0,1,2,3,4\,,
\eeq
with $g^{(0)}\equiv1$ by normalization, where $\sigma$ is the bottom component of twisted chiral primary $\Sigma$. Using eq.\,(\ref{ZtX6}), and (\ref{gnm}), we are able to compute all of them in the large volume limit ${\rm Im}\,t\sim{\rm Im}\,\tilde\tau\gg 0$\,,
\beqn
g^{(1)}\=\frac{4\xi^2 \left(1-1680\,\zeta (3)\,\xi^3\right)^2}{\left(1+3360\,  \zeta (3)\,\xi ^3\right)^2}-241920\,(\bar q+q)\,\left(\xi^3+\mc O({\xi^4})\,\right)\nonumber\\
&&\hspace{3.9cm}+12192768000\,\bar qq\,\left(\xi^4+\mc O(\xi^5)\,\right)+\mc O(q^2)+{\rm c.c.}\,,\nonumber\\\nonumber\\
g^{(2)}\=\frac{24\xi^4}{1+3360\,\zeta (3)\,\xi ^3  }+241920\,(\bar q+q\,)\left(\xi^4+\mc O({\xi^5})\right)\,\nonumber\\
&&\hspace{2.8cm}+2438553600\,\bar qq\,\left({\xi^4}+\mc O({\xi^6})\right)\,+\mc O(q^2)+{\rm c.c.}\,,\nonumber\\\nonumber\\
g^{(3)}\=\frac{144\xi^6}{\left(1-1680\, \zeta (3)\,\xi^3\right)^2}+2903040\,(\bar q+q\,)\left(\xi^6+\mc O(\xi^7)\right)\nonumber\\
&&\hspace{3.3cm}+58525286400\,\bar qq\,\left({\xi^6}+\mc O({\xi^7})\,\right)\,+\mc O(q^2)+{\rm c.c.}\,,\nonumber\\\nonumber\\
g^{(4)}\=\frac{576\xi^8}{\left(1+3360 \,  \zeta (3)\,\xi ^3\right)^2}+11612160\,(\bar q+q\,)\left(\xi^8+\mc O(\xi^9)\right)\nonumber\\
&&\hspace{3.3cm}+234101145600\,\bar qq\,\left({\xi^8}+\mc O({\xi^9})\,\right)+\mc O(q^2)+{\rm c.c.}\,,\nonumber\\
\label{X6crd}
\eeqn
up to one-instanton correction. The perturbative part of $g^{(\alpha)}$ is of closed forms, for the partition function (\ref{ZtX6}) on perturbative part is closed. 

To restore the twisted chiral ring data in original $\tilde\tau$-coordinates, we have
\beq
\widetilde g^{(\alpha)}(\tilde\tau,\bar{\tilde\tau})=\left(\frac{{\rm d}t}{{\rm d}\tilde\tau}\right)^{\alpha}{\left(\frac{{\rm d}\bar t}{{\rm d}\bar{\tilde\tau}}\right)}^{\alpha}g^{(\alpha)}\left(t(\tilde\tau),\bar t(\bar{\tilde\tau})\right)
\eeq

Now we can give some consistency checks of the eq.\,(\ref{X6crd}). First they satisfy the Toda chain eq.\,(\ref{Toda2}) as designed by the algorithm (\ref{gnm}). Second, we check the following consistency conditions (\ref{constraints}) up to four-instanton corrections, 
\beq
g^{(4)}=g^{(1)}\,g^{(3)}=\left(g^{(2)}\right)^2\,,\ \ {\rm and}\ \ \ g^{(\alpha)}=0\,,\ \ {\rm for}\ \ \alpha\geq 5\,.
\eeq
Finally it is interesting to compute the $4$-point chiral correlator (\ref{dpt}) in A-twisted topological theory on $S^2$, by use of Eq.\,(\ref{constraints}),
\beqn
\overline{\mc C^{(4)}}\,\mc C^{(4)}\=\Big|\big\langle\sigma\cdot\sigma\cdot\sigma\cdot\sigma\big\rangle_{S^2}\Big|^2=g^{(\alpha)}\,Z_A^2\nonumber\\
\=36\Big|1 + 20160\,q + 689472000\,q^2 + 24691154100480\,q^3\nonumber\\ 
\+ 903369974818590720\,q^4+\mc O(q^5)\Big|^2\nonumber
\eeqn
Therefore we have
\beqn
\mc C^{(4)}(q)\=6\Big(1 + 20160\,q + 689472000\,q^2 + 24691154100480\,q^3\nonumber\\
\!\!\!&+&\!\!\! 903369974818590720\,q^4+\mc O(q^5)\,\Big)\,,
\eeqn
which perfectly matches the results in \cite{Morrison}. It is worth of noticing that all twisted chiral ring data $g^{(\alpha)}$ are invariant respect to K\"{a}hler transformation, e.g. eq.\,(\ref{KT}), meanwhile $\mc C^{(4)}$ does depend on it. Our  choice eq.\,(\ref{KT}) is consistent with \cite{Morrison}.

\subsubsection{Landau-Ginzburg phase and localization on Higgs branch}
We can also compute the correlators in Landau-Ginzburg phase \cite{Witten}, $r\ll 0$. The integral (\ref{ZX6}) in the limit of $r\ll 0$ can be recast as \cite{Morrison2}
\beq
Z^{\rm LG}_{X_6}=\sum_{\alpha=0}^4Z^{(\alpha)}_{\rm cl}\,Z_{\rm 1-loop}^{(\alpha)}\,\overline{Z^{(\alpha)}_{\rm vortex}(z)}\, Z^{(\alpha)}_{\rm vortex}(z)
\label{ZX6B}
\eeq
with
\beqn
&&Z^{(\alpha)}_{\rm cl}={\rm e}^{4\pi\,r\cdot\frac{\alpha}{6}}=\left(\bar z z\right)^{-\frac{\alpha}{6}}\,,\nonumber\\
&&Z_{\rm 1-loop}^{(\alpha)}=\frac{(-1)^\alpha}{6}\frac{\Gamma\left(\frac{1+\alpha}{6}\right)^6}{\Gamma\left(1+\alpha\right)^2\,\Gamma\left(\frac{5-\alpha}{6}\right)^6}\,,\nonumber\\
&& Z^{(\alpha)}_{\rm vortex}(z)={}_{5}F_{4}\left(\left\{\frac{1+\alpha}{6},...,\frac{1+\alpha}{6}\right\};\,\left\{\frac{2+\alpha}{6},...,\widehat{1},...,\frac{6+\alpha}{6}\right\};\,\frac{1}{6^6z}\,\right)\,,
\nonumber
\eeqn
where $\mf q$ has been set to $\frac{1}{6}$ and $\widehat 1$ indicates the term one needs to omit. We can use eq.\,(\ref{gnm}) to compute $g_\alpha$ as before. However notice that $Z_A$ in Laudau-Ginzburg phase should be interpreted as the partition function evaluated onto Higgs branch \cite{Benini}, where $Z^{(\alpha)}_{\rm vortex}$ is the $U(1)$ vortex partition function in $\Omega$ background \cite{Shadchin}. It would be interesting if we can propose a different expression of (twisted) chiral ring data in terms of the vortex and anti-vortex partition functions.

Indeed for the $U(1)$ Abelian case, it is not hard to reformulate the twisted chiral ring data in terms of $Z^{(\alpha)}_{\rm vortex}$ and its complex conjugate. For convenience, we define the \emph{unnormalized} correlators, referred to eq.\,({\ref{gq}}), 
$$G^{(\alpha)}\equiv  \widetilde g^{(\alpha)}\, Z^{\rm LG}_{X_6}\,,$$ 
which satisfy the Toda chain equations (\ref{Toda1}). We define further
$$c_\alpha\equiv Z^{(\alpha)}_{\rm 1-loop},\ \ \ {\rm and}\ \ \ \mc F^{(\alpha)}(z)\equiv z^{-\frac{\alpha}{6}}Z^{(\alpha)}_{\rm vortex}(z)$$
and rewrite
$$Z^{\rm LG}_{X_6}=\sum_{\alpha=0}^4c_\alpha\,\overline{\mc F^{(\alpha)}}\,\mc F^{(\alpha)}=G^{(0)}\equiv \mf D_0\,.$$
Applying eq.\,(\ref{Toda1}) and a little algebra, it is easy to find
$$G^{(1)}=-\mf D_0\,\partial_{\bar\tau}\partial_\tau\log \mf D_0=-\frac{\mf D_1}{\mf D_0}\,,$$
with
$$\mf D_1=\frac{1}{2}\sum_{\alpha,\beta=0}^4c_\alpha c_\beta\left|\mc F^{(\alpha)}\partial_\tau\mc F^{(\beta)}-\mc F^{(\beta)}\partial_\tau\mc F^{(\alpha)}\right|^2\equiv\sum_{0\leq\alpha<\beta\leq 4}c_\alpha c_\beta\left|\mc W\left(\mc F^{(\alpha)},\mc F^{(\beta)}\right)\right|^2\,,$$
where 
$$\mc W\left(\mc F^{(\alpha)},\mc F^{(\beta)}\right)=\left|
\begin{array}{cc}
 \mathcal{F}^{(\alpha) } & \mathcal{F}^{(\beta) } \\
 \partial_\tau\mathcal{F}^{(\alpha) } & \partial_\tau\mathcal{F}^{(\beta) } \\
\end{array}
\right|\,,$$
is the Wronskian respect to $\mc F^{(\alpha)}$. In general, define
\beq
\mf D_n\equiv\sum_{0\leq\alpha_0<...<\alpha_{n}\leq 4}\!\!\!\!c_{\alpha_0}\cdots c_{\alpha_n}\left|\mc W(\mc F^{(\alpha_0)},...,\mc F^{(\alpha_n)})\right|^2\,,
\eeq
with 
$$\mc W\left(\mc F^{(\alpha_0)},...,\mc F^{(\alpha_n)}\right)=\left|
\begin{array}{cccc}
 \mc F^{(\alpha_0)} & \mc F^{(\alpha_1)} & ... & \mathcal{F}^{(\alpha_n)} \\
 \partial_\tau\mathcal{F}^{(\alpha_0)} & \partial_\tau\mathcal{F}^{(\alpha_1)} & ... & \partial_\tau\mathcal{F}^{(\alpha_n)} \\
 \vdots & \vdots & \ddots & \vdots \\
 \partial_\tau^n\mathcal{F}^{(\alpha_0)} & \partial_\tau^n\mathcal{F}^{(\alpha_1)} & ... & \partial_\tau^n\mathcal{F}^{(\alpha_n)} \\
\end{array}
\right|\,$$
the $n$-th Wronskian. One then can show an identity,
\beq
\partial_{\bar\tau}\partial_\tau\log \mf D_n=\frac{\mf D_{n+1}\,\mf D_{n-1}}{\mf D_n^2}\,.
\label{DnD}
\eeq 
A useful trick to derive this identity is to rewrite the Wronskian in terms of Pfaffian \cite{Hirota}, and prove it by induction on $n$. With the aid of eq.\,(\ref{DnD}), one can solve the Toda chain Eq.\,(\ref{Toda1}) as
\beq
G^{(\alpha)}=(-1)^\alpha\frac{\mf D_\alpha}{\mf D_{\alpha-1}}\,,\ \ \ {\rm for}\ \ \ \alpha=1\cdots 4
\label{GHiggs}
\eeq
In this sense, the twisted chiral ring data are expressed in \emph{closed forms}. One still needs to check if eq.\,(\ref{GHiggs}) satisfies the additional constraints (\ref{constraints}), which we indeed confirm correctly up to $\mc O\left(1/z^{10}\right)\,$. Especially for the nilpotency of twisted chiral ring data,
$$G^{(\alpha)}=0\,,\ \ \ {\rm for}\ \ \ \alpha\geq5\,,$$
are automatically guaranteed. In fact 
\beqn
\mc F^{(\alpha)}(z)&=&z^{-\frac{\alpha}{6}}Z^{(\alpha)}_{\rm vortex}(z)\nonumber\\
&=&\left(\frac{1}{z}\right)^{\frac{\alpha}{6}}\!\!{}_{5}F_{4}\left(\left\{\frac{1+\alpha}{6},...,\frac{1+\alpha}{6}\right\};\,\left\{\frac{2+\alpha}{6},...,\widehat{1},...,\frac{6+\alpha}{6}\right\};\,\frac{1}{6^6z}\,\right)\,,\nonumber
\eeqn
in the context of mirror symmetries, are the periods of the mirror manifold of $X_6$, and saturates the Picard-Fuchs equation \cite{Honma},
\beq
\left(\Theta^5-6z\prod_{k=1}^5\left(6\Theta+k\right)\right)\mc F^{(\alpha)}(z)=0\,\ \ \ {\rm with}\ \ \ \Theta\equiv z\frac{d}{dz}\,.
\label{PicardFuchs}
\eeq
Therefore, any $\mf D_\alpha$ with $\alpha\geq 5$ contains terms of $\Theta^5\mc F^{(\alpha)}$, which is linear dependent on its lower derivatives due to eq.\,(\ref{PicardFuchs}), and thus the Wronskian vanishes.

To end this subsection, we make some comments on the (twisted) chiral ring data in the formulation of localization onto Higgs branch. Firstly our derivation of eq.\,(\ref{DnD}) and (\ref{GHiggs}) is actually valid for any $U(1)$ Abelian GLSM. The $U(1)$ gauge group ensures that the (twisted) chiral ring is only generated by a single primitive generator with dimension one. The chiral ring data are thus dominated by the Toda chain eq.\,(\ref{Toda2}) \emph{universally}. On the other hand, localization formula onto Higgs branch \cite{Benini, Gomis} tells us that the partition function, 
\beq
Z_A=\sum_{\alpha=0}^nc_\alpha\,\overline{\mc F^{(\alpha)}(z)}\,\mc F^{(\alpha)}(z)\,,\eeq
is always of finite sum of factorizable building blocks, the vortex/anti-vortex partition functions, dressed up with factors of one-loop contributions as coefficients. Therefore, applying Toda chain eq.\,(\ref{Toda1}) and identity (\ref{DnD}), we can determine all chiral ring data as ratios of sum of factorizable blocks $\mc F^{(\alpha)}(z)$, see e.g. eq.\,(\ref{GHiggs}). In addition, the nilpotency of the chiral ring must be guaranteed, for $\mc F^{(\alpha)}(z)$ are in general some hypergeometric functions which will be linearly expressed by enough numbers of operator $\Theta=z\frac{d}{dz}$ acting on themselves. The argument above does not resort to any details of $\mc F^{(\alpha)}(z)$, while the constraints (\ref{constraints}) does depend on the expression of them in concrete.\\

\subsection{The determinantal Gulliksen-Neg\r{a}rd CY-threefold}
Our second example is the PAX/PAXY gauged linear sigma models introduced in \cite{Jockers} to describe the determinantal Calabi-Yau varieties. The gauge group is $U(1)\times U(2)$, associated to two $U(1)$ field strengths $\Sigma_0$ and $\Tr\,\Sigma_1$ as twisted chiral primaries. The matter fields, summarized in Table \ref{T2}, contain $8$ chiral multiplets $\Phi_a$ of gauge charge $+1$ under $U(1)_{\Sigma_0}$， $4$ chiral multiplets $P^i$ in the bifundamental representation of $U(2)_{\Sigma_1}\times U(1)_{\Sigma_0}$, and the 4 chiral multiplets $X_i$ in the antifundamental representation of $U(2)_{\Sigma_1}$, subject to the superpotential, 
$$W=\Tr\left(PA(\Phi)X\right)\,,$$
where $A(\Phi)=A^a\Phi_a$ and $A^a$ are $8$ constant $4\times 4$ matrices. 
\begin{table}[t]
\centering
\begin{tabular}{|c||c|c|c|c|}
	\hline Field & $U(1)_{\Sigma_0}$ & $U(2)_{\Sigma_1}$ & $U(1)_V$ & $U(1)_A$\\
	\hline $\Phi_a$ & $+1$ & $\mathbf 1$ & $2\mf q_\phi$ & $0$\\
	\hline $P^i$ & $-1$ & $\mathbf 2$ & $2-2\mf q_x-2\mf q_\phi$ & $0$\\
	\hline $X_i$ & $0$ & ${\bf \overline{2}}$ & $2\mf q_x$ & $0$\\
	\hline
\end{tabular}
\caption{The $U(1)_{\Sigma_0}$ and $U(2)_{\Sigma_1}$ gauge group representations, $U(1)_V$ and $U(1)_A$ R-charge of matter fields $\Phi_a$, $P_i$ and $X_i$, with $a=1,2,...,8$ and $i=1,2...,6$, in the PAX GLSM for GN-$CY_3$}
\label{T2}
\end{table}

The theory has two FI marginal couplings 
$$\tilde\tau_i=\frac{\theta_i}{2\pi}+i\,r_i\,,\ \ \ {\rm for}\ \ \ i=0,1\,,$$
corresponding to the two $U(1)$ factors of the gauge group. Therefore the K\"{a}hler moduli space is two-dimensional. The dimensions of the vertical cohomology classes, $\bigoplus_{i=0}^{3}H^{(i,i)}$, of ``PAX" $CY_3$ model are 
\beq
\{1,\,2,\,2,\,1\}\,,
\label{Hodge}
\eeq 
due to the symmetries of the cohomology ring. The twisted chiral ring thus has 6 elements with two primitive generators of dimension one.

The model has three phases respect to its FI-couplings. We restrict ourselves in the region of $r_0+2r_1\gg0$ and $r_1\gg0$ to compute the twisted chiral ring data. In this phase, it is convenient to use the linear combination of $\sigma_0$ and $\Tr\,\sigma_1$, which are the bottom components of $\Sigma_0$ and $\Tr\,\Sigma_1$ conjugate to $\tilde\tau_0$ and $\tilde\tau_1$, as the generators of the twisted chiral ring,
\beq
\chi_1\equiv\sigma_0\,,\ \ \ {\rm and}\ \ \ \chi_2\equiv\Tr\,\sigma_1-2\sigma_0\,.
\eeq
$\chi_1$ and $\chi_2$ are thus conjugate to $\tilde\tau_0+2\tilde\tau_1$ and $\tilde\tau_1$ respectively. According the ``shift" OPE basis (\ref{OPEb}), we are about to compute the following twisted chiral ring data,
\beqn
&&g^{(0)}=\left\langle\mathds 1\,\overline{\mathds 1}\,\right\rangle_{\mathbb R^2}\equiv1\,,\nonumber\\\nonumber\\
&&g^{(1)}=\left(
\begin{array}{cc}
 \big\langle\chi_1\,\overline{\chi_1}\big\rangle_{\mathbb R^2}\,, & \big\langle\chi_1\,\overline{\chi_2}\big\rangle_{\mathbb R^2} \\\\
 \big\langle\chi_2\,\overline{\chi_1}\big\rangle_{\mathbb R^2}\,, & \big\langle\chi_2\,\overline{\chi_2}\big\rangle_{\mathbb R^2} \\
\end{array}
\right)\,,\nonumber\\\nonumber\\\nonumber\\
&&g^{(2)}=\left(
\begin{array}{ccc}
 \big\langle\chi^2_1\,\overline{\chi_1^2}\big\rangle_{\mathbb R^2}\,, & \big\langle\chi^2_1\,\overline{\chi_1\chi_2}\big\rangle_{\mathbb R^2}\,, & \big\langle\chi^2_1\,\overline{\chi^2_2}\big\rangle_{\mathbb R^2} \\\\
 \big\langle\chi_1\chi_2\,\overline{\chi_1^2}\big\rangle_{\mathbb R^2}\,, & \big\langle\chi_1\chi_2\,\overline{\chi_1\chi_2}\big\rangle_{\mathbb R^2}\,, & \big\langle\chi_1\chi_2\,\overline{\chi^2_2}\big\rangle_{\mathbb R^2} \\\\
 \big\langle\chi^2_2\,\overline{\chi_1^2}\big\rangle_{\mathbb R^2}\,, & \big\langle\chi^2_2\,\overline{\chi_1\chi_2}\big\rangle_{\mathbb R^2}\,, & \big\langle\chi^2_2\,\overline{\chi^2_2}\big\rangle_{\mathbb R^2} \\
\end{array}
\right)\,,\nonumber\\\nonumber\\\nonumber\\
g^{(3)}\=\left(
\begin{array}{cccc}
 \big\langle\chi^3_1\,\overline{\chi_1^3}\big\rangle_{\mathbb R^2}\,, & \big\langle\chi^3_1\,\overline{\chi_1^2\chi_2}\big\rangle_{\mathbb R^2}\,, & \big\langle\chi^3_1\,\overline{\chi_1\chi_2^2}\big\rangle_{\mathbb R^2}\,, & \big\langle\chi^3_1\,\overline{\chi_2^3}\big\rangle_{\mathbb R^2} \\\\
\big\langle\chi^2_1\chi_2\,\overline{\chi_1^3}\big\rangle_{\mathbb R^2}\,, & \big\langle\chi^2_1\chi_2\,\overline{\chi_1^2\chi_2}\big\rangle_{\mathbb R^2}\,, & \big\langle\chi^2_1\chi_2\,\overline{\chi_1\chi_2^2}\big\rangle_{\mathbb R^2}\,, & \big\langle\chi^2_1\chi_2\,\overline{\chi_2^3}\big\rangle_{\mathbb R^2} \\\\
\big\langle\chi_1\chi^2_2\,\overline{\chi_1^3}\big\rangle_{\mathbb R^2}\,, & \big\langle\chi_1\chi^2_2\,\overline{\chi_1^2\chi_2}\big\rangle_{\mathbb R^2}\,, & \big\langle\chi_1\chi^2_2\,\overline{\chi_1\chi_2^2}\big\rangle_{\mathbb R^2}\,, & \big\langle\chi_1\chi^2_2\,\overline{\chi_2^3}\big\rangle_{\mathbb R^2} \\\\
 \big\langle\chi^3_2\,\overline{\chi_1^3}\big\rangle_{\mathbb R^2}\,, & \big\langle\chi^3_2\,\overline{\chi_1^2\chi_2}\big\rangle_{\mathbb R^2}\,, & \big\langle\chi^3_2\,\overline{\chi_1\chi_2^2}\big\rangle_{\mathbb R^2}\,, & \big\langle\chi^3_2\,\overline{\chi_2^3}\big\rangle_{\mathbb R^2} \\
\end{array}
\right)\,.\nonumber\\
\eeqn
It is worth to mention that, for $CY_3$'s, there are \emph{no} primitive generators of dimension two or three. Therefore the twisted operators, $\Tr(\sigma_1^2)$ of dimension two and $\Tr(\sigma_1^3)$, $\Tr\,\sigma_1\Tr(\sigma_1^2)$ as well as $\sigma_0\Tr(\sigma_1^2)$ of dimension three, must linearly depend on certain powers of $\chi_1$ and $\chi_2$. It can be also observed due to the symmetry (\ref{Hodge}) of the vertical cohomology of GN-$CY_3$. We thus do not include these operators but only the marginal ones. Later in the third example of the complete intersection in Grassmannian $G(2,8)$, say a $CY_4$, we do have a primitive generator of dimension two. Therefore the operator $\Tr(\sigma^2)$ in that case must be considered.
 
Now we literally follow \cite{Morrison2} to compute the partition function in the large volume limit. In the region of $r_0+2r_1\gg0$ and $r_1\gg0$, the partition function is written as,
\beqn
Z_{\rm GN}\=-\frac{1}{2}\oint\frac{\rm d\epsilon_0 d\epsilon_1 d\epsilon_2}{\left(2\pi i\right)^3}\frac{\pi^8\sin^4(\pi\epsilon_0+\pi\epsilon_1)\sin^4(\pi\epsilon_0+\pi\epsilon_2)}{\sin^8(\pi\epsilon_0)\sin^4(\pi\epsilon_1)\sin^4(\pi\epsilon_2)}\left(z_1\bar z_1\right)^{\epsilon_0}\left(z_2\bar z_2\right)^{\epsilon_1+\epsilon_2}\nonumber\\
&&\Bigg|\sum_{K_1,K_2=0}^\infty z_1^{K_1}z_2^{K_2}\sum_{k=0}^{K_2}\left(2k-K_2+\epsilon_1-\epsilon_2\right)\frac{\Gamma(1+K_1+k+\epsilon_0+\epsilon_1)^4}{\Gamma(1+K_1+\epsilon_0)^4\Gamma(1+k+\epsilon_1)^4}\nonumber\\
&&\frac{\Gamma(1+K_1+K_2-k+\epsilon_0+\epsilon_2)^4}{\Gamma(1+K_1+\epsilon_0)^4\Gamma(1+K_2-k+\epsilon_2)^4}\Bigg|^2\,,
\eeqn
with
\beq
z_1={\rm e}^{2\pi i(\tilde\tau_0+2\tilde\tau_1)}\,,\ \ \ {\rm and}\ \ \ z_2={\rm e}^{2\pi i\tilde\tau_1}\,.
\label{z1z2} 
\eeq
Evaluating the multiple residues of the integral, one finds the K\"{a}hler transformation, 
\beq
T(z_1,z_2)=1+z_1+2z_2+z_1^2+3z_2^2-54z_1^2z_2-14z_1z_2^2+351z_1^2z_2^2+...\,,
\eeq
to simplify the result, and further reads off the mirror map from the coefficients of $\log^2\bar z_1$ and $\log^2\bar z_2$,
\beqn
&&2\pi i\,t_1=\log z_1+4z_2-20z_1z_2+2z_2^2-92z_1^2z_2-72z_1z_2^2+38z_1^2z_2^2+...\,,\nonumber\\
&&2\pi i\,t_2=\log(-z_2)+4z_1+16z_1z_2+2z_1^2-128z_1^2z_2+36z_1z_2^2-1080z_1^2z_2^2+...\,.\nonumber\\
\eeqn
Solving $z_1$, $z_2$ in terms of $t_1$ and $t_2$, one arrives at the partition function up to one-instanton correction,\footnote{Our normalization is different from \cite{Morrison2} by a factor of $-\frac{1}{8\pi^3}$.}
\beqn
Z_{\rm GN}\=\frac{10}{3}\,\xi_1^{-3}+{10}\,{\xi_1^{-2}\xi_2^{-1}}+8\,{\xi_1^{-1}\xi_2^{-2}}+\frac{4}{3}\,\xi_2^{-3}+128\,\zeta(3)\nonumber\\
\+112\,q_1+384\,q_1q_2+\left(56\,q_1+192\,q_1q_2\right)\xi_1^{-1}+192\,q_1q_2\,\xi_2^{-1}+{\rm c.c.}+\mc O(q^2_1, q^2_2)\,,\nonumber\\
\eeqn
where as before we set
\beq\xi_i\equiv\frac{1}{4\pi\,{\rm Im}\,t_i}\,,\ \ \ {\rm and}\ \ \ q_i\equiv{\rm e}^{2\pi i\,t_i}\,,\ \ \ {\rm for}\ \ \ i=1,\,2\,.\eeq 
From $Z_{\rm GN}$, we can compute all correlators on $S^2$,
\beq
M_{a_\alpha \bar b_\beta}=\frac{1}{Z_{\rm GN}}\prod_{i,\bar{j}}\left(-\frac{1}{2\pi}\frac{\partial}{\partial t^{i}}\right)^{n_{i}}\left(\frac{1}{2\pi}\frac{\partial}{\partial\bar t^{\,\bar{j}}}\right)^{n^\prime_{\bar{j}}}Z_{\rm GN}\,.
\label{GNM}
\eeq
with
$$\sum_i n_i=\alpha\,,\ \ \ {\rm and}\ \ \ \sum_{\bar j}n^\prime_{\bar j}=\beta$$
The rows and columns of $M$ are indexed by the degree of operators following 
\beq
\left\{\mathds 1\,;\,\chi_1\,,\,\chi_2\,;\,\chi_1^2\,,\,\chi_1\chi_2\,,\,\chi_2^2\,;\,\chi_1^3\,,\,\chi_1^2\chi_2\,,\,\chi_1\chi_2^2\,,\,\chi_2^3\,\right\}\,.
\eeq
One can check that the $10\times 10$ matrix $M$ have ${\rm rank}(M)=6$, and the rank of its sub-matrices up to degree $\{0,1,2,3\}$ is exactly $\{1,3,5,6\}$ consistent with the dimension of dimensions of vertical cohomology (\ref{Hodge}).

Based on eq.\,(\ref{GNM}) and algorithm (\ref{gnm}), we can extract all twisted chiral ring data. For simplicity, we compute the unnormalized correlator,
\beq
G^{(\alpha)}= g^{(\alpha)} Z_{\rm GN}=G^{(\alpha)}_{\rm pert.}+G^{(\alpha)}_{\rm np.}\,. 
\eeq
by implementing eq.\,(\ref{gnm}), where we have separated them by perturbative and non-perturbative two parts. Since $G^{(\alpha)}_{\rm np.}$ are kind of lengthy even for one-instanton correction, we only present the perturbative $G_{\rm pert.}^{(\alpha)}$ here and leave $G_{\rm np.}^{(\alpha)}$ in appendix\,\ref{C}.

The (unnormalized) metric $G^{(1)}$ on $\mc M_K(\mc Y_{\rm GN})$ can be obtained by eq.\,(\ref{g1}),
\beq
G_{\rm pert.}^{(1)}=\frac{1}{D^{(1)}}\left(
\begin{array}{cc}
 N^{(1)}_{1\bar 1}\,, & N^{(1)}_{1\bar 2} \\\\
N^{(1)}_{2\bar 1}\,, & N^{(1)}_{2\bar 2} \\
\end{array}
\right)\,,
\label{GNG1}
\eeq
with
\beqn
N^{(1)}_{1\bar 1}\=56 \xi _1^4+200 \xi _2 \xi _1^3+300 \xi _2^2 \xi _1^2+200 \xi _2^3 \xi _1+50 \xi _2^4-3840 \xi _2^3 \xi _1^4 \zeta (3)-3840 \xi _2^4 \xi _1^3 \zeta (3)\nonumber\\
N^{(1)}_{2\bar 2}\=8 \xi _1^4+64 \xi _2 \xi _1^3+192 \xi _2^2 \xi _1^2+200 \xi _2^3 \xi _1+70 \xi _2^4-1536 \xi _2^3 \xi _1^4 \zeta (3)-3072 \xi _2^4 \xi _1^3 \zeta (3)\nonumber\\
N^{(1)}_{1\bar 2}\=N^{(1)}_{2\bar 1}\nonumber\\
\=16 \xi _1^4+80 \xi _2 \xi _1^3+180 \xi _2^2 \xi _1^2+160 \xi _2^3 \xi _1+50 \xi _2^4-3072 \xi _2^3 \xi _1^4 \zeta (3)-3840 \xi _2^4 \xi _1^3 \zeta (3)\nonumber\\
D^{(1)}\=\xi _1 \xi _2 \left(2 \xi _1^3+12 \xi _2 \xi _1^2+15 \xi _2^2 \xi _1+5 \xi _2^3+192 \xi _2^3 \xi _1^3 \zeta (3)\right)
\eeqn
$G^{(2)}$ are also straightforwardly computed either from eq.\,(\ref{gnm}) or (\ref{G2}),
\beq
G^{(2)}_{\rm pert.}=\frac{1}{D^{(2)}}\left(
\begin{array}{ccc}
 N^{(2)}_{11\bar 1\bar 1}\,, & N^{(2)}_{11\bar 1\bar 2}\,, & N^{(2)}_{11\bar 2\bar 2} \\\\
 N^{(2)}_{12\bar 1\bar 1}\,, & N^{(2)}_{12\bar 1\bar 2}\,, & N^{(2)}_{12\bar 2\bar 2} \\\\
 N^{(2)}_{22\bar 1\bar 1}\,, & N^{(2)}_{22\bar 1\bar 2}\,, & N^{(2)}_{22\bar 2\bar 2} \\
\end{array}
\right)\,,
\label{GNG2}
\eeq
with
\beqn
N^{(2)}_{11\bar 1\bar 1}\=200 \xi _1 \xi _2 \left(\xi _1+\xi _2\right) \left(8 \xi _1^3+18 \xi _2 \xi _1^2+15 \xi _2^2 \xi _1+5 \xi _2^3+192 \xi _2^3 \xi _1^3 \zeta (3)\right)\nonumber\\
N^{(2)}_{12\bar 1\bar 2}\=4 \xi _1 \xi _2 \big(228 \xi _1^4+800 \xi _2 \xi _1^3+1200 \xi _2^2 \xi _1^2+900 \xi _2^3 \xi _1+275 \xi _2^4\nonumber\\
\+11520 \xi _2^3 \xi _1^4 \zeta (3)+7680 \xi _2^4 \xi _1^3\zeta (3)\big)\nonumber\\
N^{(2)}_{22\bar 2\bar 2}\=16 \xi _1 \xi _2 \big(12 \xi _1^4+68 \xi _2 \xi _1^3+174 \xi _2^2 \xi _1^2+180 \xi _2^3 \xi _1+65 \xi _2^4\nonumber\\
\+1152 \xi _2^3 \xi _1^4 \zeta (3)-384 \xi _2^4 \xi _1^3 \zeta
   (3)\big)\nonumber\\
N^{(2)}_{11\bar 1\bar 2}\=N^{(2)}_{12\bar 1\bar 1}=40 \xi _1 \xi _2 \big(30 \xi _1^4+100 \xi _2 \xi _1^3+135 \xi _2^2 \xi _1^2+90 \xi _2^3 \xi _1+25 \xi _2^4\nonumber\\
\+1152 \xi _2^3 \xi _1^4 \zeta (3)+960 \xi _2^4 \xi _1^3 \zeta
   (3)\big)\nonumber\\
N^{(2)}_{11\bar 2\bar 2}\=N^{(2)}_{22\bar 1\bar 1}=160 \xi _1 \xi _2 \big(3 \xi _1^4+11 \xi _2 \xi _1^3+18 \xi _2^2 \xi _1^2+15 \xi _2^3 \xi _1+5 \xi _2^4\nonumber\\
\+288 \xi _2^3 \xi _1^4 \zeta (3)+192 \xi _2^4 \xi _1^3 \zeta
   (3)\big)\nonumber\\
N^{(2)}_{12\bar 2\bar 2}\=N^{(2)}_{22\bar 1\bar 2}=8 \xi _1 \xi _2 \big(48 \xi _1^4+200 \xi _2 \xi _1^3+390 \xi _2^2 \xi _1^2+360 \xi _2^3 \xi _1+125 \xi _2^4\nonumber\\
\+4608 \xi _2^3 \xi _1^4 \zeta (3)+1920 \xi _2^4 \xi _1^3
   \zeta (3)\big)\nonumber\\
D^{(2)}\=\left(6 \xi _1^2+10 \xi _2 \xi _1+5 \xi _2^2\right) \left(2 \xi _1^3+12 \xi _2 \xi _1^2+15 \xi _2^2 \xi _1+5 \xi _2^3-384 \xi _2^3 \xi _1^3 \zeta (3)\right)\,.\nonumber\\
\eeqn
Now before proceeding further to compute $G^{(3)}$, one will find that rank of $G^{(2)}$ is \emph{two},\footnote{We notice that, including instanton correction, the rank of $G^{(2)}$ will turn out to be three but in higher order of $q$. We believe that the rank of $G^{(2)}$ will remain \emph{two} if adding full instanton corrections.} which implies that there are only two independent operators of dimension two among $\chi_1^2$, $\chi_1\chi_2$ and $\chi_2^2$. It is surely consistent with, 
\beq
{\rm dim}H_{\rm prim}^{(2,2)}(\mc Y_{\rm GN})=2\,.
\eeq 
Therefore we have to remove one of them, and compute the inverse of $G^{(2)}$ to perform further orthogonalization. Let us, for example, remove operator $\chi_2^2$ and its corresponding row and column in $M$ from eq.\,(\ref{GNM}), and $G^{(3)}$ is given by eq.\,(\ref{gnm}),
\beq
G^{(3)}=\frac{1}{Z_{\rm GN,pert.}}\left(
\begin{array}{cccc}
 20\cdot20\,, & 20\cdot20\,, & 20\cdot20\,, & 16\cdot8 \\
 20\cdot20\,, & 20\cdot20\,, & 20\cdot16\,, & 20\cdot8 \\
 20\cdot16\,, & 20\cdot16\,, & 16\cdot16\,, & 16\cdot8 \\
 20\cdot8\,, & 20\cdot8\,, & 16\cdot8\,, & 8\cdot8 \\
\end{array}
\right)\,,
\label{GNG3}
\eeq
with
\beq
Z_{\rm GN,pert.}=\frac{10}{3}\,\xi_1^{-3}+{10}\,{\xi_1^{-2}\xi_2^{-1}}+8\,{\xi_1^{-1}\xi_2^{-2}}+\frac{4}{3}\,\xi_2^{-3}+128\,\zeta(3)\,,
\eeq
the perturbative part of partition function $Z_{\rm GN}$. Indeed one can verify that our computation (\ref{GNG3}) is independent on the choice of the removed operator.

To restore the twisted chiral ring data in $\tau$-coordinates, as before, we have
\beqn
&&\widetilde g^{(1)}_{i\bar j}(\tilde\tau,\bar{\tilde\tau})=\left(\frac{{\rm d}t_{i^\prime}}{{\rm d}\tilde\tau_i}\right){\left(\frac{{\rm d}\bar t_{j^\prime}}{{\rm d}\bar{\tilde \tau}_j}\right)}\,g^{(1)}_{i^\prime\bar j^\prime}\left(t(\tilde\tau),\bar{t}(\bar{\tilde\tau})\right)\,,\nonumber\\
&&\widetilde g^{(2)}_{ik\bar j\bar l}(\tilde\tau,\bar{\tilde\tau})=\left(\frac{{\rm d}t_{i^\prime}}{{\rm d}\tilde\tau_i}\right)\left(\frac{{\rm d}t_{k^\prime}}{{\rm d}\tilde\tau_k}\right){\left(\frac{{\rm d}\bar t_{j^\prime}}{{\rm d}\bar {\tilde\tau}_j}\right)}{\left(\frac{{\rm d}\bar t_{l^\prime}}{{\rm d}\bar{\tilde \tau}_l}\right)}\,g^{(2)}_{i^\prime k^\prime\bar j^\prime\bar l^\prime}\left(t(\tilde\tau),\bar{t}(\bar{\tilde\tau})\right)\,,\nonumber\\
&&\widetilde g^{(3)}_{ikm\bar j\bar l\bar n}(\tilde\tau,\bar{\tilde\tau})=\left(\frac{{\rm d}t_{i^\prime}}{{\rm d}\tilde\tau_i}\right)\left(\frac{{\rm d}t_{k^\prime}}{{\rm d}\tilde\tau_k}\right)\left(\frac{{\rm d}t_{m^\prime}}{{\rm d}\tilde\tau_m}\right){\left(\frac{{\rm d}\bar t_{j^\prime}}{{\rm d}\bar{\tilde \tau}_j}\right)}{\left(\frac{{\rm d}\bar t_{l^\prime}}{{\rm d}\bar{\tilde \tau}_l}\right)}{\left(\frac{{\rm d}\bar t_{n^\prime}}{{\rm d}\bar{\tilde \tau}_n}\right)}\,g^{(3)}_{i^\prime k^\prime m^\prime\bar j^\prime\bar l^\prime \bar n^\prime}\left(t(\tilde\tau),\bar{t}(\bar{\tilde\tau})\right)\,.\nonumber\\
\label{taut}
\eeqn

Now we give some consistency checks of above results. First the rank of $G^{(3)}$ is one, which implies that all $\chi_1^3$, $\chi_1^2\chi_2$, $\chi_1\chi_2^2$ and $\chi_2^3$ are linear dependent, corresponding to the unique top element of twisted chiral ring of $H^{(3,3)}(\mc Y_{\rm GN})$. Further applying eq.\,(\ref{gnm}), we have
$$G^{(\alpha)}_{\rm pert.}=0\,,\ \ \ {\rm for}\ \ \ \alpha\geq4\,,$$
showing the nilpotency of the twisted chiral ring. We next check that, using eq.\,(\ref{GNG1}), (\ref{GNG2}) and (\ref{GNG3}), the constraints eq.\,(\ref{constraints2}) and (\ref{constraints3}) are also satisfied as designed. More interestingly, we can read off the topological correlators in A-twisted theory. For example, let $\alpha=d$ in eq.\,(\ref{constraints2}), we have
\beq
G^{(3)}\,Z_{\rm GN}=\mc C_{a_3\,0}^{(d)}\,\overline{\mc C^{(d)}_{b_3\,0}}\,.
\eeq
Comparing to eq.\,(\ref{GNG3}), the diagonal entries of eq.\,(\ref{GNG3}) imply that
$$\big\langle\chi^3_1\big\rangle_{S^2,{\rm pert.}}\!\!\!\!\!=20\,,\ \ \ \big\langle\chi^2_1\chi_2\big\rangle_{S^2,{\rm pert.}}\!\!\!\!\!=20\,,\ \ \ \big\langle\chi_1\chi^2_2\big\rangle_{S^2,{\rm pert.}}\!\!\!\!\!=16\,,\ \ \ {\rm and}\ \ \ \big\langle\chi^3_2\big\rangle_{S^2,{\rm pert.}}\!\!\!\!\!=8\,,$$
which are the classical intersection numbers of hyperplanes in $G(2,4)\supset\mc Y_{\rm GN}$. 

All above checks are surely correct after including instanton corrections. We perform up to $3$-instanton computation. Especially we can compute the topological correlators in the A-twist theory \cite{Morrison2} up to the order of $3$-instantons,
\beqn
\big\langle\chi^3_1\big\rangle_{S^2}\=4 \big(5+14q_1+14q_1^2+14q_1^3+48q_1q_2+1792q_1^2q_2+1296q_1^3q_2+14q_1q_2^2\nonumber\\
\+5136q_1^2q_2^2+155358q_1^3q_2^2+1792q_1^2q_2^3+357312q_1^3q_2^3
+\mc O(q_1^4,q_2^4)\,\big)\,,\nonumber\\\nonumber\\
\big\langle\chi^2_1\chi_2\big\rangle_{S^2}\=4\big(5+48q_1q_2+896q_1^2q_2+432q_1^3q_2+28q_1q_2^2+5136q_1^2q_2^2+103572q_1^3q_2^2
\nonumber\\
\+2688q_1^2q_2^3+357312q_1^3q_2^3+\mc O(q_1^4,q_2^4)\,\big)\,,\nonumber
\eeqn
\beqn
\big\langle\chi_1\chi^2_2\big\rangle_{S^2}\=16\big(1+12q_1q_2+112q_1^2q_2+36q_1^3q_2+14q_1q_2^2+1284q_1^2q_2^2+17262q_1^3q_2^2\nonumber\\
\+1008q_1^2q_2^3+89328q_1^3q_2^3+\mc O(q_1^4,q_2^4)\,\big)\,,\nonumber\\\nonumber\\
\big\langle\chi^3_2\big\rangle_{S^2}\=8(1+24q_1q_2+112q_1^2q_2+24q_1^3q_2+56q_1q_2^2+2568q_1^2q_2^2+23016q_1^3q_2^2\nonumber\\
\+3024q_1^2q_2^3+178656q_1^3q_2^3+\mc O(q_1^4,q_2^4)\,\big)\,.
\eeqn
One can further apply eq.\,(\ref{taut}) to obtain these chiral correlators in $\tilde\tau$-coordinates. The results also match with those in \cite{Park} up to a K\"{a}hler transformation, see more details in appendix \ref{C}.\\


\subsection{The complete intersection in Grassmannian: $X_{1^8}\subset G(2,8)$}
Our final example is the complete intersection of eight hyperplanes with degree one, a Calabi-Yau fourfold $X_{1^8}$, in Grassmannian $G(2,8)$ \cite{Honma}, see also \cite{Hori}. It is endowed with a GLSM description with a $U(2)$ gauge group. The matter content are summarized in Table\,\ref{T3}, and constrained by a superpotential 
$$W=\sum_{\substack{a,i,j=8\\ \alpha,\beta=1,2}}A^a_{ij}P_a\Phi_\alpha^i\Phi_\beta^j\epsilon^{\alpha\beta}\,,$$
where $A^a$ are eight $8\times 8$ constant anti-symmetric matrices, and $\epsilon^{\alpha\beta}$ is the $SU(2)\subset U(2)_\Sigma$ invariant antisymmetric tensor.
\begin{table}[t]
\centering
\begin{tabular}{|c||c|c|c|c|}
	\hline Field & $U(1)\subset U(2)_\Sigma$ &  $SU(2)\subset U(2)_\Sigma$ & $U(1)_V$ & $U(1)_A$\\
	\hline  $\Phi^i$ & $+1$ & $\mathbf 2$ & $2\mf q$ & $0$\\
	\hline  $P_a$ & $-2$ & $\mathbf 1$ & $2-4\mf q$ & $0$\\
	\hline
\end{tabular}
\caption{The $U(2)_\Sigma$ gauge representation, $U(1)_V$ and $U(1)_A$ R-charge of matter fields $P_a$, $\Phi^i$ for $a,i=1,2...,8$.}
\label{T3}
\end{table}
The dimensions of the vertical cohomology classes, $\bigoplus_{i=0}^{4}H^{(i,i)}(X_{1^8})$, are listed below,
\beq
\{1,\,1,\,2,\,1,\,1\}\,.
\label{Hodge2}
\eeq
Therefore its twisted chiral ring is generated by two primitive generators, the marginal twisted primary $\psi$, and further a primitive generator $\chi$ of degree two,
\beq
\psi\equiv\Tr\,\sigma\,,\ \ \ {\rm and}\ \ \ \chi\equiv\Tr\,\left(\sigma^2\right)\,,
\eeq
where $\sigma$ is the bottom component of the twisted chiral multiplet $\Sigma$. We therefore need to compute the following twisted chiral ring data,
\beqn
&&g^{(0)}=\left\langle\mathds 1\,\overline{\mathds 1}\,\right\rangle_{\mathbb R^2}\equiv1\,,\ \ \ \ \hspace{2.8cm} g^{(1)}=\left\langle\psi\,\overline\psi\,\right\rangle_{\mathbb R^2}\,,\nonumber\\\nonumber\\
&&g^{(2)}=\left(
\begin{array}{cc}
 \big\langle\psi^2\,\overline{\psi^2}\big\rangle_{\mathbb R^2}\,, & \big\langle\psi^2\,\overline{\chi}\big\rangle_{\mathbb R^2} \\\\
\big\langle\chi\,\overline{\psi^2}\big\rangle_{\mathbb R^2}\,, & \big\langle\chi\,\overline{\chi}\big\rangle_{\mathbb R^2} \\
\end{array}
\right)\,,\ \ \ \ g^{(3)}=\left(
\begin{array}{cc}
 \big\langle\psi^3\,\overline{\psi^3}\big\rangle_{\mathbb R^2}\,, & \big\langle\psi^3\,\overline{\psi\chi}\big\rangle_{\mathbb R^2} \\\\
\big\langle\psi\chi\,\overline{\psi^3}\big\rangle_{\mathbb R^2}\,, & \big\langle\psi\chi\,\overline{\psi\chi}\big\rangle_{\mathbb R^2} \\
\end{array}
\right)\,,\nonumber\\\nonumber\\\nonumber\\
&&g^{(4)}=\left(
\begin{array}{ccc}
 \big\langle\psi^4\,\overline{\psi^4}\big\rangle_{\mathbb R^2}\,, & \big\langle\psi^4\,\overline{\psi^2\chi}\big\rangle_{\mathbb R^2}\,, & \big\langle\psi^4\,\overline{\chi^2}\big\rangle_{\mathbb R^2} \\\\
\big\langle\psi^2\chi\,\overline{\psi^4}\big\rangle_{\mathbb R^2}\,, & \big\langle\psi^2\chi\,\overline{\psi^2\chi}\big\rangle_{\mathbb R^2}\,, & \big\langle\psi^2\chi\,\overline{\chi^2}\big\rangle_{\mathbb R^2} \\\\
 \big\langle\chi^2\,\overline{\psi^4}\big\rangle_{\mathbb R^2}\,, & \big\langle\chi^2\,\overline{\psi^2\chi}\big\rangle_{\mathbb R^2}\,, & \big\langle\chi^2\,\overline{\chi^2}\big\rangle_{\mathbb R^2} \\
\end{array}
\right)\,.\nonumber\\\nonumber
\label{X8tcrd}
\eeqn
Since now there exists a primitive generator of dimension two, we have to deform the original GLSM by an additional twisted superpotential as eq.\,(\ref{twisted superpotential}),
\beq
\widetilde W_{\rm deform.}+{\rm h.c.}=\frac{1}{2}\left(\tilde\tau_2\Tr\,\Sigma^2-\bar{\tilde\tau}_2\Tr\,\overline\Sigma^2\right)\,,
\eeq
and obtain the deformed partition function $Z_{X_{1^8}}$ by use of eq.\,(\ref{main}),
\beqn
Z_{X_{1^8}}\=\frac{1}{2}\!\!\!\sum_{m_1,m_2\in \mathbb Z}{\rm e}^{-i\theta(m_1+m_2)}\int_{-\infty}^\infty\frac{{\rm d}\sigma_1}{2\pi}\frac{{\rm d}\sigma_2}{2\pi}\,{\rm e}^{-4\pi i r(\sigma_1+\sigma_2)}\,{\rm e}^{-2\pi\tilde\tau_2\sum_{j=1}^2\left(\sigma_j+i\frac{m_j}{2}\right)^2-{\rm c.c.}}\nonumber\\
\X\left(\frac{(m_1-m_2)^2}{4}+(\sigma_1-\sigma_2)^2\right)\,\left(\frac{\Gamma\left(1-2\mf q+i(\sigma_1+\sigma_2)+\frac{1}{2}(m_1+m_2)\right)}{\Gamma\left(2\mf q-i(\sigma_1+\sigma_2)+\frac{1}{2}(m_1+m_2)\right)}\right)^8
\nonumber\\
\X\left(\frac{\Gamma\left(\mf q-i\sigma_1-\frac{1}{2}m_1\right)\,\Gamma\left(\mf q-i\sigma_2-\frac{1}{2}m_2\right)}{\Gamma\left(1-\mf q+i\sigma_1-\frac{1}{2}m_1\right)\,\Gamma\left(1-\mf q+i\sigma_2-\frac{1}{2}m_2\right)\,}\right)^8\,,
\label{ZX81}
\eeqn
where $\tilde\tau=\frac{\theta}{2\pi}+ir$ is the marginal parameter as before, and $\tilde\tau_2$ is the irrelevant parameter introduced to probe the twisted chiral primary $\Tr\,\Sigma^2$ of degree two. We will calculate $Z_{X_{1^8}}$ in the CY geometric phase $r\gg 0$, and thus set
$$\mf q-i\sigma_i\equiv\kappa_i\,,$$
and close the contour in the left half-planes of $\kappa_i$. During the evaluation, we will keep\footnote{The correct R-charge should be $\mf q\rightarrow \frac{1}{2}^-$. However we check up to the order of two-instanton, the twisted chiral ring data is independent on $\mf q$.} $\mf q\rightarrow0^+$ to simplify the computation. With $\mf q=0$ and the standard method in \cite{Morrison2, Honma}, one can rewrite eq.\,(\ref{ZX81}) as a sum of multi-residues,
\beqn
Z_{X_{1^8}}\=-\frac{1}{2}\oint\frac{\rm d\epsilon_1 d\epsilon_2}{\left(2\pi i\right)^2}\,\frac{\pi^8\sin^8(\pi\epsilon_1+\pi\epsilon_2)}{\sin^8(\pi\epsilon_1)\,\sin^8(\pi\epsilon_2)}\,\left(z\bar z\right)^{-\epsilon_1-\epsilon_2}\left(U\,V\right)^{\epsilon_1^2+\epsilon_2^2}\\
\X\left(\sum_{J=0}^\infty z^{J}\sum_{j=0}^{J}U^{j^2-2j\epsilon_1+(J-j)^2-2(J-j)\epsilon_2}\frac{\left(2j-J-\epsilon_1+\epsilon_2\right)\Gamma(1+J-\epsilon_1+\epsilon_2)^8}{\Gamma(1+j-\epsilon_1)^8\Gamma(1+J-j-\epsilon_2)^8}\right)\nonumber\\
\X\left(\sum_{L=0}^\infty \bar z^{L}\sum_{l=0}^{L}V^{l^2-2l\epsilon_1+(L-l)^2-2(L-l)\epsilon_2}\frac{\left(2l-L-\epsilon_1+\epsilon_2\right)\Gamma(1+L-\epsilon_1+\epsilon_2)^8}{\Gamma(1+l-\epsilon_1)^8\Gamma(1+L-l-\epsilon_2)^8}\right)\,,\nonumber
\label{ZX82}
\eeqn 
with
\beq
z={\rm e}^{2\pi i\tilde\tau}\,,\ \ \ U={\rm e}^{2\pi\tilde\tau_2}\,\ \ \ {\rm and}\ \ \ V={\rm e}^{-2\pi\bar{\tilde\tau}_2}\,.
\eeq
There are two technical issues to clarify before evaluating eq.\,(\ref{ZX82}). First there also exists a mirror map $t=f(\tilde\tau)$ to simplify $Z_{X_{1^8}}$ in large volume limit. However, since now that we include irrelevant parameters in the partition function, there seems no straightforward way to see how operator $\chi$ transforms respect to the change of coordinates from $``\tilde\tau"$ to $``t"$, while it does transform tensorially as a section living on the vector bundle over $\mc M_{K}(\mc Y_{X_{1^8}})$. Therefore, we directly evaluate the integral in the $\tilde\tau$-coordinates. Secondly we can find a ``K\"{a}hler transformation", 
\beq
T(z,U)=1 - 6 z U + 256 z^2 U^2 - 22 z^2 U^4 + 2 z U \log U + 
 8 z^2 U^4 \log U+...\,,
\label{KTX8}
\eeq
to simplify the computation as we did before. But note that the variables $U$ and $V$ are actually not complex conjugate to each other, $T(z,U)$ and $\overline{T}(\bar z,V)$ are thus not either. Recall that the twisted chiral ring data are invariant respect to K\"{a}hler transformations. We checked our result with and without use of the ``K\"{a}hler transformation" $T(z,U)$, and fortunately the results agree up to two-instanton. 

With eq.\,(\ref{KTX8}), we evaluate
\beq
Z_{X_{1^8}}\longrightarrow\frac{Z_{X_{1^8}}}{T(z,U)\,\overline{T}(\bar z,V)}\,,
\eeq
and spell out the perturbative part of $Z_{X_{1^8}}$\,,
\beq
Z_{X_{1^8}}=\frac{11}{2}\,\xi^{-4}+24\log\left(U\,V\right)\,\xi^{-2}+672\,\zeta(3)\,\xi^{-1}+10\log\left(U\,V\right)+Z_{\rm np.}\left(z,\bar z,U,V\right)\,,
\eeq
where 
\beq
\xi\equiv\frac{1}{4\pi\,{\rm Im}\,\tilde\tau}\,\ \ \ {\rm and}\ \ \ z={\rm e}^{2\pi i \tilde\tau}\,.
\eeq
and $Z_{\rm np.}\left(z,\bar z,U,V\right)$ denotes the non-perturbative contributions that is too lengthy to present. From $Z_{X_1^8}$ and eq.\,(\ref{bigM}), we compute all correlators on $S^2$,
\beq
M_{a_\alpha \bar b_\beta}=\frac{1}{Z_{X_{1^8}}}\left(-\frac{1}{2\pi}\frac{\partial}{\partial\tilde\tau}\right)^{n_1}\left(-\frac{1}{2\pi}\frac{\partial}{\partial\tilde\tau_2}\right)^{n_2}\left(\frac{1}{2\pi}\frac{\partial}{\partial\bar{\tilde\tau}}\right)^{m_1}\left(\frac{1}{2\pi}\frac{\partial}{\partial\bar{\tilde\tau}_2}\right)^{m_2}Z_{X_{1^8}}\Big|_{\tilde\tau_2=\bar{\tilde\tau}_2=0}\,,
\eeq 
with $n_1+2n_2=\alpha$ and $m_1+2m_2=\beta$. We index the rows and columns of $M$ by the degrees of operators as,
\beq
\left\{\mathds 1\,;\,\psi\,;\,\psi^2\,,\,\chi\,;\,\psi^3\,,\,\psi\chi\,;\,\psi^4\,,\,\psi^2\chi\,,\,\chi^2\,\right\}\,.
\eeq
One can check that the rank of $M$ is $6$, and that of its sub-matrices up to the degrees of operators $\{0,1,2,3,4\}$ are precisely $\{1,2,4,5,6\}$ matching with the dimensions of vertical cohomology classes (\ref{Hodge2}) of ${X_{1^8}}$.

Implementing algorithm (\ref{gnm}), we find up to one-instanton,
\beqn
g^{(1)}\=4 \xi ^2\left(\frac{ F_1(\xi)}{F_2(\xi)}\right)^2+\left(-80\xi^2(z+\bar z)+\mc O(\xi^3)\right)+\mc O\left(z^2,\bar z^2\right)\,,\nonumber\\\nonumber\\
g^{(2)}\=\frac{1}{F_2(\xi)}\left(
\begin{array}{cc}
 264\xi^4\,, & 96\xi^4 \\\\
 96\xi^4\,, & 40\xi^4 \\
\end{array}
\right)+g^{(2)}_{\rm 1-inst.}+\mc O\left(z^2,\bar z^2\right)\,,\nonumber\\\nonumber\\
g^{(3)}\=\frac{1}{F^2_1(\xi)}\left(
\begin{array}{cc}
 17424\xi^6\,, & 6336\xi^6 \\\\
 6336\xi^6\,, & 2304\xi^6 \\
\end{array}
\right)+g^{(3)}_{\rm 1-inst.}+\mc O\left(z^2,\bar z^2\right)\,,\nonumber
\eeqn
\beqn
g^{(4)}\=\frac{1}{F^2_2(\xi)}\left(
\begin{array}{ccc}
 69696\xi^8\,, & 25344\xi^8\,, & 10560\xi^8 \\\\
 25344\xi^8\,, & 9216\xi^8\,, & 3840\xi^8 \\\\
 10560\xi^8\,, & 3840\xi^8\,, & 1600\xi^8 \\
\end{array}
\right)+g^{(4)}_{\rm 1-inst.}+\mc O\left(z^2,\bar z^2\right)\,,\nonumber\\
\eeqn
where
\beqn
F_1(\xi)\,\=\,11-672\zeta(3)\,\xi^3\,,\ \ \ \ \ \ \ \ \ F_2(\xi)=11 + 1344\zeta(3)\, \xi^3\,,\nonumber\\\nonumber\\
g^{(2)}_{\rm 1-inst.}\=\,-\frac{32}{11}\xi^4\left(
\begin{array}{cc}
 435 z-11514\bar z z +{\rm c.c.}\,, &  162 z +210 \bar z-11088 \bar z z  \\\\
 210 z + 162\bar z-11088 \bar z z\,, & 77 z-2674 \bar z z +{\rm c.c.}  \\
\end{array}
\right)+\mc O(\xi^5)\,,\nonumber\\\nonumber\\\nonumber\\
g^{(3)}_{\rm 1-inst.}\=\,-\frac{64}{121}\xi^6\left(
\begin{array}{cc}
 23265 z-994050\bar z z +{\rm c.c.}\,, &  8460 z +10296 \bar z-879840 \bar z z  \\\\
 10296 z + 8460\bar z-879840 \bar z z\,, & 3744 z-194688 \bar z z +{\rm c.c.}  \\
\end{array}
\right)\nonumber\\
\+\mc O(\xi^7)\,,\nonumber\\
g^{(4)}_{\rm 1-inst.}\=\,-\frac{512}{121}\xi^8\left(
\begin{array}{ccc}
 g^{(4)}_{1\bar1}  & g^{(4)}_{1\bar2}  & g^{(4)}_{1\bar3}  \\\\
 g^{(4)}_{2\bar1}  & g^{(4)}_{2\bar2}  & g^{(4)}_{2\bar3}  \\\\
 g^{(4)}_{3\bar1}  & g^{(4)}_{3\bar2}  & g^{(4)}_{3\bar3}  \\
\end{array}
\right)+\mc O(\xi^9)\,,\nonumber\
\eeqn
\beqn
g^{(4)}_{1\bar1}\=14355z+14355\bar z-1513800\bar z z\,,\ \ g^{(4)}_{1\bar2}=\overline{g^{(4)}_{2\bar1}}=5220z+6138\bar z-647280\bar z z\,,\nonumber\\
g^{(4)}_{1\bar3}\=\overline{g^{(4)}_{3\bar1}}=2175z+2541\bar z-267960\bar z z\,,\ \ g^{(4)}_{2\bar2}=2232z+2232\bar z-276768\bar z z\,,\nonumber\\
g^{(4)}_{2\bar3}\=\overline{g^{(4)}_{3\bar2}}=930z+924\bar z-114576\bar z z\,,\ \ g^{(4)}_{3\bar3}=385z+385\bar z-47432\bar z z\,.\nonumber
\eeqn
In the evaluation, the rank of $g^{(3)}$ is one as expected. Therefore we have to remove, for example, the row and column corresponding to ``$\psi\chi$" to define $\widetilde M$ in eq.\,(\ref{tildeM}) and compute $g^{(4)}$.

Now we give some consistency checks. First the nilpotency of twisted chiral ring is confirmed up to two-instanton order, i.e.
\beq
\big\langle\psi^m\chi^n\,\overline{\psi^p\chi^q}\big\rangle_{\mathbb R^2}=0\, \ \ \ {\rm for}\ \ \ m+2n=p+2q\geq5\,.
\eeq
Second the constraints eq.\,(\ref{constraints2}) get satisfied. Let $\alpha=d$ in eq.\,(\ref{constraints2}), we find the A-twisted chiral correlators
\beqn
&&\big\langle\psi\cdot\psi\cdot\psi\cdot\psi\big\rangle_{S^2}=132 - 13920z+ 1912032 z^2\,,\nonumber\\
&&\big\langle\psi\cdot\psi\cdot\chi\big\rangle_{S^2}=48 - 5952z + 791424z^2\,,\nonumber\\
&&\big\langle\chi\cdot\chi\big\rangle_{S^2}=20 - 2464 z + 327776 z^2\,.
\eeqn
The last chiral correlator can be normalized, by a K\"{a}hler transformation, to certain constants as in standard topological field theories. However because of the lack of knowledge on how the operator $\chi$ transforms under mirror map, we cannot reproduce the results of chiral correlators in large volume limit as we did in last subsection. It would be interesting to investigate this point further.\\\\

\section{Discussions}
In this paper, we have provided a general method to extract (twisted) chiral ring data directly from deformed partition functions in two-dimensional \ntt\ SCFT. In the context of Calabi-Yau complex moduli, the method is endowed with explicit geometric interpretations. In the examples, we also developed alternative formulas, via localization onto Higgs branch, for (twisted) chiral ring data on the complete intersections in toric varieties.

There would be several interesting directions deserving further studies. First, as we have seen, in the case of complete intersections in toric varieties, the (twisted) chiral ring data can be formulated in terms of sum of factorized vortex and anti-vortex partition functions, or say the periods of their mirror manifolds. It would be instructive to see how this formalism could be generalized to the case when there are more marginal generators and additional primitive generators of dimension greater than one. On the other hand, deforming the partition function by irrelevant operators could help us extract additional information on primitive generators of dimension greater than one. In this sense the deformed partition function is the generating function of all (twisted) chiral ring data. Therefore if one has (twisted) chiral ring data as input data, it would be possible to reconstruct the deformed partition function reciprocally. Combining these two observations, one may reformulate the partition function of a theory $\mc S$ in terms of the periods of its mirror $\widetilde{\mc S}$ with certain patterns, as people already knew or conjectured for threefolds \cite{Morrison2}, fourfolds \cite{Honma} and perturbative part of partition functions of general $n$-folds \cite{Morrison3}. 

Secondly it has been known since long time ago that $tt^*$-equations of (twisted) chiral correlators were valid even for off-critical theories \cite{Vafa}. It would be interesting to extract these correlators directly from the partition functions of off-critical theories. Correspondingly, for critical theories with center charge $c\geq 3$, we have an explanation on the counter terms and operators mixing from the perspective of supergravity, see section \ref{3.3}. It would be very nice to extend this picture to off-critical theories, e.g. minimal \ntt\ SCFT perturbed by relevant operators, so that we could have a better understanding how the correlators from $S^2$ to $\mathbb R^2$ are related to each other.

At last, we should mention that the work was originally motivated from an attempt to understand semi-local vortex strings and $2d/4d$ correspondence and so on. It has been well known and studied on the correspondence between the BPS spectrum in $4d$ \ntwo\ gauge theories on the bulk and that in $2d$ \ntt\ GLSM for vortex strings \cite{Dorey, Tong1, Tong2, Shifman1, Park2}. Especially, for $U(2)$ gauge theory with 4 hypermultiplets, the corresponding worldsheet theory, describing the low-dynamics of critical semi-local vortex string, is a non-compact Calabi-Yau threefold, the resolved conifold $\mc O(-1)\oplus\mc O(-1)$ \cite{Shifman2, Shifman3}. It is straightforward to apply our algorithm (\ref{gnm}) to non-compact CY cases, so long as one could correctly resolve the singularities in evaluation of the partition functions \cite{Honma}. It would be curious to see if our computations on twisted chiral ring data have interesting implications or applications in the thin string regime, $r\sim 0$ \cite{Shifman4}. We expect to answer some of these questions in subsequent works.\\\\

\acknowledgments
I would like to thank Misha Shifman who led me to this interesting topic, and Arkady Vainshtein and Misha for helpful discussion in the early stage of the work. I am also grateful to Gang Yang for many valuable suggestions, comments and careful proofreading of the draft. This work is supported in part by the Chinese Academy of Sciences (CAS) Hundred-Talent Program and by Project 11647601 supported by National Natural Science Foundation of China.
\\\\

\begin{appendix}
\section{Notations and $\mathfrak{su}(2|1)$ superalgebra on $S^2$}
\label{A}
Most of our notations follow from \cite{Benini}. For self-consistency, we list some of them, and briefly discuss how to obtain the $\mathfrak{su}(2|1)$ superalgebra on $S^2$.\\

\noindent{\bf Gamma matrices:}
$$\gamma_\mu=(\sigma_1,\,\sigma_2),\ \ \ \sigma_{1,2}\ \ {\rm are\ \  Pauli\ \ matrices},\ \ {\rm for\ \ \mu=1,2}$$
$$\{\gamma_\mu,\,\gamma_\nu\}=2\delta_{\mu\nu},\ \ \ \gamma_3=-i\gamma_1\gamma_2=\sigma_3,\ \ \ \gamma_{\pm}\equiv\frac{1}{2}(1\pm\gamma_3)$$\\
\noindent{\bf Charge conjugation:}
$$\mc C\equiv\sigma_2={\left(
\begin{array}{cc}
 0 & -i \\
 i & 0 \\
\end{array}
\right)}=\mc C^\dagger=-\mc C^T\ \ $$
satisfying
$$\mc C^2=1,\ \ \ \mc C\gamma_i\mc C^{-1}=-\gamma_i^T,\ \ {\rm for\ }i=1,2,3$$\\
{\bf Spinors:}\\
Throughout the paper, we take Killing spinors 
$$\epsilon\equiv\left(
\begin{array}{c}
 \epsilon_+ \\
 \epsilon_- \\
\end{array}
\right)\,,\ \ \ 
\tilde\epsilon\equiv\left(
\begin{array}{c}
 \tilde{\epsilon }_+ \\
 \tilde\epsilon _- \\
\end{array}
\right)\,,$$
as two \emph{independent} $\mathbb C$-valued spinors. For convenience we also define
$$\zeta\equiv\gamma_+\epsilon+\gamma_-\tilde\epsilon=\left(
\begin{array}{c}
 \epsilon_+ \\
 \tilde\epsilon_- \\
\end{array}
\right)\,,\ \ \ 
\tilde\zeta\equiv\gamma_+\tilde\epsilon+\gamma_-\epsilon=\left(
\begin{array}{c}
 \tilde{\epsilon }_+ \\
 \epsilon _- \\
\end{array}
\right)\,.$$
For fermionic fields $\psi$ and $\bar\psi$,
$$\psi\equiv\left(
\begin{array}{c}
 \psi_+ \\
 \psi_- \\
\end{array}
\right)\,,\ \ \ 
\bar\psi\equiv\left(
\begin{array}{c}
 \bar\psi_+ \\
 \bar\psi_- \\
\end{array}
\right)\,,$$
they are considered as two independent two-component Grassmannian spinors and thus anticommuting. Given two spinors $\epsilon$ and $\psi$, the (Euclidean) Lorentz scalar is defined
$$\epsilon\cdot\psi\equiv\epsilon^T\mc C\psi\ \ {\rm or}\ \ \epsilon^\dagger\psi\,.$$\\
{\bf Supersymmetries on curved space:}\\
We first spell out the supersymmetries of (twisted) chiral multiplets on flat space.\footnote{The superalgebra quoted here is from \cite{Benini}. Curious readers who care about its relation with eq.\,(\ref{superalgebraG}) may consult the appendix of \cite{Gomis}.} For the chiral multiplet $\Phi=(\phi,\ \psi,\ \mc O)$ with dimension $\omega$,
\beqn
&&\delta\phi=\tilde{\epsilon}\cdot\psi\nonumber\\
&&\delta\psi=i\gamma^\mu\epsilon D_\mu\phi+\tilde{\epsilon}\mc O\nonumber\\
&&\delta \mc O=i\epsilon\cdot\gamma^\mu D_\mu\psi\nonumber
\eeqn
and the twisted chiral multiplet $\Sigma=(\tilde\sigma,\,\tilde\lambda,\,\widetilde{\mc O})$ with dimension $\tilde\omega$
\beqn
&&\delta\tilde\sigma=\zeta\cdot\tilde\lambda\,,\nonumber\\
&&\delta\tilde\lambda=i\gamma^\mu\tilde\zeta\,D_\mu\tilde\sigma+\zeta\,\widetilde{\mc O}\nonumber\\
&&\delta\widetilde{\mc O}=i\tilde\zeta\cdot\gamma^\mu D_\mu\tilde\lambda\,,\nonumber
\eeqn
where $D_\mu$ is ordinary partial derivative so far. To place the fields on curved space, $\epsilon$ and $\tilde\epsilon$ are spacetime dependent. The above transformations are also required Weyl covariant, see also \cite{Gomis}. Therefore they receive compensations respect to their dimensions,
\beqn
&&\delta\phi=\tilde{\epsilon}\cdot\psi\nonumber\\
&&\delta\psi=i\gamma^\mu\epsilon D_\mu\phi+\tilde{\epsilon}\mc O+i\omega\gamma^\mu D_\mu\epsilon\,\phi\nonumber\\
&&\delta \mc O=i\epsilon\cdot\gamma^\mu D_\mu\psi+i\omega D_\mu\epsilon\cdot\gamma^\mu\psi\nonumber
\eeqn
and
\beqn
&&\delta\tilde\sigma=\zeta\cdot\tilde\lambda\,,\nonumber\\
&&\delta\tilde\lambda=i\gamma^\mu\tilde\zeta\,D_\mu\tilde\sigma+\zeta\,\widetilde{\mc O}+i\tilde\omega\gamma^\mu D_\mu\tilde\zeta\,\tilde\sigma\nonumber\\
&&\delta\widetilde{\mc O}=i\tilde\zeta\cdot\gamma^\mu D_\mu\tilde\lambda+i\tilde\omega D_\mu\tilde\zeta\cdot\gamma^\mu \tilde\lambda\,,\nonumber
\eeqn
where $D_\mu$ is necessary to be improved as covariant derivative respect to the spin connections of curved space.\\

\noindent{\bf Killing spinor equations:}\\
On $S^2$ with radius $R$, there are two different ways to put \ntt\ supersymmetries, $\mathfrak{su}(2|1)_A$ and $\mathfrak{su}(2|1)_B$. They can be realized by imposing different Killing equations on spinors $\epsilon$ and $\tilde\epsilon$. For $\mathfrak{su}(2|1)_A$,
$$D_\mu\epsilon=\frac{i}{2R}\gamma_\mu\epsilon\,,\ \ \ \ D_\mu\tilde\epsilon=\frac{i}{2R}\gamma_\mu\tilde\epsilon\,,$$
and for $\mathfrak{su}(2|1)_B$, we require
$$D_\mu\epsilon=\frac{i}{2R}\gamma_\mu\tilde\epsilon\,,\ \ \ \ D_\mu\tilde\epsilon=\frac{i}{2R}\gamma_\mu\epsilon\,,$$
or in terms of $\zeta$ and $\tilde\zeta$
$$D_\mu\zeta=\frac{i}{2R}\gamma_\mu\tilde\zeta\,,\ \ \ \ D_\mu\tilde\zeta=\frac{i}{2R}\gamma_\mu\zeta\,.$$
Applying the above equation, we can obtain $\mathfrak{su}(2|1)_A$ and $\mathfrak{su}(2|1)_B$ superalgebra of (twisted) chiral fields respectively, i.e. eq\,.(\ref{sttc},\, \ref{stc})\\

\section{Irrelevant deformations in GLSM with $U(N)$ gauge group}
\label{B}
In this section, we show how to introduce irrelevant deformations for twisted chiral rings, or say a type of generic twisted superpotentials, in GLSM with $U(N)$ gauge group, see also section \ref{3.2}.

We first recall the \ntt\ SUSY of vector multiplet $V=(A_\mu,\ \sigma_1,\ \sigma_2,\ \lambda,\ \bar{\lambda},\ D)$ 
on flat space, 
\beqn
\delta A_\mu\=-\frac{i}{2}\,\tilde{\epsilon}\cdot\gamma_\mu\lambda-\frac{i}{2}\,\epsilon\cdot\gamma_\mu\bar{\lambda}\nonumber\\
\delta\sigma_1\=\frac{1}{2}\,\tilde{\epsilon}\cdot\lambda-\frac{1}{2}\,\epsilon\cdot\bar{\lambda},\ \ \ \ \delta\sigma_2=-\frac{i}{2}\,\tilde{\epsilon}\cdot\gamma_3\lambda-\frac{i}{2}\,\epsilon\cdot\gamma_3\bar{\lambda}\nonumber\\
\delta\lambda\=i\gamma_3\epsilon(G+i[\sigma_1,\,\sigma_2])-\epsilon D+i\gamma^\mu\epsilon D_\mu\sigma_1-\gamma_3\gamma^\mu\epsilon D_\mu\sigma_2\nonumber\\
\delta\bar{\lambda}\=i\gamma_3\tilde{\epsilon}(G-i[\sigma_1,\,\sigma_2])+\bar{\epsilon} D-i\gamma^\mu\tilde{\epsilon}D_\mu\sigma_1-\gamma_3\gamma^\mu\tilde{\epsilon}D_\mu\sigma_2\nonumber\\
\delta D\=-\frac{i}{2}\,\tilde{\epsilon}\cdot\gamma^\mu D_\mu\lambda+\frac{i}{2}\,\epsilon\cdot\gamma^\mu D_\mu\bar{\lambda}+\frac{i}{2}[\tilde{\epsilon}\cdot\lambda,\, \sigma_1]+\frac{i}{2}[\epsilon\cdot\bar{\lambda},\, \sigma_1]\nonumber\\
\+\frac{1}{2}[\tilde{\epsilon}\cdot\gamma_3\lambda,\, \eta]-\frac{1}{2}[\epsilon\cdot\gamma_3\bar{\lambda},\, \sigma_2]\nonumber
\eeqn
where
$$G\equiv\frac{1}{2}\epsilon^{\mu\nu}(\partial_\mu A_\nu-\partial_\nu A_\mu-i[A_\mu,\ A_\nu])\,,$$
is the $2d$ field strength. In the case of $U(1)$ vector multiplet $V$, one can define a twisted chiral multiplet $\Sigma=(\sigma,\,\tilde\lambda,\,\widetilde{\mc O})$,
\beqn
&&\sigma\equiv\sigma_1+i\sigma_2\nonumber\\
&&\tilde\lambda\equiv\left(
\begin{array}{c}
 \lambda _+ \\
 -\bar{\lambda }_- \\
\end{array}
\right)\nonumber\\
&&\widetilde{\mc O}\equiv -D+iG\,,\nonumber
\eeqn
$\Sigma$ serving as the superfield strength of $V$. Its superalgebra can be found from that of vector multiplet,
\beqn
&&\delta\sigma=\zeta\cdot\tilde\lambda\,,\nonumber\\
&&\delta\tilde\lambda=i\gamma^\mu\tilde\zeta\,D_\mu\sigma+\zeta\,\widetilde{\mc O}\nonumber\\
&&\delta\widetilde{\mc O}=i\tilde\zeta\cdot\gamma^\mu D_\mu\tilde\lambda\,.\nonumber
\eeqn
For non-Abelian vector multiplet, one can find
\beqn
&&\delta\sigma=\zeta\cdot\tilde\lambda\,,\nonumber\\
&&\delta\tilde\lambda=i\gamma^\mu\tilde\zeta\,D_\mu\sigma+\zeta\,\widetilde{\mc O}-\frac{i}{2}\gamma_3\zeta[\sigma,\,\bar{\sigma}]\nonumber\\
&&\delta\widetilde{\mc O}=i\tilde\zeta\cdot\gamma^\mu D_\mu\tilde\lambda-\frac{i}{2}[\tilde\zeta\cdot\gamma_3\bar{\tilde\lambda},\,\sigma]-\frac{i}{2}[\zeta\cdot\gamma_3\tilde\lambda,\,\bar{\sigma}]\,.\nonumber
\eeqn
where $\bar{\sigma}\equiv\sigma_1-i\sigma_2$ and $\bar{\tilde\lambda}\equiv(\bar\lambda_+,\,-\lambda_-)^T$. The superalgebra of $\Sigma=(\sigma,\,\tilde\lambda,\,\widetilde{\mc O})$ does \emph{not} close anymore. In addition, we actually want gauge invariant twisted chiral multiplet. We therefore consider taking trace of above multiplet $\Sigma$. Indeed $\Tr\left(\Sigma\right)$ is a gauge invariant twisted chiral multiplet. If our gauge group contains $U(1)$ factors, it reproduce the result of Abelian case. For non-Abelian gauge group, one can consider further $\frac{1}{2}\Tr\left(\Sigma^2\right)$, $\left(\Tr(\Sigma)\right)^2$,..., etc.. One can check by above algebra that they are all gauge invariant twisted chiral operators. Here $\frac{1}{2}\Tr\left(\Sigma^2\right)$, for example, has components,
$$\frac{1}{2}\Tr\left(\Sigma^2\right)=\left(\,\frac{1}{2}\Tr(\sigma^2),\,\Tr(\tilde\lambda\sigma),\,\Tr(\widetilde{\mc O}\sigma-\frac{1}{2}\tilde\lambda\cdot\tilde\lambda)\,\right)\,,$$
whose superalgebra,
\beqn
&&\delta\left(\frac{1}{2}\Tr\left(\sigma^2\right)\right)=\zeta\cdot\Tr\left(\tilde\lambda\sigma\right)\,,\nonumber\\
&&\delta\left(\Tr\left(\tilde\lambda\sigma\right)\right)=i\gamma^\mu\tilde\zeta\,D_\mu\left(\frac{1}{2}\Tr\left(\sigma^2\right)\right)+\zeta\left(\Tr\left(\widetilde{\mc O}-\frac{1}{2}\tilde\lambda\cdot\tilde\lambda\right)\,\right)\nonumber\\
&&\delta\left(\Tr\left(\widetilde{\mc O}\sigma-\frac{1}{2}\tilde\lambda\cdot\tilde\lambda\right)\,\right)=i\tilde\zeta\cdot\gamma^\mu D_\mu\left(\Tr\left(\tilde\lambda\tilde\sigma\right)\right)\,,\nonumber
\eeqn
shows it indeed a \ntt\ twisted chiral multiplet.
In the case of GLSM with $U(N)$ gauge group, the set $\big\{\Tr(\Sigma),\,\Tr(\Sigma^2),\,\Tr(\Sigma^3),\,...,\,\Tr(\Sigma^N)\big\}$ exhausts all primitive generators of the twisted chiral ring. One can further put them all on $S^2$ with respect to their dimensions as discussed in appendix \ref{A}, see also section \ref{3.2}.\\

\section{One-instanton correction to twisted chiral ring data}
\label{C}
In this appendix, we collect the non-perturbative contribution, up to one-instanton, of twisted chiral ring data $G^{(\alpha)}$ in GN-$CY_3$ in case of need. The full perturbative expression of $G^{(\alpha)}$ are given in eq.\,(\ref{GNG1}), (\ref{GNG2}) and (\ref{GNG3}). Meanwhile, the one instanton corrections to $G^{(\alpha)}$ are expanded in terms of $\mc O(\xi_2)$ while fixing $\xi_1$,
\beq
G_{\rm 1-inst.}^{(1)}=\left(
\begin{array}{cc}
I^{(1)}_{1\bar1}\, & I^{(1)}_{1\bar2} \\\\
I^{(1)}_{2\bar1}\, &  I^{(1)}_{2\bar2} \\
\end{array}
\right)+\mc O(\xi_2)
\eeq
with
\beqn
I^{(1)}_{1\bar1}\=56 q_1-960 q_1 q_2+56 \bar q_1-960 \bar q_1 \bar q_2\nonumber\\
I^{(1)}_{1\bar2}\=\overline{I^{(1)}_{1\bar2}}=-384 q_1 q_2-960 \bar q_1\bar q_2\nonumber\\
I^{(1)}_{2\bar2}\=-384 q_1 q_2-384 \bar q_1 \bar q_2\,,
\eeqn

\beq
G^{(2)}_{\rm 1-inst.}=\xi_2\left(
\begin{array}{ccc}
 I^{(2)}_{11\bar 1\bar 1}\,, & I^{(2)}_{11\bar 1\bar 2}\,, & I^{(2)}_{11\bar 2\bar 2} \\\\
 I^{(2)}_{12\bar 1\bar 1}\,, & I^{(2)}_{12\bar 1\bar 2}\,, & I^{(2)}_{12\bar 2\bar 2} \\\\
 I^{(2)}_{22\bar 1\bar 1}\,, & I^{(2)}_{22\bar 1\bar 2}\,, & I^{(2)}_{22\bar 2\bar 2} \\
\end{array}
\right)+\mc O(\xi_2^2)\,,
\eeq
with
\beqn
I^{(2)}_{11\bar1\bar1}\=-\frac{280}{3}q_1+1280q_1q_2+\frac{784}{3}q_1\bar q_1-896q_1q_2\bar q_1+6144q_1q_2\bar q_1\bar q_2+{\rm c.c.}\nonumber\\
I^{(2)}_{11\bar1\bar2}\=\overline{I^{(2)}_{12\bar1\bar1}}=-56q_1+960q_1q_2+1280\bar q_1\bar q_2-896q_1\bar q_1\bar q_2+12288q_1q_2\bar q_1\bar q_2\nonumber\\
I^{(2)}_{11\bar2\bar2}\=\overline{I^{(2)}_{22\bar1\bar1}}=-384 q_1 q_2+1280 \bar q_1 \bar q_2-896q_1\bar q_1\bar q_2+12288q_1q_2\bar q_1\bar q_2\nonumber\\
I^{(2)}_{12\bar1\bar2}\=960q_1q_2+960\bar q_1\bar q_2+12288q_1q_2\bar q_1\bar q_2\nonumber\\
I^{(2)}_{12\bar 2\bar2}\=\overline{I^{(2)}_{22\bar 1\bar2}}=384q_1q_2+960\bar q_1\bar q_2+12288q_1q_2\bar q_1\bar q_2\nonumber\\
I^{(2)}_{22\bar 2\bar2}\=384q_1q_2+384\bar q_1\bar q_2+12288q_1q_2\bar q_1\bar q_2\,,
\eeqn
and
\beq
G^{(3)}_{\rm 1-inst.}=\xi^3_2\left(
\begin{array}{cccc}
 I^{(3)}_{111\bar1\bar1\bar1}\,, & I^{(3)}_{111\bar1\bar1\bar2}\,, &I^{(3)}_{111\bar1\bar2\bar2}\,, & I^{(3)}_{111\bar2\bar2\bar2} \\\\
 I^{(3)}_{112\bar1\bar1\bar1}\,, & I^{(3)}_{112\bar1\bar1\bar2}\,, &I^{(3)}_{112\bar1\bar2\bar2}\,, & I^{(3)}_{112\bar2\bar2\bar2} \\\\
 I^{(3)}_{122\bar1\bar1\bar1}\,, & I^{(3)}_{122\bar1\bar1\bar2}\,, &I^{(3)}_{122\bar1\bar2\bar2}\,, & I^{(3)}_{122\bar2\bar2\bar2} \\\\
I^{(3)}_{222\bar1\bar1\bar1}\,, & I^{(3)}_{222\bar1\bar1\bar2}\,, &I^{(3)}_{222\bar1\bar2\bar2}\,, & I^{(3)}_{222\bar2\bar2\bar2}\\
\end{array}
\right)+\mc O(\xi_2^4)\,,
\eeq
with
\beqn
I^{(3)}_{111\bar1\bar1\bar1}\=840q_1+2880q_1q_2+1176q_1\bar q_1+8064q_1q_2\bar q_1+13824q_1q_2\bar q_1\bar q_2+{\rm c.c.}\nonumber\\
I^{(3)}_{111\bar1\bar1\bar2}\=\overline{I^{(3)}_{112\bar1\bar1\bar1}}=840q_1+2880q_1q_2+2880\bar q_1\bar q_2+8064q_1\bar q_1\bar q_2+27648q_1q_2\bar q_1\bar q_2\nonumber\\
I^{(3)}_{111\bar1\bar2\bar2}\=\overline{I^{(3)}_{122\bar1\bar1\bar1}}=672q_1+2304q_1q_2+2880\bar q_1\bar q_2+8064q_1\bar q_1\bar q_2+27648q_1q_2\bar q_1\bar q_2\nonumber\\
I^{(3)}_{111\bar2\bar2\bar2}\=\overline{I^{(3)}_{222\bar1\bar1\bar1}}=336q_1+1152q_1q_2+2880\bar q_1\bar q_2+8064q_1\bar q_1\bar q_2+27648q_1q_2\bar q_1\bar q_2\nonumber\\
I^{(3)}_{112\bar1\bar1\bar2}\=2880 q_1 q_2+2880\bar q_1\bar q_2+27648q_1q_2\bar q_1\bar q_2\nonumber\\
I^{(3)}_{112\bar1\bar2\bar2}\=\overline{I^{(3)}_{122\bar1\bar1\bar2}}=2304 q_1 q_2+2880\bar q_1\bar q_2+27648q_1q_2\bar q_1\bar q_2\nonumber\\
I^{(3)}_{112\bar2\bar2\bar2}\=\overline{I^{(3)}_{222\bar1\bar1\bar2}}=1152 q_1 q_2+2880\bar q_1\bar q_2+27648q_1q_2\bar q_1\bar q_2\nonumber\\
I^{(3)}_{122\bar1\bar2\bar2}\=2304 q_1 q_2+2304\bar q_1\bar q_2+27648q_1q_2\bar q_1\bar q_2\nonumber\\
I^{(3)}_{122\bar2\bar2\bar2}\=\overline{I^{(3)}_{222\bar1\bar2\bar2}}=1152 q_1 q_2+2304\bar q_1\bar q_2+27648q_1q_2\bar q_1\bar q_2\nonumber\\
I^{(3)}_{222\bar2\bar2\bar2}\=1152 q_1 q_2+1152\bar q_1\bar q_2+27648q_1q_2\bar q_1\bar q_2\,.
\eeqn
Together with the perturbative part of $G^{(3)}$ from eq.\,(\ref{GNG3}), by use of eq.\,(\ref{constraints2}), one can obtain the topological correlators in the A-twist theory to the order of one-instanton, 
\beqn
\big\langle\chi^3_1\big\rangle_{S^2}(t_1,t_2)\=20+56q_1+192q_1q_2\,,\nonumber\\
\big\langle\chi^2_1\chi_2\big\rangle_{S^2}(t_1,t_2)\=20+192q_1q_2\,,\nonumber\\
\big\langle\chi_1\chi^2_2\big\rangle_{S^2}(t_1,t_2)\=16+192q_1q_2\nonumber\\
\big\langle\chi^3_2\big\rangle_{S^2}(t_1,t_2)\=8+192q_1q_2\,.
\label{c1}
\eeqn
As a simple consistency check, we can use mirror map to obtain Yukawa couplings in $\tau$-coordinates \cite{Candelas}, in a similar fashion of eq.\,(\ref{taut})\,,
\beq
\mc C^{(3)}_{ijk}(\tilde\tau)=\left(\frac{{\rm d}t_{l}}{{\rm d}\tilde\tau_i}\right)\left(\frac{{\rm d}t_{m}}{{\rm d}\tilde\tau_j}\right)\left(\frac{{\rm d}t_{n}}{{\rm d}\tilde\tau_k}\right)\mc C^{(3)}_{lmn}\left(t(\tilde\tau)\right)\,,\ \ \ {\rm with}\ \ \ i,j,k,l,m,n=1,2\,.
\label{Yukawa}
\eeq
Applying eq.\,(\ref{Yukawa}) to eq.\,({\ref{c1}}), we arrive at 
\beqn
\big\langle\chi^3_1\big\rangle_{S^2}(\tilde\tau_1,\tilde\tau_2)\=20+296z_1-208z_1z_2\,,\nonumber\\
\big\langle\chi^2_1\chi_2\big\rangle_{S^2}(\tilde\tau_1,\tilde\tau_2)\=20+128z_1+80z_2+304z_1z_2\,,\nonumber\\
\big\langle\chi_1\chi^2_2\big\rangle_{S^2}(\tilde\tau_1,\tilde\tau_2)\=16+32z_1+160z_2-160z_1z_2\nonumber\\
\big\langle\chi^3_2\big\rangle_{S^2}(\tilde\tau_1,\tilde\tau_2)\=8+192z_2-768z_1z_2\,,
\eeqn
with $z_i={\rm e}^{2\pi i \tilde\tau_i}$. Since the chiral correlators are determined up to a K\"{a}hler transformation, one can verify that, accompanied by a transformation 
$$T(\tilde\tau_1,\tilde\tau_2)=1+2z_1-4z_2-28z_1z_2\,,$$
\beq
\mc C^{(3)}_{ijk}(\tilde\tau_1,\tilde\tau_2)\longrightarrow T(\tilde\tau_1,\tilde\tau_2)\, \mc C^{(3)}_{ijk}(\tilde\tau_1,\tilde\tau_2)
\eeq
reproduce the result of \cite{Park} to one-instanton order.\\\\

\end{appendix}


\end{document}